\documentclass[twocolumn,tighten,twocolappendix]{aastex7}
\usepackage{CJK}
\usepackage{enumitem}
\usepackage{amsmath}
\usepackage{hyperref}

\hypersetup{linkcolor=red,citecolor=blue,filecolor=cyan,urlcolor=blue}
\shorttitle{An $\eta$ Carinae-like Giant Eruption in SBS 0335-052 E}
\shortauthors{PENG et al.}
\begin{document}
\begin{CJK*}{UTF8}{gbsn}

% \title{When Stars Mimic Monsters: Luminous Blue Variables in SBS 0335$-$052 E}
\title{When Stars Mimic Monsters: Spectral Evidence for an $\eta$ Carinae-like Giant Eruption in SBS 0335$-$052 E}

\author[orcid=0000-0003-3467-6810]{Zixuan Peng (彭子轩)}
\affiliation{Department of Physics, Broida Hall, University of California at Santa Barbara, Santa Barbara, CA 93106, USA}
\email[show]{zixuanpeng@ucsb.edu}  
\correspondingauthor{Zixuan Peng (彭子轩)}

\author[orcid=0000-0001-9189-7818]{Crystal L. Martin}
\affiliation{Department of Physics, Broida Hall, University of California at Santa Barbara, Santa Barbara, CA 93106, USA}
\email{cmartin@physics.ucsb.edu}

\author[orcid=0000-0002-5721-0709]{Jiamu Huang (黄嘉沐)}
\affiliation{Department of Physics, Broida Hall, University of California at Santa Barbara, Santa Barbara, CA 93106, USA}
\email{jiamu_huang@ucsb.edu}

\author[orcid=0000-0001-5847-7934]{Nikolaus Z.\ Prusinski}
\affiliation{Cahill Center for Astronomy and Astrophysics, California Institute of Technology, \\ 1216 E. California Boulevard, MC 249-17, Pasadena, CA 91125, USA}
\email{nik@astro.caltech.edu}

\author[orcid=0000-0001-9446-6853]{Chenliang Huang (黄辰亮)}
\affiliation{Shanghai Astronomical Observatory, Chinese Academy of Sciences, Shanghai 200030, People's Republic of China}
\email{huangcl@shao.ac.cn}

\author[orcid=0000-0002-1945-2299]{Zhuyun Zhuang}
\affiliation{Cahill Center for Astronomy and Astrophysics, California Institute of Technology, \\ 1216 E. California Boulevard, MC 249-17, Pasadena, CA 91125, USA}
\email{zzhuang@astro.caltech.edu}

\author[orcid=0000-0001-8237-1515]{Yuan Li (李远)}
% \author[orcid=0000-0003-3467-6810]{Zixuan Peng}
\affiliation{Department of Physics and Astronomy, Texas A\&M University, College Station, TX 77843-4242, USA}
\email{yuanli@tamu.edu}

\author[orcid=0000-0001-9195-7390]{Tin Long Sunny Wong}
\affiliation{Department of Physics, Broida Hall, University of California at Santa Barbara, Santa Barbara, CA 93106, USA}
\affiliation{The Observatories of the Carnegie Institution for Science, Pasadena, CA 91101, USA}
\email{tinlongsunny@ucsb.edu}

\author[orcid=0000-0001-7593-9205]{Jiayang Yang}
\affiliation{Department of Physics, Broida Hall, University of California at Santa Barbara, Santa Barbara, CA 93106, USA}
\email{jiayangyang@ucsb.edu}

\author[orcid=0000-0002-7054-4332]{Joseph F. Hennawi}
\affiliation{Department of Physics, Broida Hall, University of California at Santa Barbara, Santa Barbara, CA 93106, USA}
\affiliation{Leiden Observatory, Leiden University, Niels Bohrweg 2, NL-2333 CA Leiden, the Netherlands}
\email{joe@physics.ucsb.edu}

% \author[]{ET AL}
% \affiliation{}
% \email{}

% \author[orcid=0000-0000-0000-0002,gname=Bosque, sname='Sur America']{Forrest Sur Am\'{e}rica} 
% \altaffiliation{Las Campanas Observatory}
% \affiliation{Universidad de Chile, Department of Astronomy}
% \email{fakeemail2@google.com}

% \author[gname=Savannah,sname=Africa]{S. Africa}
% \affiliation{South African Astronomical Observatory}
% \affiliation{University of Cape Town, Department of Astronomy}
% \email{fakeemail3@google.com}

% \author{River Europe}
% \affiliation{University of Heidelberg}
% \email{fakeemail4@google.com}

% \author[0000-0000-0000-0003,sname=Asia,gname=Mountain]{Asia Mountain}
% \altaffiliation{Astrosat Post-Doctoral Fellow}
% \affiliation{Tata Institute of Fundamental Research, Department of Astronomy}
% \email{fakeemail5@google.com}

% \author[0000-0000-0000-0004]{Coral Australia}
% \affiliation{James Cook University, Department of Physics}
% \email{fakeemail6@google.com}

% \author[gname=IceSheet]{Penguin Antarctica}
% \affiliation{Amundsen–Scott South Pole Station}
% \email{fakeemail7@google.com}

% \collaboration{all}{The Terra Mater collaboration}


%%%%% revised version 3 %%%%%
\begin{abstract}
SBS\,0335$-$052\,E is an extremely low-metallicity ($Z\sim0.04\,Z_{\odot}$) blue compact dwarf galaxy. An active galactic nucleus has been proposed to explain the broad H$\alpha$ emission and near-infrared (NIR) time variability in super star clusters 1 and 2 (SSCs 1\&2). 
% , a region with very-high-ionization (e.g, [Fe\,{\footnotesize V}]) and ultra-luminous X-ray (ULX) emission.
However, Peng et al. discovered broad wings in the forbidden [O\,{\footnotesize III}]\,$\lambda5007$ emission (up to $\sim5\,000\,\rm{km\,s^{-1}}$), challenging the broad-line region interpretation.
%This AGN interpretation is challenged given the detections of broad wings in forbidden [O\,{\footnotesize III}] $\lambda %5007$ emission (up to $\sim 5\,000\ \rm{km\,s^{-1}}$).
% To better characterize the outflow properties and the ionizing source, we present new KCWI/KCRM integral-field spectroscopy of this region covering $\sim 3600-9500\, \rm{\AA}$.
% ($\sim 3600-9500\, \rm{\AA}$)
We present new KCWI/KCRM integral-field spectroscopy to directly compare spectra across multiple SSCs. 
The nebula surrounding SSCs 1\&2 shows unique features. 
The Ly$\beta$-pumped O\,{\footnotesize I}\,$\lambda8446$ emission constrains $\tau_{\rm\,Ly\alpha}\sim10^8$. 
% the Ly$\alpha$ optical depth to
Multiple ionization states of iron are detected from Fe$^{+}$ to Fe$^{+4}$.
%The elevated [Fe\,{\footnotesize II}]\,$\lambda5262$/[Fe\,{\footnotesize II}]\,$\lambda8617$ ratio indicates dense gas 
Stellar photoionization models 
can reproduce the [Fe\,{\footnotesize III}]/[Fe\,{\footnotesize II}] and [Fe\,{\footnotesize IV}]/[Fe\,{\footnotesize III}] line ratios at high density ($n_e\sim10^6\,\rm{cm^{-3}}$), but they fail to account for most of the [Fe\,{\footnotesize V}] emission.
% [Fe\,{\footnotesize IV}] and 
% at $\sim10^{16}$\,cm.
The broad H$\alpha$ wings exhibit an exponential profile; 
the asymmetric wings extend from $\sim-5\,000\,\rm{km\,s^{-1}}$ to $\sim10\,000\,\rm{km\,s^{-1}}$.
Thomson scattering in a radially expanding medium provides a good fit with
$v_w\sim200\,\rm{km\,s^{-1}}$, optical depth $\tau_e\sim10$, and an outer to inner radius of 10.
%medium of size $\sim10^{17}$\,cm.
%SSCs 1\&2 are also nitrogen-enriched by $\gtrsim0.1$ dex relative to the other SSCs.
% Along with a high [Fe\,{\footnotesize II}] $\lambda5262$/[Fe\,{\footnotesize II}] $\lambda8617$ ratio, these features trace a high-density medium; statistical equilibrium requires densities of at least $n_e \sim 10^6$ $\rm{cm^{-3}}$ to reproduce this ratio. 
% The detected O\,{\footnotesize I} $\lambda8446$ pumped emission and elevated [Fe\,{\footnotesize II}] $\lambda5262$/[Fe\,{\footnotesize II}] $\lambda8617$ ratio trace high-density gas ($n_e \sim 10^6$ $\rm{cm^{-3}}$) in SSCs 1\&2,
% which are also nitrogen-enriched by $\gtrsim 0.1$ dex relative to other SSCs. 
% We also detect [Fe\,{\footnotesize V}] emission, which requires a high-energy radiation field with $\gtrsim55$ eV photons. 
% detected broad wings in the forbidden [O\,{\footnotesize III}] $\lambda 5007$ emission (up to $\sim 5\,000\ \rm{km\,s^{-1}}$) rule out the AGN interpretation.
% These new measurements rule out a Type I AGN, while a Type II AGN interpretation remains awkward. 
% The Type-1 AGN interpretation remains questionable given the presence of broad wings in the forbidden [O\,{\footnotesize III}] emission. 
 % formed by pre-eruption stellar winds
Enhanced N/O and potentially depleted Fe/O ratios are consistent with CNO-cycled ejecta from massive stars and with dust formation, respectively.
We propose that mass loss from a massive star interacting with its circumstellar medium drives a shock that powers the NIR variability, the luminous X-ray point source, and the [Fe\,{\footnotesize V}] emission.
%In this picture, dense gas and warm dust reside in a partially ionized region of the CSM, accounting for the NIR variability.
%while nitrogen enrichment arises from ejections of CNO-cycled material.
%The asymmetric broad H$\alpha$ wings result from electron scattering within the expanding, optically thick ($\tau_e \sim 10$) CSM. 
%To produce the luminosities of [Fe\,{\footnotesize V}] and the luminous X-ray point source, the shock interaction requires a giant eruption similar to $\eta$\,Carinae.
%while [Fe\,{\footnotesize II}] and [Fe\,{\footnotesize IV}] emission may arise from gas photoionized by a binary system.
% The giant eruption in $\eta$ Carinae provides a plausible picture, wherein [Fe\,{\footnotesize II}] and [Fe\,{\footnotesize IV}] emission may arise from gas photoionized by a binary system. 
% High-cadence observations can further test the binary interpretation by identifying periodic [Fe\,{\footnotesize IV}] emission and distinguishing it from AGN stochasticity.
% This CSM hosts abundant warm dust that accounts for the corresponding NIR excess.
% This interpretation can be further tested by monitoring variability patterns to distinguish them from AGN stochasticity or by searching for evidence of binarity.
% Our results highlight how shock interaction between a giant eruption and a dense CSM can mimic signatures commonly attributed to an AGN in low-metallicity dwarf galaxies.
% If this picture is correct, continued monitoring of this region should reveal the binary orbital period. 
If confirmed, the proposed stellar eruption would be a distant example of an $\eta$ Carinae-like giant eruption, and the 
first in an ultra-low metallicity environment.
% second most distant known example, surpassed only by Godzilla in the Sunburst Arc.
%Our results highlight how the combination of high-density indicators, NIR variability, and electron-scattering wings can reveal one of the most distant giant eruptions in a low-metallicity galaxy.
\end{abstract}

% \end{CJK*}
\section{Introduction} 

Unambiguously identifying active galactic nucleus (AGN) activity in low-metallicity dwarf galaxies is highly challenging.
% $-$regarded as key laboratories for understanding the early phases of star and black hole formation$-$
Common diagnostics include detecting very-broad (VB) wings (FWHM $\gtrsim 1\,000\,\rm km\,s^{-1}$) in permitted lines such as H$\alpha$ \citep[e.g.,][]{Harikane_2023, Salehirad_2025, Juodzbalis_2025},
very-high-ionization rest-UV and optical emission lines (e.g., [Ne\,{\footnotesize V}]; \citealp{Chisholm_2024, Hernandez_2025}), and
stochastic photometric variability across a wide wavelength range, from X-ray to optical/UV and infrared emission, on timescales of days to years \citep[e.g.,][]{Ulrich_1997, Shappee_2014}.

However, these diagnostics are not definitive indicators of AGN activity and can also arise from non-AGN processes.
For example, any corresponding VB component in forbidden transitions such as the [O\,{\footnotesize III}] $\lambda\lambda 4959,5007$ doublet may be too faint to detect (typically $\lesssim 10\%$ of the total line flux; \citealp[e.g.,][]{Peng_2025}).
This can misclassify purely star-forming galaxies with high-velocity outflows as broad-line AGNs.
% , particularly when the [O\,{\footnotesize III}] signal-to-noise ratio (S/N) is insufficient to detect a VB wing (often demanding S/N $\gtrsim 200$; \citealp{Peng_2025}).
Moreover, hard stellar radiation fields may mimic the ionizing signatures of AGNs in metal-poor environments \citep[e.g.,][]{Richardson_2025},
and similar multi-wavelength time variability can also arise from a $\eta$-Carinae like system with a giant eruption phase 
(i.e., a significant mass-loss episode ejecting $\sim 10-40\, M_{\odot}$ over $10-40$ years; \citealp{DH_2012, Davidson_2020, WB_2020}).

\defcitealias{Hatano_2026}{H26}
Nearby low-metallicity dwarf galaxies with high signal-to-noise ratio, spatially-resolved spectroscopy and multi-wavelength observations provide key laboratories for offering critical insight into disentangling extreme star formation from AGN activity.
In this study, we focus on a local ($z = 0.0135$) blue compact dwarf galaxy \object{SBS 0335-052 E} (J0337-0502 hereafter), which has a low stellar mass ($M_{\ast} \sim 10^7\,M_{\odot}$), exceptionally low gas-phase metallicity ($Z \sim 0.03 - 0.05\,Z_{\odot}$), and high specific star formation rate ($\log\,{\rm sSFR}\,(\rm yr^{-1}) \sim -7.40$; see Table \ref{table:j0337_global_properties} for a summary of its main global properties).

\setlength{\tabcolsep}{6pt}
\begin{deluxetable}{lccccccccc}
\tablenum{1}
\tablecaption{Global Physical Properties of J0337$-$0502 Collected from the Literature}
\tablehead{
\colhead{$z$} 
& \colhead{$\log(M_\ast)$}
& \colhead{$\log(\rm{SFR})$}
& \colhead{$\log(\rm{sSFR})$}
& \colhead{$\log(\dot{\Sigma}_{\ast})$}
& \colhead{12 + $\log(\rm{O/H})$}
& \colhead{$E(B-V)$}
& \colhead{$\log(L_{\rm bol})$}
& \colhead{$R_{\rm core}$}
\\
\colhead{} 
& \colhead{($M_\odot$)}
& \colhead{($M_\odot\,\rm{yr^{-1}}$)}
& \colhead{($\rm{yr^{-1}}$)}
& \colhead{($M_\odot\,\rm{yr^{-1}\,kpc^{-2}}$)}
& \colhead{}
& \colhead{}
& \colhead{($\rm{erg\,s^{-1}}$)}
& \colhead{($\rm{kpc}$)}
}
\startdata
0.01352 
& 7.06$_{-0.21}^{+0.24}$ 
& -0.32$_{-0.11}^{+0.07}$ 
& -7.37$_{-0.22}^{+0.26}$ 
& -0.42$_{-0.84}^{+0.43}$ 
& $7.46\pm0.04$ 
& $0.053\pm0.006$
& $43.08$\tablenotemark{a}
& $\sim1.40$\tablenotemark{b}
\enddata
\tablenotetext{a}{
% \textbf{
This value is derived from the best-fit \texttt{CIGALE} SED model \citepalias{Hatano_2026} and is roughly consistent with the bolometric luminosity estimated from the total H$\alpha$ emission of all SSCs \citep[i.e., $L_{\rm bol} \sim (1-2)\times10^2 \, L_{\rm H\alpha} \sim (4.5-9.0)\times10^{42}\,\rm{erg \, s^{-1}}$;][]{Kennicutt_2012, Calzetti_2013}, where the H$\alpha$ luminosities of the individual SSCs are summarized in Table \ref{table:ssc_properties}.
% }
}
\tablenotetext{b}{
% \textbf{
This value refers to the radius of the high-SB ``core'' region, within which the H$\alpha$ surface brightness drops rapidly from $\mathrm{SB}_{\mathrm{H}\alpha} \sim 2\times10^{-14}$ to $2\times10^{-16}\,\rm erg\,s^{-1}\,cm^{-2}\,arcsec^{-2}$ over the first $5\arcsec$ \citep[dominated by the central SSCs;][]{Herenz_2023}.
% }
}
\tablecomments{Compiled from Tables 5 and 6 in \cite{Berg_2022}, Table A2 in \cite{Xu_2022}, and Section 4.1.1 in \citetalias{Hatano_2026}. For reference, the solar oxygen abundance is 12 + $\log(\mathrm{O/H})_{\odot}=8.69$ \citep{Asplund_2021}.}
\label{table:j0337_global_properties}
\end{deluxetable}

% \defcitealias{Hatano_2026}{H26}
% \citet[][hereafter H26]{Hatano_2026} recently propose the presence of an active massive BH in a local ($z = 0.0135$) blue compact dwarf galaxy \object{SBS 0335-052 E} (J0337-0502 hereafter), which has a low stellar mass ($M_{\ast} \sim 10^7\,M_{\odot}$), exceptionally low gas-phase metallicity ($Z \sim 0.03 - 0.05\,Z_{\odot}$), and high specific star formation rate ($\log\,{\rm sSFR}\,(\rm yr^{-1}) \sim -7.40$; see Table \ref{table:j0337_global_properties} for a summary of its main global properties).
% Their argument is primarily based on the observed near-infrared (NIR) variability in the W1 (3.4 $\mu$m) and W2 (4.6 $\mu$m) bands (i.e., increased by $\Delta f_{\rm{W1}} \sim 0.1$ and $\Delta f_{\rm{W2}} \sim 0.7$ mJy, respectively, from $\sim$ 57800 to 59600 MJD) in the super star clusters 1 and 2 (SSCs 1\&2; see Figure \ref{fg:j0337_hst_plot}). 

\begin{figure*}[htb]
\includegraphics[width=0.99\linewidth]{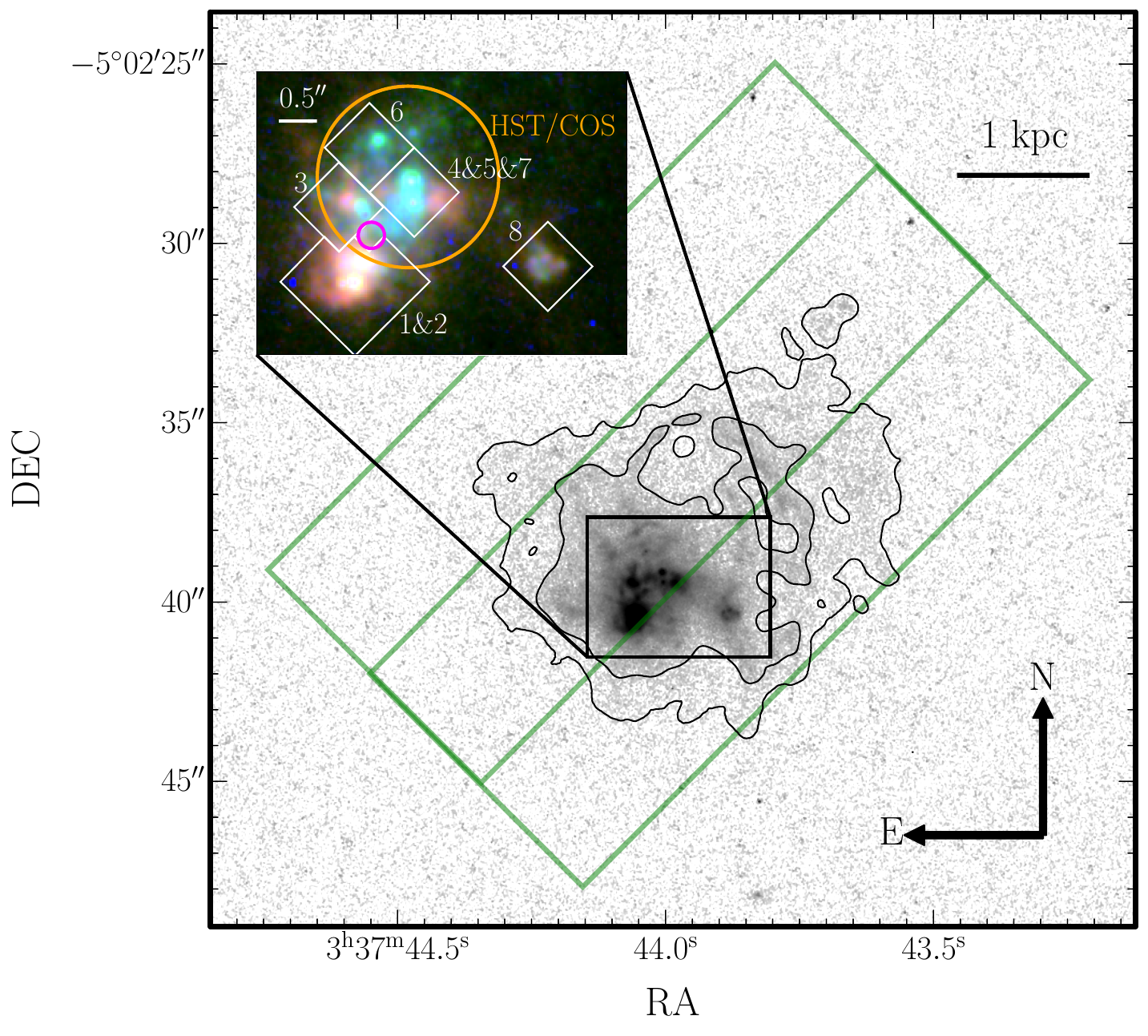}
\caption{The two KCWI/KCRM observation pointings (green rectangles) are overlaid on the \textit{HST} FR656N image of J0337-0502. 
% The third pointing (dashed), observed during the 12- and 18-degree twilight, is excluded from the analysis due to rapid sky background changes and infrequent sky chopping, which prevent valid sky subtraction.
% The dashed pointing is not used in this analysis. 
% because it was observed during the 12- and 18-degree twilight periods, resulting in suboptimal sky background subtraction (Section \ref{sec:observation_and_data}). 
The zoom-in panel shows the color-composite image (purple: F220W; blue: F330W; green: F435W; red: FR656N), which indicates the positions of the SSCs, including the luminous X-ray point source \citep[magenta circle;][]{Thuan_2004}, the apertures used to extract the 1D spectra, and the $2\farcs5$ \textit{HST}/COS circular aperture. 
The black solid contours demarcate H$\alpha$ isophotes at SB$_{\rm{H}\alpha}$ $\simeq$ \{2.5, 5.0\} $\times 10^{-16} \ \rm{erg \ s^{-1} \ cm^{-2} \ arcsec^{-2}}$.
% The key physical properties of this galaxy are summarized in the lower right corner based on \citet{Berg_2022} and this work. 
} 
\label{fg:j0337_hst_plot}
\end{figure*}

\citet[][hereafter H26]{Hatano_2026} recently propose the presence of an active massive BH in its super star clusters 1 and 2 (SSCs 1\&2; see Figure \ref{fg:j0337_hst_plot}), primarily based on the observed VB H$\alpha$ component (extending to $\sim10\,000\ \mathrm{km\,s^{-1}}$) and the near-infrared (NIR) variability in the W1 (3.4 $\mu$m) and W2 (4.6 $\mu$m) bands (i.e., increased by $\Delta f_{\rm{W1}} \sim 0.1$ and $\Delta f_{\rm{W2}} \sim 0.7$ mJy, respectively, from $\sim$ 57800 to 59600 MJD).

This region also appears as a point source in \textit{JWST}/MIRI \citep{Mingozzi_2025} and NICMOS 1.6$\mu$m \citep{Thompson_2009} observations, with detections of very-high-ionization lines such as He\,{\footnotesize II} $\lambda4686$ \citep{Kehrig_2018}. 
It uniquely exhibits dust, H$_2$, and thermal radio emission \citep{Johnson_2009} that are not detected in the other SSCs. 
The luminous X-ray point source ($\sim 3\times10^{39}\,\rm{erg\,s^{-1}}$ in the $0.5-10$ keV band; \citealp{Thuan_2004}) may also originate from this region. Although the X-ray source is located $\sim0\farcs3 - 0\farcs7$ north of SSCs 1\&2 \citep[magenta circle in Figure \ref{fg:j0337_hst_plot};][]{Prestwich_2013}, previous studies suggest that 
% it may be associated with SSC 2 \citep{Thuan_2004, Kehrig_2018}, as 
this position offset can be attributed to the $\sim0\farcs42$ rms astrometric uncertainty \citep{Thuan_2004, Kehrig_2018}.

Nevertheless, the existence of an AGN remains questionable for the following reasons.
% \begin{enumerate}[leftmargin=*] 
% \item 
% and \cite{Mingozzi_2025}
First, \cite{Peng_2025} report a VB component ($v_{\rm max}\sim5\,000~\rm{km~s^{-1}}$) in the forbidden [O\,{\footnotesize III}] $\lambda 5007$ emission in SSCs 1\&2. 
% (see Appendix \ref{sec:esi_ssc1a2_o3_ha} for the observed broad wings in these two lines from the ESI spectrum of super star clusters 1 and 2 (SSCs 1\&2)).
% , with a FWHM of $\sim 1500 \ \rm{km \ s^{-1}}$.
This finding is inconsistent with a Type 1 AGN interpretation, because the broad-line region (BLR) in such systems typically has electron densities of $n_e \gtrsim 10^8 \ \rm{cm^{-3}}$, which would strongly collisionally de-excite the forbidden [O\,{\footnotesize III}] $\lambda 5007$ emission.
% \item 
The observed $v_{\rm max}$ also exceeds those typically measured in local Type 2 AGNs \citep[$\lesssim 2\,000\,\rm{km\,s^{-1}}$; e.g.,][]{Zakamska_2003, Harrison_2014, Woo_2016}.
Second, assuming the VB H$\alpha$ component originates in the BLR, \citetalias{Hatano_2026} estimate the BH mass to be $M_{\rm BH}\sim10^{6}-10^{8}\,M_{\odot}$. 
This estimate exceeds the value predicted by the local black-hole-mass-stellar-mass ($M_{\rm{BH}}-M_{\ast}$) relation by $\sim4-6$  dex \citep{Reines_2015} at the stellar mass of SSCs 1\&2 \citep[$M_{\ast} \sim 10^6 \ M_{\odot}$;][]{Reines_2008}.
% , where \citetalias{Hatano_2026} identify the VB H$\alpha$ component. 
Furthermore, it is $\gtrsim 2$ dex above the high-redshift $M_{\rm{BH}}-M_{\ast}$ relation derived for AGN and ``little red dots'' at $z \sim 2 - 9$ \citep{Pacucci_2023, II_2024, Juodzbalis_2025}.
% \item Chandra observations indicate spatially extended X-ray emission \citep[e.g.,][]{Thuan_2004}, rather than the point-source distribution typically seen in Type-1 AGNs, supporting instead a starburst-driven superbubble scenario \citep[e.g.,][]{Herenz_2023}.
% \end{enumerate}

Therefore, alternative physical mechanisms are required to account for the extreme broad wings, which are difficult to reconcile with an AGN framework.

Broad wings in strong emission lines can arise from galactic winds, in which the acceleration may be driven either by radiation pressure from central massive stars \citep[e.g.,][]{Murray_2005, Murray_2011} or by the mechanical energy injected by core-collapse supernovae \citep[CCSNe; e.g.,][]{Thompson_2016, Fielding_2022}, although such models typically predict maximum velocities of $v_{\rm max} \lesssim (1-2)\times10^{3}\ \rm km\,s^{-1}$.
% Nevertheless, the typical maximum velocities, $v_{\rm max}\lesssim1-2\times10^{3}\ \rm km\,s^{-1}$, predicted by either radiatively or mechanically driven galactic wind models cannot reproduce the extremely large $v_{\rm max}$ observed in H$\alpha$ and [O\,{\footnotesize III}] $\lambda5007$ emission of SSCs 1\&2.
% The observed broad-wing maximum velocities, $v_{\rm{max}}$, far exceed the typical $v_{\rm{max}}\sim1000-2000\ \rm{km\,s^{-1}}$ predicted by both single-phase \citep{Thompson_2016} and multiphase \citep{Fielding_2022} supernova-driven (i.e., driven by the mechanical energy of core-collapse supernovae) galactic wind models.

% \textbf{
% The observed broad wings may instead arise from electron scattering at high column density, rather than from bulk Doppler motions. 
% In this scenario, even though the thermal velocity of the electrons is only $v_{\rm th} \simeq 670\, (T_{e}/10^{4}\,\rm K)^{1/2}\ \rm km\,s^{-1}$, multiple scattering events can significantly broaden the emergent line profile. 
% For an electron-scattering optical depth $\tau_e$, the typical number of scatterings scales as $N\sim\tau_e^{2}$, leading to a characteristic broadening of the FWHM by a factor of $N^{1/2}\sim\tau_e$ \citep{CF_2017}. 
% A key observational signature of such an optically thick, electron-scattering outflow is an exponential form ($\propto e^{-|x|}$) of the broad-wing line profile \citep{Davidson_2020}.
% }

% \textbf{
The observed broad wings may also arise from Thomson scattering at high column density, rather than from bulk Doppler motions.
% A key observational signature of such an optically thick, electron-scattering outflow is an exponential form ($\propto e^{-|x|}$) of the broad-wing line profile \citep[see, e.g., ][]{Davidson_2020}.}
Such Thomson-scattering wings exhibit an exponential profile ($\propto e^{-|x|}$; \citealp{Davidson_2020}) and are frequently observed in the dense circumstellar medium (CSM) associated with
% of $\eta$ Carinae \citep[e.g.,][]{Davidson_1995, Hillier_2001} and, more generally, 
Type IIn supernovae \citep[SNe IIn; e.g., see references in][]{CF_2017}, which are characterized by strong shock interactions between the stellar/SN ejecta and a dense CSM.
SNe IIn may originate from either genuine supernovae \citep[e.g.,][]{Chugai_2001, Fransson_2002} or supernova imposters (i.e., the star survives) like the giant eruption phase of $\eta$ Carinae \citep[e.g.,][]{Davidson_1995, Hillier_2001}. 
Such giant eruptions are also often discussed in the context of super-Eddington outflows \citep[e.g.,][]{Quataert_2016}.
% and are observationally related to supernova impostors. 
% \citep[e.g.,][]{DH_2012, Davidson_2020}.
% Throughout this work, we adopt the term ``giant eruption'' as our primary terminology.
We note that although $\eta$ Carinae is frequently cited as a luminous blue variable (LBV) in the literature, it is significantly more luminous than typical LBVs, which are usually called S Doradus-type variables \citep{DH_2012, Davidson_2020}.
% Similar exponential wings have also been proposed to arise naturally from electron scattering in dense gas cocoons surrounding rapidly accreting supermassive black holes in LRDs \citep[e.g,][]{Rusakov_2006, Sneppen_2026}. 
% In contrast, electron scattering is not expected to be the dominant mechanism shaping emission-line profiles in a typical AGN in the local universe \citep{GK_2012, KZ_2013}.
% In the latter case, however, feedback from the central engine is expected to rapidly clear the dense cocoon, allowing the system to transition into a typical Type-1 AGN in the local universe; consequently, the electron-scattering wings should disappear.
% Previous studies of the giant eruption of $\eta$ Carinae \citep[e.g.,][]{Davidson_1995, Hillier_2001, Mehner_2015, Davidson_2020} have demonstrated that the high-velocity, asymmetric broad wings arise from Thomson scattering by free electrons. 
% Photons can acquire large Doppler shifts through multiple scattering events before escaping from the optically thick CSM. 
% In addition, the radial expansion of the CSM preferentially shifts photons toward longer wavelengths, producing the observed red excess in the broad wings \citep[][]{HC_2018, Davidson_2020}.
% }

% \textbf{
Moreover, in $\eta$ Carinae, the dense ``Homunculus'', a bipolar CSM formed primarily from the bulk ejecta of its giant eruption and responsible for the Thomson-scattering wings, absorbs most of the UV photons emitted by the luminous, hot star through dust and re-radiates this energy in the infrared \citep{Davidson_2001, HM_2012}.
% Moreover, previous studies \citep{Davidson_2001, HM_2012} argue that most of the UV photons emitted by the luminous, hot star for $\eta$ Carinae were absorbed by dust in the ``Homunculus'', the bipolar CSM formed primarily from the bulk ejecta of its giant eruption, and subsequently re-radiated in the infrared.
The continuous shock interaction between the giant-eruption ejecta and the dense CSM powered sustained infrared thermal emission light curves that evolved slowly over $\sim1840-1860$ \citep[][]{Smith_2018}.
% , analogous to the NIR excess plateau observed in SSCs 1\&2.
The Homunculus now exhibits a double-shell structure \citep[e.g.,][]{Ishibashi_2003}, in which most of the mass ($\gtrsim 10\, M_{\odot}$) resides in a denser, geometrically thin outer shell traced by H$_2$ emission (hosting cold dust at $\sim140$ K), while $\sim10\%$ of the mass is contained in a geometrically thicker, inner shell of partially ionized gas traced by [Fe\,{\footnotesize II}] emission (hosting warm dust at $\sim200$ K; \citealp[e.g.,][]{DS_2000, Smith_2006}). 
% During its giant eruption, the continuous interaction between the expanding stellar ejecta and the dense CSM powers sustained infrared thermal emission light curves that evolve slowly over several years (e.g., analogous to the NIR excess plateau observed in SSCs 1\&2).
% Moreover, SNe IIn may exhibit more warm dust emission than other SN type, owing to efficient post-shock dust formation either within the stellar/SN ejecta or in the cold dense shell (CDS) between the forward and reverse shocks \citep[e.g,][]{Fransson_2002, Smith_2009, Kokubo_2019}.
% For example, \cite{HM_2012} argue that most of the UV photons emitted by the extremely luminous, hot star for $\eta$ Carinae were absorbed by dust formed during its giant eruption 
% % from $\sim1840-1860$ 
% and subsequently re-radiated in the infrared.
% Alternatively, the dense CSM produced by intense pre-SN mass loss may already contain substantial amounts of pre-existing dust \citep[e.g.,][]{Fox_2011, Fransson_2014}. 
% In either case, the continuous interaction between the shock and CSM can power sustained infrared thermal emission light curves that evolve slowly over several years, analogous to the NIR excess plateau observed in SSCs 1\&2.
% efficient post-shock dust formation is a common outcome of strong CSM interaction \citep[e.g,][]{Smith_2009, Fransson_2014, Kokubo_2019}. 
% }

% \textbf{
In this work, we propose that an $\eta$ Carinae-like giant eruption 
% \citep[i.e., a significant mass loss episode with $\dot{M} \gtrsim 0.1 \, M_{\odot} \, \rm{yr^{-1}}$ lasting for $10-40$ years;][]{Davidson_2012} 
can naturally reproduce both the broad H$\alpha$ wings and the NIR excess plateau observed in SSCs 1\&2, motivated by their unusual spectral similarity to $\eta$ Carinae.
% Such giant eruptions are often discussed in the context of super-Eddington outflows and are observationally related to supernova impostors. 
% % \citep[e.g.,][]{DH_2012, Davidson_2020}.
% % Throughout this work, we adopt the term ``giant eruption'' as our primary terminology.
% We note that although $\eta$ Carinae is frequently cited as a LBV in the literature, it is significantly more luminous than typical LBVs, which are usually called S Doradus-type variables \citep{DH_2012, Davidson_2020}.
% }
% \textbf{
% In this work, we focus on the unusual spectral features of SSCs 1\&2 in J0337-0502 and propose an alternative physical mechanism to explain these features based on an $\eta$ Carinae-like giant-eruption (or giant-outburst) phase, i.e., a significant mass loss episode.
% }
We also summarize the nebular properties of other SSCs for comparison (see Table \ref{table:ssc_properties} for key properties of each SSC compiled from the literature). 
Detailed analysis of the large-scale outflow kinematics and mass, momentum, and energy outflow rates will be presented in future work.

This work is organized as follows. 
In Section \ref{sec:observation_and_data}, we briefly summarize the configuration setup of our KCWI/KCRM observation of J0337-0502 and outline the essential steps employed to reduce the final mosaic cubes. 
We then derive the nebular properties of each SSC in Section \ref{sec:nebular_prop}, which are critical for inferring the potential signatures of giant eruption discussed in Section \ref{sec:lbv_signs}. 
In Section \ref{sec:discussions}, we provide estimations of the CSM properties during the giant eruption phase, discuss the AGN possibilities, and assess the role of Ly$\alpha$ radiation pressure in SSCs 1\&2.
Finally, we summarize our results and the corresponding implications in Section \ref{sec:conclusion}.
Throughout this work, we adopt the solar abundance for each element from \citet{Asplund_2021} and assume a Flat $\Lambda$CDM cosmology with $\Omega_m = 0.3$, $\Omega_{\Lambda} = 0.7$, and $H_0 = 70 \ \rm{km \ s^{-1} \ Mpc^{-1}}$.

\begin{figure*}[htb]
\includegraphics[width=1\linewidth]{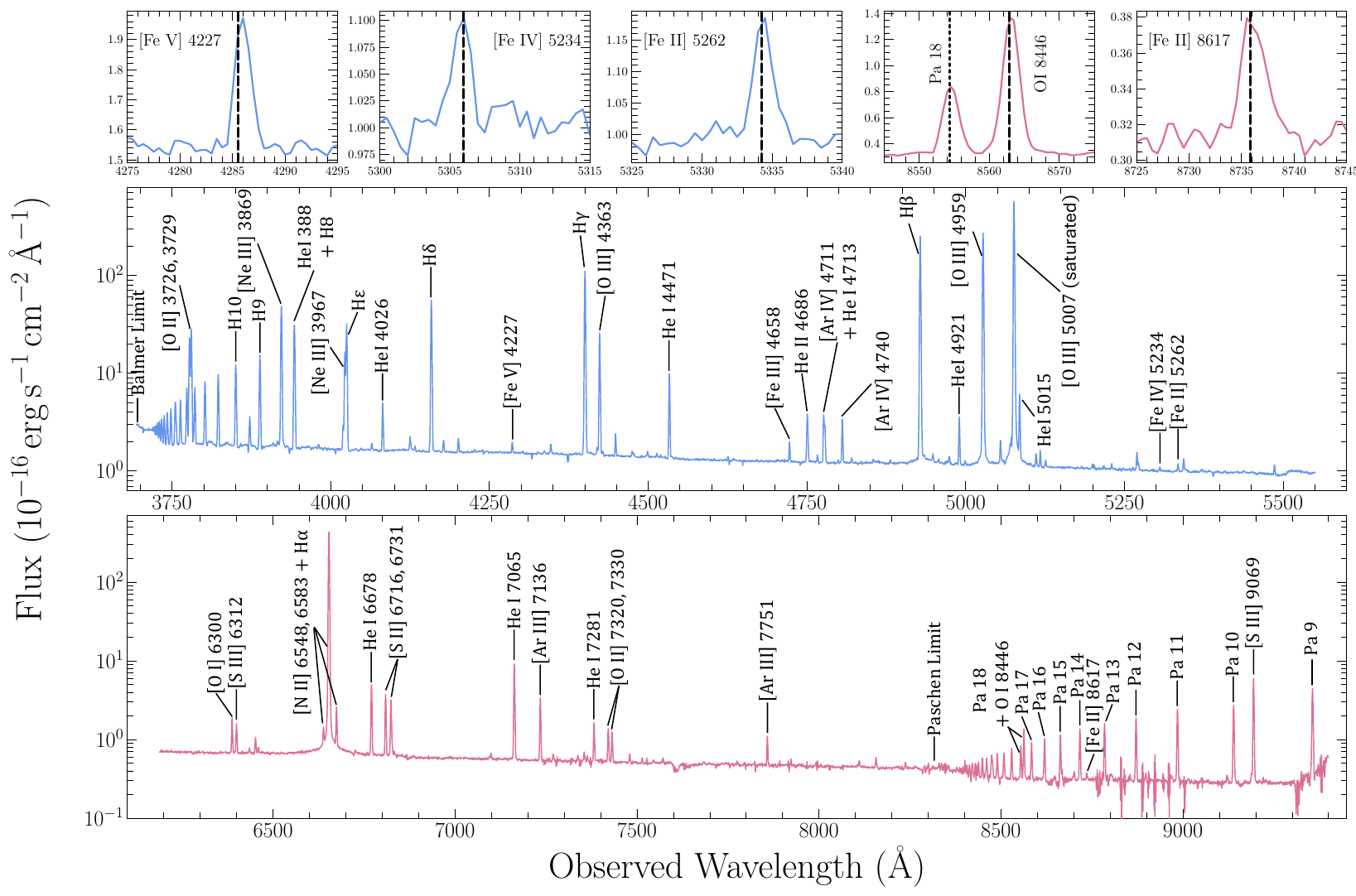}
\caption{Extracted 1D spectra of KCWI (blue; second row) and KCRM (red; third row) for SSCs 1\&2. 
First-row panels provide zoomed views around the detections of high-density gas indicators (Section \ref{subsec:high_den}), such as O\,{\footnotesize I} $\lambda8446$, [Fe\,{\footnotesize II}] $\lambda5262$, and [Fe\,{\footnotesize II}] $\lambda8617$, as well as the high and very-high ionization lines like [Fe\,{\footnotesize IV}] $\lambda5234$ and [Fe\,{\footnotesize V}] $\lambda4227$ (Section \ref{sec:nebular_prop}).
} 
\label{fg:j0337_1and2_spectrum}
\end{figure*}

% \begin{figure*}[htb]
% \includegraphics[width=1.0\linewidth]{figures/J0337-0502_master_plot_spec.pdf}
% \includegraphics[width=0.5\linewidth]{figures/J0337-0502_master_plot_Greys_r_best_hst_panel.pdf}
% \includegraphics[width=0.5\linewidth]{figures/j0337_ssc1a2_cartoon_plot.pdf}
% \caption{\textit{Top:} Extracted 1D spectra of KCWI (blue) and KCRM (red) from SSCs 1\&2 (shown in the \textit{bottom left} panel). The inset panels provide zoomed views around the detections of high-density CSM indicators, such as O\,{\footnotesize I} $\lambda8446$ and [Fe\,{\footnotesize II}] $\lambda8617$, as well as the [Fe\,{\footnotesize V}] $\lambda4227$ line from the very high-ionization zone (Section \ref{sec:nebular_prop}).
% \textit{Bottom Left:} The three KCWI/KCRM observation pointings (each marked with an 8\farcs4$\times$20\farcs4 green rectangle) are overlaid on the \textit{HST} FR656N image of J0337. 
% The dashed pointing is not used in this analysis. 
% % because it was observed during the 12- and 18-degree twilight periods, resulting in suboptimal sky background subtraction (Section \ref{sec:observation_and_data}). 
% The zoom-in panel shows the position of SSCs 1\&2 and the aperture used to extract the 1D spectrum presented in this Letter.
% \textit{Bottom Right:} The proposed LBV scenario for SSCs 1\&2 \citep[similar to the $\eta$-Carinae picture][]{Smith_2013, Smith_2018}, which could potentially explain the observed signatures (see Section \ref{sec:lbv_signs} for details).
% } 
% \label{fg:j0337_master_plot}
% \end{figure*}
% \smallskip

\setlength{\tabcolsep}{3pt}
\begin{deluxetable}{lcccc}
  \tablenum{2}
  \tablecaption{Physical Properties of each SSC in J0337-0502 Collected from the Literature}
    \tablehead{
    \colhead{Region}
    & \colhead{Age} 
    & \colhead{Stellar Mass} 
    & \colhead{$R_{\rm H\,{\footnotesize II}}$} 
    & \colhead{$L_{\rm H\alpha}$} 
    \\
    \colhead{} & 
    \colhead{(Myr)} & 
    \colhead{($M_{\odot}$)} & 
    \colhead{($\rm pc$)} &
    \colhead{($\rm 10^{38} \, erg \, s^{-1}$)} 
% \colhead{($\rm{km \ s^{-1}}$)} & 
% \colhead{($\rm{km \ s^{-1}}$)} & 
% \colhead{($\rm{km \ s^{-1}}$)} & 
  }
  \startdata
        SSC 1 & 3.0 & $4.7\times10^5$  & $11\pm11$  & $207 \pm 2$   \\
        SSC 2 & 3.0 & $3.7\times10^5$  & $10\pm10$  & $134 \pm 1$   \\
        SSC 3 & 7.0 & $7.1\times10^5$  & $29\pm27$  & $23.2 \pm 0.6$   \\
        SSC 4 & 11.0 & $1.1\times10^6$  & $33\pm27$  & $9.7 \pm 0.4$   \\
        SSC 5 & 13.0 & $2.9\times10^6$  & $39\pm30$  & $9.2 \pm 0.4$   \\
        SSC 6 & 11.0 & $2.6\times10^5$  & $21\pm21$  & $2.6 \pm 0.2$   \\
        SSC 7 & 4.0 & $9.4\times10^4$  & $15\pm10$ & $42 \pm 4$ \\
        SSC 8 & 8.0 & $2.2\times10^5$  & $21\pm10$ & $13 \pm 1$   \\
    \enddata
  % \startdata
  %       SSC 1 & 3.0 & $4.7\times10^5$  & $11.3\pm11.5$  & $207 \pm 2$   \\
  %       SSC 2 & 3.0 & $3.7\times10^5$  & $10.4\pm9.7$  & $134 \pm 1$   \\
  %       SSC 3 & 7.0 & $7.1\times10^5$  & $29.1\pm26.6$  & $23.2 \pm 0.6$   \\
  %       SSC 4 & 11.0 & $1.1\times10^6$  & $32.9\pm26.9$  & $9.7 \pm 0.4$   \\
  %       SSC 5 & 13.0 & $2.9\times10^6$  & $38.6\pm29.5$  & $9.2 \pm 0.4$   \\
  %       SSC 6 & 11.0 & $2.6\times10^5$  & $20.6\pm21.49$  & $2.6 \pm 0.2$   \\
  %       SSC 7 & 4.0 & $9.4\times10^4$  & $24\pm16$ & $42 \pm 4$ \\
  %       SSC 8 & 8.0 & $2.2\times10^5$  & $25\pm10$ & $13 \pm 1$   \\
  %   \enddata
\tablecomments{Age, stellar mass, H\,{\footnotesize II} region Str\"{o}mgren radius ($R_{\rm H\,{\footnotesize II}}$), and H$\alpha$ luminosity ($L_{\rm H\alpha}$, assuming a luminosity distance of 58 Mpc) for each SSC available in the literature \citep{Thompson_2009, Adamo_2010}. 
The Str\"{o}mgren radius of SSC 7 and SSC 8 are derived using $R_{\rm H\,{\footnotesize II}} = \left(3\,Q_{\rm LyC}/4\,\pi\,n_{\rm H}^2\,\alpha_B\,f\right)^{1/3}$.
Consistent with \citet{Adamo_2010}, we derive $Q_{\rm LyC}$ from the \texttt{STARBURST99} $Z=0.0004$ model \citep{Leitherer_1999}, assume a gas filling factor of $f = 0.01$, and adopt the Case B recombination coefficient $\alpha_B$ at $20\,000$ K.
Similar to \citet{Adamo_2010}, $n_{\rm H}$ of SSC 7 (SSC 8) is assumed to range from $10^3$ to $10^4$ $\rm cm^{-3}$ ($5\times10^2$ to $10^3$ $\rm cm^{-3}$).
The $L_{\rm H\alpha}$ of SSC 7 and SSC 8 are converted from their Pa$\alpha$ luminosities \citep{Thompson_2009} using the Case B recombination ratio (i.e., $\rm{H}\alpha /\rm{Pa}\alpha\simeq 9.15 - 9.72$ for $T_e \in [15\, 000, 20\,000]$ K and $n_e \in [10, 5\,000] \ \rm{cm^{-3}}$; \citealp{pyneb}).
\label{table:ssc_properties}}
\end{deluxetable}

\section{Observation and Data Reduction} 
\label{sec:observation_and_data}

\subsection{KCWI/KCRM Observation}\label{subsec:kcwi_kcrm_observation}

We observed J0337-0502 on 2023 September 18 with the Keck Cosmic Web Imager \citep[KCWI, the blue channel;][]{Morrissey_2018} and the Keck Cosmic Reionization Mapper (KCRM, the red channel) mounted on the Keck~II telescope using the Small Slicer. 
The blue (red) channel was configured with the BL (RL) grating centered at $4600\,\text{\AA}$ ($7700\,\text{\AA}$), providing a usable spectral range of $\sim 3650$ to $5550\,\text{\AA}$ ($\sim 6200$ to $9500\,\text{\AA}$). 
The combination of narrow slits and the BL or RL grating resulted in an average spectral resolution of $R \sim 3600$ for the blue channel and $R \sim 2350$ for the red channel. 
The seeing determined from the FWHM of the point-spread function (PSF) of the standard star \object{G191B2B} was around $0\farcs9$ in the blue channel and $0\farcs8$ in the red channel. 
The airmass ranged from $1.17$ to $1.21$ during the observation.

We obtained exposures of $1320\,\mathrm{s}$ and $300\,\mathrm{s}$ for the blue channel and $4\times300\,\mathrm{s}$ for the red channel at two pointings (shown in Figure \ref{fg:j0337_hst_plot}).
The initial exposure focused on the compact starburst, and then we offset approximately half the slicer width to the north, 
% and south, 
each subtending an 8\farcs4 $\times$ 20\farcs4 field of view (FoV).
% The third pointing (dashed), observed during the 12- and 18-degree twilight, is excluded from the analysis due to rapid sky background changes and infrequent sky chopping, which prevent valid sky subtraction.
% The third pointing (dashed green rectangle) is observed during the 12- and 18-degree twilight when the sky background changes too quickly to allow for valid sky subtraction due to our infrequent sky chopping; therefore, this pointing is not used for analysis in this work.
Since J0337-0502 exhibits spatially extended emissions (e.g., [O\,{\footnotesize III}] $\lambda 5007$ and H$\alpha$) across the entire FoV, we chopped the slicer to a blank field and conducted off-field sky exposures of $660\,\mathrm{s}$ for the blue channel and $2\times300\,\mathrm{s}$ for the red channel between the on-target exposures.

% For the blue arm (KCWI “BLUE” camera), we used the BL grating with a central wavelength of $\sim4599.9\,\text{\AA}$, providing a usable spectral range of about $3694$ to $5678\,\text{\AA}$. In the red arm (KCWI “RED” camera), we employed the RL grating centered at $\sim7699.9\,\text{\AA}$, covering approximately $6188$ to $9481\,\text{\AA}$. From calibration arc lines, we estimate a spectral resolution of $R \sim 3600$ (FWHM $\sim2.5\,\text{\AA}$), though the resolution may vary slightly across the field and wavelength range.

% Standard calibration data (bias frames, continuum bars, arc lamps, and flat fields) were taken throughout the night. 

\subsection{Data Reduction}\label{subsec:data_reduction}
We used the modified version of the KCWI Data Extraction and Reduction Pipeline (\texttt{KCWIDRP}; \citealt{Neill+23}) pipeline\footnote{\url{https://github.com/yuguangchen1/KCWI_DRP/tree/KCWIKit}} for basic data processing, including overscan and bias subtraction, scattered-light subtraction, flat-fielding, geometric rectification, flux calibration, and corrections for differential atmospheric refraction. 
The default sky-subtraction stage of \texttt{KCWIDRP} generates a 2D model of the sky spectrum using b-splines, which are insufficient for wavelength regions densely populated with sky lines. Therefore, we skip this stage and instead employ a method based on principal component analysis (PCA) to improve sky subtraction (see below).
We then leveraged the \texttt{KSkyWizard} package (Z. Zhuang et al., in prep.)\footnote{\url{https://github.com/zhuyunz/KSkyWizard}}to iteratively improve the fitting of the inverse sensitivity curve and apply telluric correction of the red channel based on the \texttt{tellfit} function of \texttt{PypeIt} \citep{pypeit:zenodo}. 
Residuals between the flux-calibrated standard star spectra and the \texttt{KCWIDRP} reference spectra are within $\pm$3\%.
% $\pm$2\% and $\pm$3\% for the blue and the red channels, respectively.

Based on the \texttt{LA Cosmic} \citep[LAplacian Cosmic ray detection;][]{LAcosmic} algorithm, \texttt{KCWIDRP} utilizes the \texttt{astroscrappy} package \citep{curtis_mccully_2018_1482019} for cosmic-ray (CR) removal, which can efficiently identify CR-contaminated pixels in the blue channel where few sky lines are present.
However, this edge detection algorithm, which uses the Laplacian kernel, incorrectly classifies most sky lines as CR in the red channel. 
For our science exposures with four consecutive frames, we could create the median frame and use sigma-clipping (3-sigma) to remove CR from these target frames. 
This sigma-clipping strategy does not apply to our sky exposures consisting of only two consecutive frames or to the standard star exposures taken during the 12- and 18-degree twilight, preventing us from computing the median frame accurately.
Therefore, we trained a \texttt{Cosmic-CoNN} model \citep[a deep-learning framework for cosmic ray detections designed for both spectroscopic and imaging data;][]{Xu_cosmic_conn} on the KCRM datasets with at least three consecutive exposures, which include both our observational data and the KCRM commissioning data (Peng et al. in prep.). 

% The default sky-subtraction stage of \texttt{KCWIDRP} generates a 2D model of the sky spectrum using b-splines that are insufficient for wavelength regions densely populated with sky lines. 
To improve sky subtraction, we utilize the modified version of the Zurich Atmosphere Purge (\texttt{ZAP}) package\footnote{\url{https://github.com/jasonpeng17/zap_for_kcwi}} \citep{ZAP}. 
The GUI implementation of this modified ZAP version is available in \texttt{KSkyWizard} and discussed in Peng et al. (in prep.). 
% We refer readers to Appendix \ref{sec:modified_zap} for details on this modified \texttt{ZAP} version.

After subtracting the sky background from the reduced data cubes obtained from \texttt{KCWIDRP}, the cubes were aligned, stacked, and resampled into a mosaic cube with $0\farcs29 \times 0\farcs29 \times 0.5\,\text{\AA}$ voxels using the \texttt{KCWIKIT}\footnote{\url{https://github.com/yuguangchen1/KcwiKit}} package (\citealt{KCWIKIT_2024}; see \citealt{Chen+21} and \citealt{Prusinski+25} for additional details) with a drizzle factor of 1.0. 
Since H$\gamma$, H$\beta$, and the [O\,{\footnotesize III}] $\lambda\lambda 4959, 5007$ doublet are saturated in the regions around SSCs 1\&2 for the $1320\,\mathrm{s}$ exposure frames of the blue channel (more than 60\,000 counts in the raw 2D frames), we replaced these saturated pixels with corresponding pixels from the $300\,\mathrm{s}$ exposure frames.
% spaxels with corresponding spaxels from the $300\,\mathrm{s}$ exposures in a wavelength region of $\sim 5 - 10 \ \rm{\AA}$ centered on these emission lines. 
[O\,{\footnotesize III}] $\lambda 5007$ is saturated in both $1320\,\mathrm{s}$ and $300\,\mathrm{s}$ exposures, so we only measure [O\,{\footnotesize III}] $\lambda 4959$ for subsequent analysis and use the intrinsic ratio of the [O\,{\footnotesize III}] doublet \citep[2.98;][]{SZ00} to estimate the [O\,{\footnotesize III}] $\lambda 5007$ flux. 

To accurately locate each SSC within our reduced cubes, we aligned both the \textit{HST} images (proposal ID 10575; PI: G. \"{O}stlin) and the KCRM narrow-band H$\alpha$ image to the DECaLS r-band image \citep{decals_16, desi_2019}. 
We then used the SSC apertures defined in Figure \ref{fg:j0337_hst_plot}, which satisfy Nyquist sampling, to extract the corresponding 1D spectra from the data cubes (see Figure \ref{fg:j0337_1and2_spectrum} for the spectra of SSCs 1\&2).

% We use the uniform priors for the \texttt{BAGPIPES} parameters across the allowed range.

\section{Nebular Properties of SSCs}\label{sec:nebular_prop}
% \textbf{
In this section, we summarize the measured key nebular properties, including electron density and temperature ($n_e$, $T_e$), ionization parameter ($\log U$), and gas-phase abundances (oxygen, nitrogen, and iron), for all SSCs shown in Figure \ref{fg:j0337_hst_plot}.
The methods used to derive these nebular properties are described in Appendix \ref{sec:appendix_ssc_neb_prop} in detail.
% } 
We then combine these measurements (summarized in Table \ref{table:nebular_properties_ssc}) with other observed nebular properties from the literature to highlight the distinguishing nebular characteristics of SSCs 1\&2 compared to other SSCs in Section \ref{sec:lbv_signs}. 
% In Section \ref{sec:lbv_signs}, we discuss how these properties support the identification of SSCs 1\&2 as the hosts of giant eruption candidate(s) and explore additional giant eruption signatures.
% \subsection{Emission Line Fittings}\label{subsec:emis_line_fit}
% \subsection{Properties from Line Fittings}\label{subsec:line_fit}

Line ratios are key constraints for deriving the nebular properties in our analysis.
We employ the \texttt{Python} package \texttt{VerEmisFitting} \citep{veremisfit, Peng_2025} to model each emission line's profile. Both single-Gaussian and multi-component models are utilized--for example, a double-Gaussian plus Lorentzian (dGL) model for H$\alpha$ as shown in Section~\ref{subsec:asymm_broad_balmer}--to fit the observed line profiles.
% We refer readers to \cite{Peng_2025} for the fitting approaches and constraints adopted in this letter.
% \textbf{
It is noteworthy that the observed broad-wing line profile of H$\alpha$ cannot be well described by a Lorentzian model (i.e., $\propto x^{-2}$), but is instead better characterized by the wings of a Thomson-scattering profile (i.e., $\propto e^{-|x|}$).
% \textbf{
This is a key signature of an optically thick outflow produced by a giant eruption, such as that observed in $\eta$ Carinae, and more generally by SNe IIn 
\citep[see Section \ref{subsec:asymm_broad_balmer} for details;][]{Davidson_1995, Chugai_2001, Hillier_2001, Dessart_2009, Humphreys_2012, Mehner_2015, Davidson_2020}.
% }
% }

% \subsection{Extinction}\label{subsec:extinction}
% To correct for Galactic foreground extinction, we obtain $E(B-V)_{\rm{MW}} = 0.0402 \pm 0.002$ from the dust maps of \cite{schlafly_measuring_2011} and apply the correction using the Galactic extinction curve interpolated from Table 3 of \cite{Fitzpatrick_1999}. 
% We then model the intrinsic dust attenuation using the SMC extinction law \citep{Gordon_2003}.
% The SNR‑weighted intrinsic color excess, $E(B-V)_{\rm int}$, is derived by comparing the observed recombination line ratios—specifically, $\rm H\delta/\rm H\gamma$, $\rm H9/\rm H\gamma$, $\rm H10/\rm H\gamma$, and Paschen lines Pa\,9 through Pa\,18 relative to H$\gamma$ to the theoretical Case B ratios from \citet{storey_recombination_1995}.
% These chosen relatively high-order Balmer and Paschen lines are free from contamination by nearby strong lines or obvious electron‑scattering broad wings that are seen in H$\alpha$ and H$\beta$ (Section \ref{subsec:asymm_broad_balmer}).
% The measured $E(B-V)_{\rm int}$ value for SSCs 1\&2 is $0.06 \pm 0.01$, whereas the remaining SSCs are consistent with zero internal reddening (i.e., $E(B-V)_{\rm int} = 0$). 
% % The zero $E(B-V)_{\rm int}$ value for other SSCs is confirmed by our full-spectrum fittings using \texttt{BAGPIPES} (see Appendix \ref{appendix:full_spec_fit} for details).

\defcitealias{berg_characterizing_2021}{B21}
\subsection{$n_e$, $T_e$, and $\log U$ Constraints of Different Ionization Zones}\label{subsec:ne_Te}

% \textbf{
We characterize the physical conditions of the ionized gas by separating the nebulae into multiple ionization zones following \citet[][hereafter B21; see Appendices \ref{subsec:ionization_zone} and \ref{subsec:logU_constraint} for the diagnostics used to define each ionization zone]{berg_characterizing_2021}.
This approach should be regarded as a simplified parameterization of an intrinsically clumpy and multiphase nebular medium, because each ``zone'' may in reality comprise a collection of gas condensations with a wide range of physical conditions.
% }
% $n_e$, $T_e$, and $\log U$ are derived using  methodological described in Appendix \ref{subsec:ionization_zone}.

\begin{figure}
\includegraphics[width=1\linewidth]{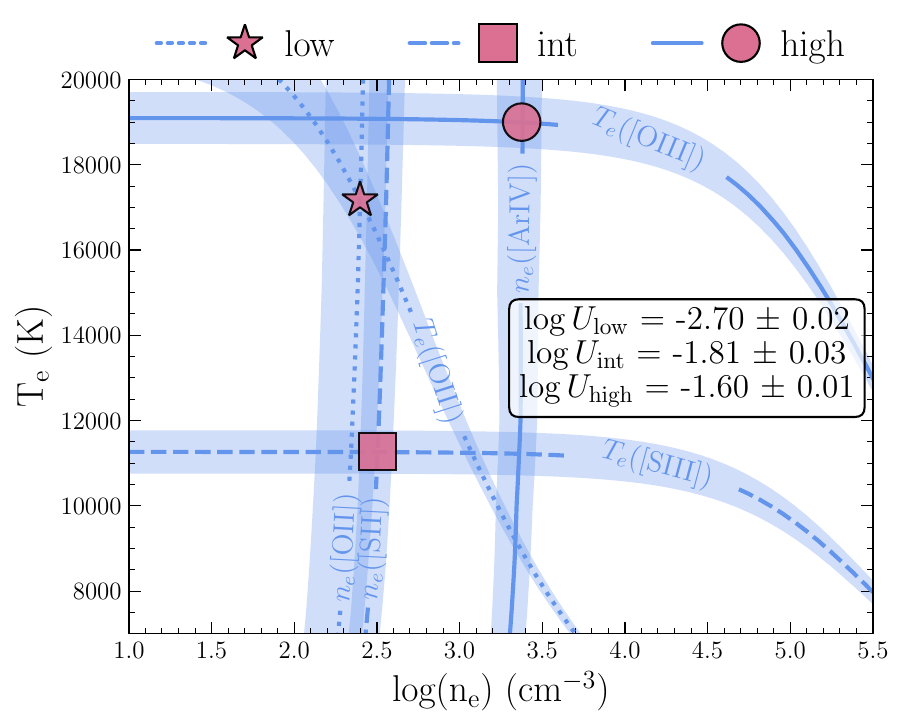}
\caption{$T_e$ and $n_e$ solutions for SSCs 1\&2. 
The best-fit $T_e$ and $n_e$ for each ionization zone are defined as the intersection of the temperature and density diagnostics.
% the low (dotted), intermediate (dashed), and high (solid) ionization regions are indicated by the corresponding markers, with the associated $\log U$ value shown in the legend. 
The estimated $\log U$ value for each ionization zone is also shown in the panel.
% The estimated $\log U$ values for the low-to-intermediate (``low''; [S\,{\footnotesize III}]/[S\,{\footnotesize II}]), intermediate-to-high (``int''; [O\,{\footnotesize III}]/[O\,{\footnotesize II}]), and high-to-very-high (``high''; [Ar\,{\footnotesize IV}]/[Ar\,{\footnotesize III}]) ionization zones are also shown in the panel.
% Oxygen, sulphur, argon
% diagnostics are depicted in green, grey, and red colors, respectively, with shades representing the 1-$\sigma$ confidence band.
} 
\label{figure:ne_Te_j0337_1and2}
\end{figure}

% \textbf{
% Figure \ref{figure:ne_Te_j0337_1and2} shows the derived $T_e$, $n_e$, and $\log U$ values using \texttt{PyNeb} for the low ($14.5 - 20.6 \ \rm{eV}$), intermediate ($23.3 - 34.8 \ \rm{eV}$), and high ($35.1 - 54.9 \ \rm{eV}$ ionization zones for SSCs 1\&2 (see Appendix \ref{sec:appendix_ssc_neb_prop} for other SSCs).
% $T_e$ within the high-ionization zone are uniformly high, with $T_e({\rm [O\,{\footnotesize III}]}) \simeq (1.8-1.9)\times10^4$ K across all SSCs, 
% while lower-ionization zones show cooler gas, $T_e({\rm [S\,{\footnotesize III}]}) \sim (1.1-1.9)\times10^4$ K and $T_e({\rm [O\,{\footnotesize II}]}) \sim (1.3-1.7)\times10^4$ K, albeit with larger uncertainties. 
% $n_e$ inferred from low-ionization diagnostics are modest, with $n_e({\rm [S\,{\footnotesize II}]}) \sim 10-350\ \rm cm^{-3}$ and $n_e({\rm [O\,{\footnotesize II}]}) \sim 80-250\ \rm cm^{-3}$, whereas very-high ionization gas traced by [Ar\,{\footnotesize IV}] indicates systematically denser gas, $n_e({\rm [Ar\,{\footnotesize IV}]}) \sim (2-4)\times10^3\ \rm cm^{-3}$.
Figure \ref{figure:ne_Te_j0337_1and2} shows the derived $T_e$, $n_e$, and $\log U$ values using \texttt{PyNeb} for the low ($14.5 - 20.6 \ \rm{eV}$), intermediate ($23.3 - 34.8 \ \rm{eV}$), and high ($35.1 - 54.9 \ \rm{eV}$) ionization zones for SSCs 1\&2 (see Appendix \ref{sec:appendix_ssc_neb_prop} for other SSCs).
$T_e$ within the high-ionization zone reaches $T_e({\rm [O\,{\footnotesize III}]}) = 19\,000\pm600$ K, 
while lower-ionization zones show cooler gas, with $T_e({\rm [S\,{\footnotesize III}]})= 11\,000\pm500$ K and $T_e({\rm [O\,{\footnotesize II}]})= 17\,200\pm1600$ K.
% , albeit with larger uncertainties. 
$n_e$ inferred from low-ionization diagnostics are modest, with $n_e({\rm [S\,{\footnotesize II}]})= 320\pm70\ \rm cm^{-3}$ and $n_e({\rm [O\,{\footnotesize II}]})= 250\pm100\ \rm cm^{-3}$, whereas the high- to very-high (VH; $>54\,\rm{eV}$) ionization gas traced by [Ar\,{\footnotesize IV}] indicates systematically denser gas, with $n_e({\rm [Ar\,{\footnotesize IV}]})= 2\,400\pm700\ \rm cm^{-3}$.
The estimated $\log U$ values for the low-to-intermediate (``low''; [S\,{\footnotesize III}]/[S\,{\footnotesize II}]), intermediate-to-high (``int''; [O\,{\footnotesize III}]/[O\,{\footnotesize II}]), and high-to-VH (``high''; [Ar\,{\footnotesize IV}]/[Ar\,{\footnotesize III}]) ionization zones span $-2.70$ to $-1.60$.
% }

\setlength{\tabcolsep}{6pt} % Default value: 6pt
% \vspace{-0.5cm} % reduce the spacing by 0.5cm
\begin{deluxetable*}{lcccccc}
  \tablenum{3}
  \tablecaption{Spectroscopically‐Derived Nebular Properties of SSCs}
  \tablehead{
    \colhead{Property} 
    & \colhead{Zone} 
    & \colhead{SSCs\,1\&2} 
    & \colhead{SSC\,3} 
    & \colhead{SSCs\,4\&5\&7} 
    & \colhead{SSC\,6} 
    & \colhead{SSC\,8}
  }
  \startdata
  $T_e$([Fe\,{\footnotesize V}])$_{\rm{CL}}$\,(K) & VH & $\sim 23\,000$ & \nodata & \nodata & \nodata & \nodata \\
  $T_e$([O\,{\footnotesize III}])\,(K) & H    & $19\,000 \pm 600$ & $18\,200 \pm 170$ & $19\,500 \pm 200$ & $18\,400 \pm  180$ & $18\,300 \pm 380$ \\
  $T_e$([S\,{\footnotesize III}])\,(K) & I   & $11\,000 \pm 500$ & $15\,900 \pm 330$ & $15\,800 \pm 610$ & $19\,400 \pm 690$ & $1\,5200 \pm 2\,200$ \\
  $T_e$([O\,{\footnotesize II}])\,(K) & L   & $17\,200 \pm 1\,600$  & $16\,100 \pm 2\,700$ & $15\,700 \pm 3\,800$ & $15\,700 \pm 5\,000$ & $1\,3000 \pm 3\,900$\\
  $T_e$([Fe\,{\footnotesize II}])$_{\rm{CL}}$\, (K) & PI & $\sim 7\,500$ & \nodata & \nodata & \nodata & \nodata \\
  $n_e$([Ar\,{\footnotesize IV}])\,(cm$^{-3}$) & VH & $2\,400 \pm 700$ & $2\,200 \pm 420$ & $2\,800 \pm 950$ & $2\,600 \pm 870$ & $4\,100 \pm 2\,900$ \\
  $n_e$([S\,{\footnotesize II}])\,(cm$^{-3}$)     & L &  $320 \pm 70$ & $180 \pm 50$ & $150 \pm 70$ & $250 \pm 90$ & $10 \pm 10$\\
  $n_e$([O\,{\footnotesize II}])\,(cm$^{-3}$)     & L &  $250 \pm 100$ & $180 \pm 20$ & $140 \pm 30$ & $90 \pm 90$ & $80 \pm 20$\\
  $n_e$([Fe\,{\footnotesize II}])\,(cm$^{-3}$) & PI & $\sim 10^6$ & $\sim 10^6$ & \nodata & \nodata & \nodata \\  
  $\log U_{\rm low}$ ([S\,{\footnotesize III}]/[S\,{\footnotesize II}]) & I/L & $-2.70 \pm 0.02$ & $-2.61 \pm 0.01$ & $-2.52 \pm 0.01$ & $-2.55 \pm 0.01$ & $-2.79 \pm 0.01$\\
  $\log U_{\rm int}$ ([O\,{\footnotesize III}]/[O\,{\footnotesize II}]) & H/I  & $-1.81 \pm 0.03$ & $-1.77 \pm 0.01$ & $-1.71 \pm 0.01$ & $-1.71 \pm 0.01$ & $-2.02 \pm 0.01$\\ 
  $\log U_{\rm high}$ ([Ar\,{\footnotesize IV}]/[Ar\,{\footnotesize III}]) & VH/H & $-1.60 \pm 0.01$ & $-1.60 \pm 0.01$ & $-1.52 \pm 0.02$ & $-1.58 \pm 0.01$ & $-1.90 \pm 0.03$ \\
  $\log U_{\rm ave}$ & L/I/H & $-1.67 \pm 0.01$ & $-1.67 \pm 0.01$ & $-1.59 \pm 0.02$ & $-1.64 \pm 0.01$ & $ -2.08 \pm 0.03$\\
  12 + log(O/H) & All & $7.32 \pm 0.02$ & $7.38 \pm 0.01$ & $7.30 \pm 0.01$ & $7.38 \pm 0.06$ & $7.21 \pm 0.01$ \\
  12 + log(N/H) \citepalias{berg_characterizing_2021} & All & $6.30 \pm 0.02$ & $6.20 \pm 0.02$ & $6.21 \pm 0.02$ & 
 $6.13 \pm 0.07$ & $6.10 \pm 0.03$ \\
  12 + log(N/H) \citep{PC_1969} & All & $6.02 \pm 0.03$ & $5.91 \pm 0.02$ & $5.82 \pm 0.02$ & $ 5.86 \pm 0.07$ & $5.67 \pm 0.03$ \\
  12 + log(N/H) \citep{Esteban_2020} & All & $6.11 \pm 0.03$ & $6.00 \pm 0.02$ & $5.90 \pm 0.02$ & $5.94 \pm 0.07$ & $5.78 \pm 0.03$ \\
  % 12 + log(Fe/H) & All & $5.72 \pm 0.04$ & \nodata & \nodata & \nodata & \nodata \\
  % 12 + log(Fe/H) & All & $5.78 \pm 0.06$ & \nodata & \nodata & \nodata & \nodata \\
  12 + log(Fe/H) & All & $5.81 \pm 0.06$ & \nodata & \nodata & \nodata & \nodata \\
  12 + log(Fe/H) \citepalias{berg_characterizing_2021} & All & $6.02 \pm 0.03$ & $6.03 \pm 0.02$ & $6.20 \pm 0.04$ & $6.01 \pm 0.08$ & $5.93 \pm 0.05$ \\
  log(N/O) \citepalias{berg_characterizing_2021} & All & $-1.02 \pm 0.02$ & $-1.17 \pm 0.01$ & $-1.09 \pm 0.02$ & $-1.25 \pm 0.03$ & $-1.11 \pm 0.03$ \\
  log(N/O) \citep{PC_1969} & All & $-1.29 \pm 0.02$ & $-1.46 \pm 0.01$ & $-1.48 \pm 0.02$ & $-1.52 \pm 0.03$ & $-1.54 \pm 0.03$ \\
  log(N/O) \citep{Esteban_2020} & All & $-1.21 \pm 0.02$ & $-1.38 \pm 0.01$ & $-1.39 \pm 0.02$ &  $-1.43 \pm 0.03$ & $-1.44 \pm 0.03$ \\
  % log(Fe/O) & All & $-1.60 \pm 0.04$ & \nodata & \nodata & \nodata & \nodata \\
  % log(Fe/O) & All & $-1.55 \pm 0.06$ & \nodata & \nodata & \nodata & \nodata \\
  log(Fe/O) & All & $-1.52 \pm 0.06$ & \nodata & \nodata & \nodata & \nodata \\
  log(Fe/O) \citepalias{berg_characterizing_2021} & All & $-1.30 \pm 0.03$ & $-1.34 \pm 0.02$ & $-1.10 \pm 0.04$ & $-1.37 \pm 0.04$ & $-1.28 \pm 0.05$ \\
  $E(B-V)_{\rm{int}}$ & All & $0.06 \pm 0.01$ & 0 & 0 & 0 & 0 \\
  \enddata
\tablecomments{Electron temperatures, densities, $\log U$, element abundances, and intrinsic color excesses for the SSCs in J0337-0502, derived using the methodologies described in Section \ref{subsec:ne_Te} and Appendix \ref{sec:appendix_ssc_neb_prop}. 
Column 2 specifies each property's ionization zone(s), where PI = partially ionized, L = low, I = intermediate, H = high, VH = very high, and All = all ionization zones. \label{table:nebular_properties_ssc}}
\end{deluxetable*}

\begin{figure}[htb]
\centering
\includegraphics[width=1.0\linewidth]{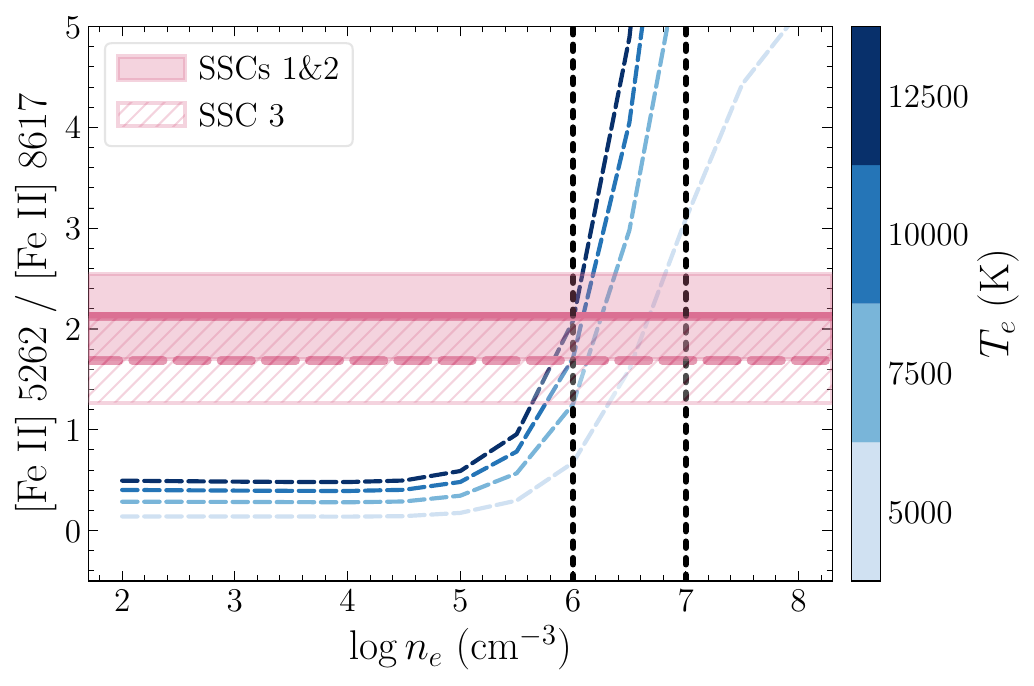}
\caption{The [Fe\,{\footnotesize II}] $\lambda 5262$/[Fe\,{\footnotesize II}] $\lambda 8617$ ratio for SSCs 1\&2 (solid red) and SSC 3 (dashed red).
Predictions of the [Fe\,{\footnotesize II}] $\lambda5262$/[Fe\,{\footnotesize II}] $\lambda8617$ ratio across different $n_e$ and $T_e$ are shown as dashed bluish lines. 
The rough $n_e$ constraints for gas traced by [Fe\,{\footnotesize II}] and [Fe\,{\footnotesize III}] in the Weigelt knots of $\eta$ Carinae are shown as dotted lines.
% primary (B-star) and secondary (O-star) stars of $\eta$ Carinae \textbf{\citep[cf.][]{Verner_2005}} are shown as dashed and dot-dashed lines, respectively.
% The [Fe\,{\footnotesize II}] ratio reveals that a high-density CSM exists in SSCs 1\&2.
% \textit{Right:} Comparison between the observed O\,{\footnotesize I} $\lambda8446$/H$\alpha$ and H$\alpha$/H$\beta$ ratios and \texttt{CLOUDY} models (see Section \ref{subsec:high_den} for details). 
% The best-matched model indicates high optical depths in Ly$\alpha$ ($\tau_{\rm{Ly}\alpha} \sim 10^8$) and H$\alpha$ ($\tau_{\rm{H}\alpha} \sim 0.1$).
} 
\label{fg:j0337_high_den}
\end{figure}

Moreover, a key result of this work is the identification of an extremely dense partially ionized (PI) zone ($\sim 8 - 15 \ \rm{eV}$),
traced by [Fe\,{\footnotesize II}] emission, in SSCs~1\&2.
Figure \ref{fg:j0337_high_den} compares the observed [Fe\,{\footnotesize II}] $\lambda 5262$/[Fe\,{\footnotesize II}] $\lambda 8617$ ratios for SSCs 1\&2 and SSC 3 with theoretical predictions computed at various $n_e$ and $T_e$ values using \texttt{PyNeb} \citep{pyneb_2015}. 
SSCs 1\&2 are the only clusters that have $> 3 \sigma$ detections in both [Fe\,{\footnotesize II}] lines.
SSC 3 only has a $\sim 2.3 \sigma$ ($\sim 2.8 \sigma$) detection for [Fe\,{\footnotesize II}] $\lambda 5262$ ([Fe\,{\footnotesize II}] $\lambda 8617$).
At $T_e$ ranges from 5\,000 to 12\,500 K, we constrain  $10^6 \lesssim n_e \lesssim 10^7 \ \rm{cm^{-3}}$ for both SSCs 1\&2 and SSC 3 in the PI region 
(compare to the densities inferred for gas traced by [Fe\,{\footnotesize II}] and [Fe\,{\footnotesize III}] in the ``Weigelt knots'' of $\eta$ Carinae; \citealp{Mehner_2010, Hamann_2012}, see Figure \ref{fg:j0337_high_den}]).
If we use the ion-fraction-weighted temperature of [Fe\,{\footnotesize II}] from the best-fitting \texttt{CLOUDY} models (see Section \ref{subsec:high_den} for details) to estimate $T_e$ ([Fe\,{\footnotesize II}])$_{\rm{CL}}$ (e.g., $\sim 7\,500 \ \rm{K}$ for SSCs 1\&2) in the PI zone, the lower end of this $n_e$ constraint is favored (i.e., $n_e \sim 10^6 \ \rm{cm^{-3}}$).
% Similar to $\eta$ Carinae \citep{Verner_2005}, we constrain $10^6 \lesssim n_e \lesssim 10^7 \ \rm{cm^{-3}}$ for both SSCs 1\&2 and SSC 3 in the partially-ionized region. 
Based on their higher [Fe\,{\footnotesize II}] $\lambda 5262$/[Fe\,{\footnotesize II}] $\lambda 8617$ ratio, SSCs 1\&2 are $\sim 0.2$ dex denser than SSC 3 within the PI zone, but remain consistent with SSC 3 within 1$\sigma$.
% Similar to the VH ionization zone, we use the ion‐weighted temperature of [Fe\,{\footnotesize II}] from the best‐fitting \texttt{CLOUDY} models to estimate $T_e$ ([Fe\,{\footnotesize II}])$_{\rm{CL}}$ (e.g., $\sim 7\,500 \ \rm{K}$ for SSCs 1\&2) in the partially ionized zone.

Notably, although $n_e$ in the neutral zone cannot be directly constrained, the detection of O\,{\footnotesize I} $\lambda8446$ in all SSCs, the Bowen Resonance Fluorescence line of Ly$\beta$, suggests an optically-thick region where Ly$\alpha$ trapping can effectively pump the $n = 2$ level in neutral hydrogen gas. 
The O\,{\footnotesize I} $\lambda8446$/H$\alpha$ ratio can therefore serve as a tracer of optical depth \citep[e.g.,][]{Netzer_1976, Johansson_2005}. 
We find that O\,{\footnotesize I} $\lambda8446$/H$\alpha$ reaches its highest value in SSCs 1\&2 ($\sim2.5\times10^{-3}$), followed by SSC 3 and SSC 8 ($\sim1.8-2.0\times10^{-3}$), and decreases further in SSCs 4 to 7 ($\sim1.2-1.3\times10^{-3}$). This trend is consistent with the high-density sequence independently inferred from the [Fe\,{\footnotesize II}] diagnostics, and also with the increasing stellar population age listed in Table \ref{table:ssc_properties}. 
In this picture, radiative and mechanical feedback from massive stars can progressively disperses the dense neutral gas within the SSCs.
We therefore use the O\,{\footnotesize I} $\lambda8446$/H$\alpha$ ratio to estimate $\tau_{\rm Ly\alpha}$ (and the corresponding hydrogen column density $N_{\rm H}$) and examine the spatial correspondence between O\,{\footnotesize I} and [Fe\,{\footnotesize II}] emission in Section \ref{subsec:high_den}.

\subsection{Gas-phase Abundances of $N$, $O$, and $Fe$}\label{subsec:abundance}
To determine the gas-phase ionic abundances of N, O, and Fe, we adopt the corresponding $T_e$ and $n_e$ measurements for each ionization zone described in Section \ref{subsec:ne_Te}. 
The specific emission lines and ionization correction factors (ICFs) used to derive each abundance are summarized in Appendix \ref{subsec:abundance_details}.

% We consider only the first two ionization stages (O$^+$ and O$^{+2}$) for oxygen, with their corresponding ionic abundances derived from the [O\,{\footnotesize II}] $\lambda\lambda3726,29$ doublet and [O\,{\footnotesize III}] $\lambda4959$ emission line, because contributions from O$^0$ and O$^{+3}$ are negligible in galaxies with physical properties (e.g., $\log U_{\rm{ave}}$ and gas-phase metallicity) similar to J0337-0502 \citepalias[e.g., J1044+0353;][]{berg_characterizing_2021}.
% \begin{align}
%     \label{eq:3.1}
%     \rm{\frac{O}{H} = \frac{O^{+} + O^{+2}}{H^{+}}} 
% \end{align}
We consider the first two ionization stages of oxygen ($\rm O^{+}$ and $\rm O^{+2}$) to derive the oxygen abundance. The derived oxygen abundance expressed as 12 + log(O/H) ranges from $7.21 \pm 0.01$ for SSC 8 to $7.38 \pm 0.06$ for SSC 6.
Notably, the oxygen abundance for SSC 6 is measured for the first time in this work, 
while the measurements for the other SSCs are consistent with previous studies \citep[e.g.,][]{Thuan_2004, Papaderos_2006, Izotov_2009}. 
%$7.34 \pm 0.10$

For nitrogen, we consider only the first ionization stage ($\rm{N}^{+}$) and apply three different $\mathrm{ICF}(\rm{N}^{+})$ to estimate the nitrogen abundance.
We find that SSCs 1\&2 exhibit a $\log(\rm{N/O})$ value that is $\sim0.1-0.2$ dex higher than those of other SSCs, regardless of the choice of $\rm{ICF(N^{+})}$.
We account for systematic uncertainties in $\rm{ICF(N^{+})}$, which can vary $\log(\rm{N/O})$ by up to 0.3 dex, when comparing SSCs 1\&2 to other high-$\log(\mathrm{N/O})$ galaxies in the literature in Section \ref{subsec:no_feo_analysis}.
We also note that although \citet{Izotov_2009} adopt an ICF(N$^+$) that is a function of $\mathrm{O}^+/(\mathrm{O}^+ + \mathrm{O}^{2+})$, as in \citealt{PC_1969} and \citealt{Esteban_2020}, their derived $\log(\mathrm{N/O})$ for SSCs 1\&2 is $\sim0.1-0.2$ dex lower than those obtained using the latter ICF prescriptions.

% \setlength{\tabcolsep}{3pt}
% \begin{deluxetable}{lccccccc}
%   \tablenum{2}
%   \tablecaption{Derived Properties of J0337-0502 SSCs 1\&2}
%     \tablehead{
%     \colhead{$L$ (primary)} 
%     & \colhead{$L$ (secondary)} 
%     & \colhead{$L_{\rm H\alpha}$ (narrow)} 
%     & \colhead{$L_{\rm H\alpha}$ (broad)} 
%     & \colhead{$L_{\rm H\alpha}$ (VB)} 
%     & \colhead{$L_{\rm H\alpha}$ (VB, model)} 
%     & \colhead{$\tau_{\rm e}$ (CSM)} 
%     & \colhead{$\Delta R$ (CSM)} 
%     \\
%     \colhead{($L_{\odot}$)} & 
%     \colhead{($L_{\odot}$)} & 
%     \colhead{($\rm 10^{38} \, erg \, s^{-1}$)} &
%     \colhead{($\rm 10^{38} \, erg \, s^{-1}$)} &
%     \colhead{($\rm 10^{38} \, erg \, s^{-1}$)} &  
%     \colhead{($\rm 10^{38} \, erg \, s^{-1}$)} & 
%     \colhead{} & 
%     \colhead{($\rm{cm}$)} 
% % \colhead{($\rm{km \ s^{-1}}$)} & 
% % \colhead{($\rm{km \ s^{-1}}$)} & 
% % \colhead{($\rm{km \ s^{-1}}$)} & 
%   }
%   \startdata
%         $10^{5.70 \pm 0.17}$ & $10^{6.22 \pm 0.43}$  & $\left(414 \pm 3\right)$  & $\left(32 \pm 2\right)$ & $\left(11.5 \pm 0.7\right)$ & $7.23$ & $\sim 10$ & $10^{17} \ \rm{cm}$  \\
%     \enddata
% \tablecomments{
% \label{table:ssc1a2_properties}}
% \end{deluxetable}

Because we detect Fe at $>3\sigma$ significance in multiple ionization states—Fe$^{+}$ (PI), Fe$^{+2}$ (low-intermediate), Fe$^{+3}$ (intermediate-high), and Fe$^{+4}$ (VH)—we include all four stages when determining the gas-phase iron abundance in SSCs 1\&2.
This approach is not applicable to the other SSCs: for SSCs 3 to 7 we adopt ICF(Fe$^{+2}$+Fe$^{+4}$), while for SSC 8 we use ICF(Fe$^{+2}$).
The derived $\log(\rm{Fe/O})$ value for SSCs 1\&2 is \(\sim0.1-0.3\) dex lower than those of other SSCs. 
However, this comparison is subject to systematic uncertainties in the adopted Fe ICFs. 
For example, if we instead adopt the \citetalias{berg_characterizing_2021} ICF(Fe$^{+2}$+Fe$^{+4}$) for SSCs 1\&2, we find that the $\log(\rm{Fe/O})$ of SSCs 1\&2 is comparable to those of SSCs 3, 6, and 8 (see Table \ref{table:nebular_properties_ssc}). 
% \begin{align}
%     \label{eq:3.3}
%     \rm{\frac{Fe}{H} = \frac{Fe^{+} + Fe^{+2} + Fe^{+3} + Fe^{+4}}{H^{+}}}
% \end{align}
% The derived iron abundance expressed as 12 + log(Fe/H) is $5.73 \pm 0.09$.

\section{Signatures of Giant Eruption in SSCs 1\&2} \label{sec:lbv_signs}

As summarized in Section \ref{sec:nebular_prop}, SSCs 1\&2 are distinct from the other SSCs in that they are the only clusters exhibiting: (1) $>3\sigma$ detections of [Fe\,{\footnotesize II}] emission lines, implying $n_e\sim10^6 \ \mathrm{cm^{-3}}$ in the PI zone (see Sections \ref{subsec:high_den} and \ref{subsec:iron_lines} for the stellar and nebular conditions required to produce the observed [Fe\,{\footnotesize II}] emission); 
and (2) nitrogen enrichment 
% coupled with iron depletion 
relative to the remaining SSCs (see Section \ref{subsec:no_feo_analysis} for the possible origin). 
% \citetalias{Hatano_2026} also argue that the WISE variability of NIR emission should originate from SSCs 1\&2.
When combined with their detections of 
% NIR variability \citepalias{Hatano_2026}, 
warm dust \citep{Hunt_2014}, radio emission \citep{Johnson_2009}, H$_2$ emission \citep{Thompson_2009}, and Thomson-scattering broad wings (Section \ref{subsec:asymm_broad_balmer}),
these characteristics strongly motivate a detailed investigation of the underlying source of excitation in SSCs 1\&2.
% their point‐source nature \citep{Thompson_2009} and 
% Therefore, SSCs 1\&2 (see Figure \ref{fg:j0337_hst_plot}) are the most likely hosts of the giant-eruption candidate.
% as indicated by 
% % their point-source nature \citep{Thompson_2009} and the 
% the detections of key giant eruption signatures such as 
% % O\,{\footnotesize I} (this work), 
% [Fe\,{\footnotesize II}] \citep{Mingozzi_2025}, warm dust ($\sim 400-500$ K; \citealp{Hunt_2014}; \citetalias{Hatano_2026}), radio emission \citep{Johnson_2009},  $\rm{H}_2$ \citep{Thompson_2009},  nitrogen enrichment (this work), and Thomson-scattering emission-line wings (this work).
% and iron depletion onto dust (this work). 
% Therefore, in this work, we focus on the potential giant eruption signatures of SSCs 1\&2 in J0337-0502, while providing a summary of the nebular properties of other SSCs for comparison (see Table \ref{table:ssc_properties} for key properties of each SSC compiled from the literature). 
% Detailed analysis of the large-scale outflow kinematics and mass, momentum, and energy outflow rates will be presented in future work.

\begin{figure*}[htb]
\centering
\includegraphics[width=1.0\linewidth]{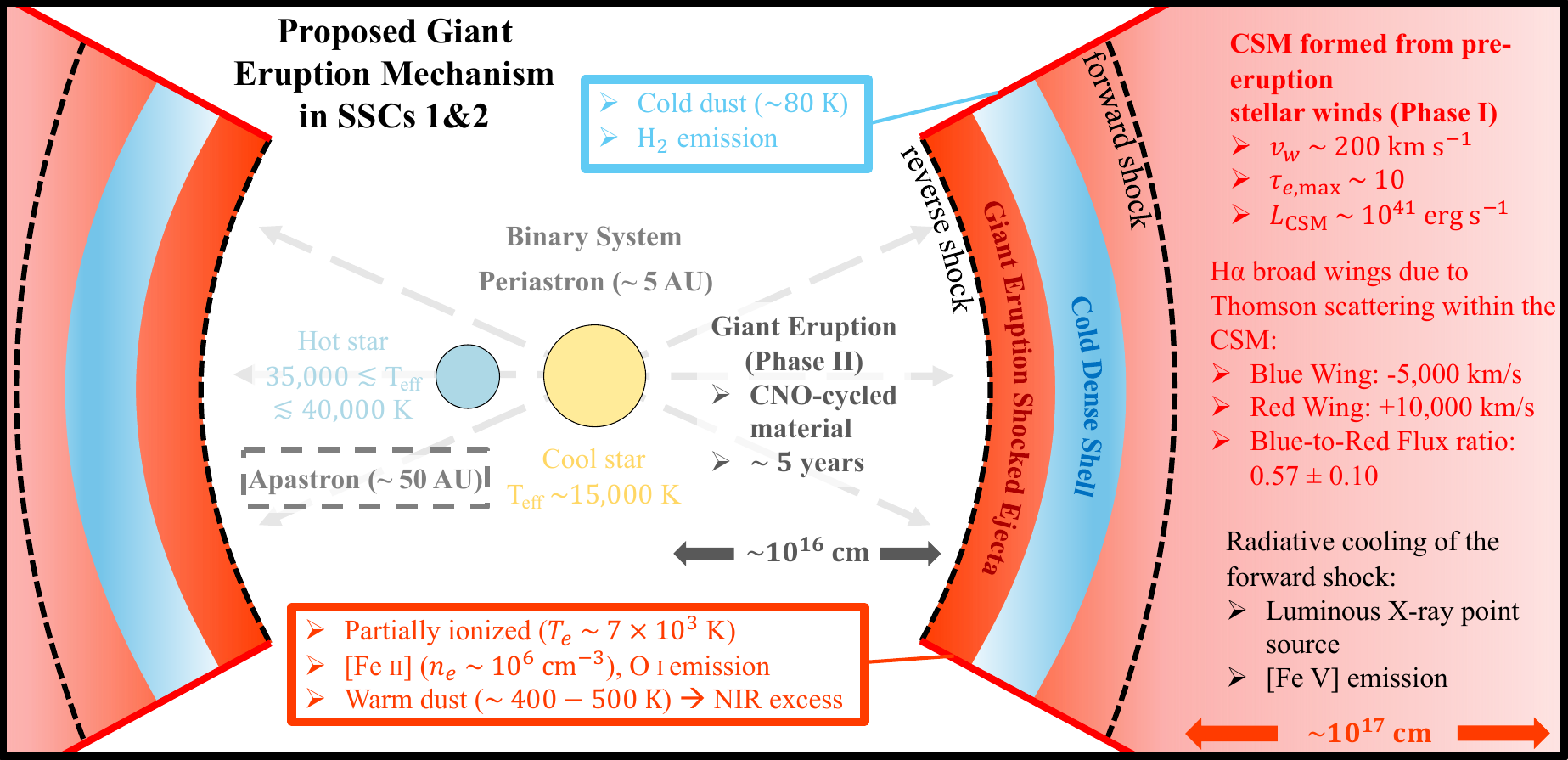}
\caption{
% Illustration of the potential geometry of the binary system near periastron and apastron (adapted from the $\eta$ Carinae picture proposed in \citealt{Mehner_2011phd} and \citealt{Gull_2022}).
Illustration of the proposed giant eruption mechanism (adapted from the $\eta$ Carinae picture; \citealp[e.g.,][]{Mehner_2011phd, Davidson_2020, Gull_2022}) in SSCs 1\&2. 
The high-density CSM initially forms from pre-eruption stellar winds (Phase I). 
Subsequent shock interaction between the giant eruption ejecta (Phase II; near periastron if the system is binary) and the CSM naturally accounts for the observed spectroscopic signatures discussed in Section \ref{sec:lbv_signs}.
We illustrate the proposed picture in Section \ref{subsec:picture}.
} 
\label{fg:lbv_cartoon}
\end{figure*}

Since these spectral features are also observed in $\eta$ Carinae (e.g., \citealt{Davidson_1982, Davidson_1986, Duncan_1995, Hillier_2001, DH_2012, Davidson_2020}), 
we propose that these unusual features in SSCs 1\&2 arise during an $\eta$ Carinae-like giant-eruption phase. 
% (or giant-outburst) phase, i.e., a significant mass loss episode.
% from efficient dust formation in the dense circumstellar medium (CSM)
% luminous blue variable (LBV)
% Such giant eruptions are often discussed in the context of super-Eddington outflows and are observationally related to ``supernova impostors''.
% Throughout this work, we adopt the term ``giant eruption'' as our primary terminology.
We calibrate the relevant physical mechanisms using the exceptionally well-studied giant eruption of $\eta$ Carinae as a reference case. 

% not necessarily ``typical'' of all LBVs \citep[i.e., far more luminous than the common type of S Doradus
% variability;][]{DH_2012, Davidson_2020}.
% Motivated by the exceptionally well-studied giant eruption of $\eta$ Carinae as a reference case for calibrating the relevant physical mechanisms (while noting that $\eta$~Car is not necessarily ``typical'' of all LBVs), we interpret the observed NIR variability in J0337$-$0502 in terms of rapid dust formation associated with CSM interaction.
% In this picture, dust condensation is favored in radiatively cooled, dense post-shock gas produced when fast giant-eruption ejecta collide with slower pre-eruption winds, rather than requiring dust formation in a steady-state wind.} \citep[e.g.,][]{Davidson_2020}.
% In this work, we propose that the observed NIR variability in J0337-0502 arises from efficient dust formation in the dense circumstellar medium (CSM) during a luminous blue variable (LBV) outburst phase, i.e., a significant mass loss episode.
% Similar to the prototypical LBV $\eta$ Carinae, the rapid dust formation may result from the shock interaction between the fast-outflowing LBV ejecta and the pre-outburst stellar wind material that forms the CSM \citep[][]{Smith_2013, Smith_2018}.

In this section, we first present a coherent physical picture based on an $\eta$ Carinae-like giant eruption that can naturally account for the unusual spectral features observed in SSCs 1\&2 (Section \ref{subsec:picture}). 
We then provide detailed analyses of each key feature in Sections \ref{subsec:high_den} through \ref{subsec:asymm_broad_balmer} (and also in Section \ref{subsec:csm_analytical_model}).

\begin{figure*}[htb]
\includegraphics[width=1.0\linewidth]{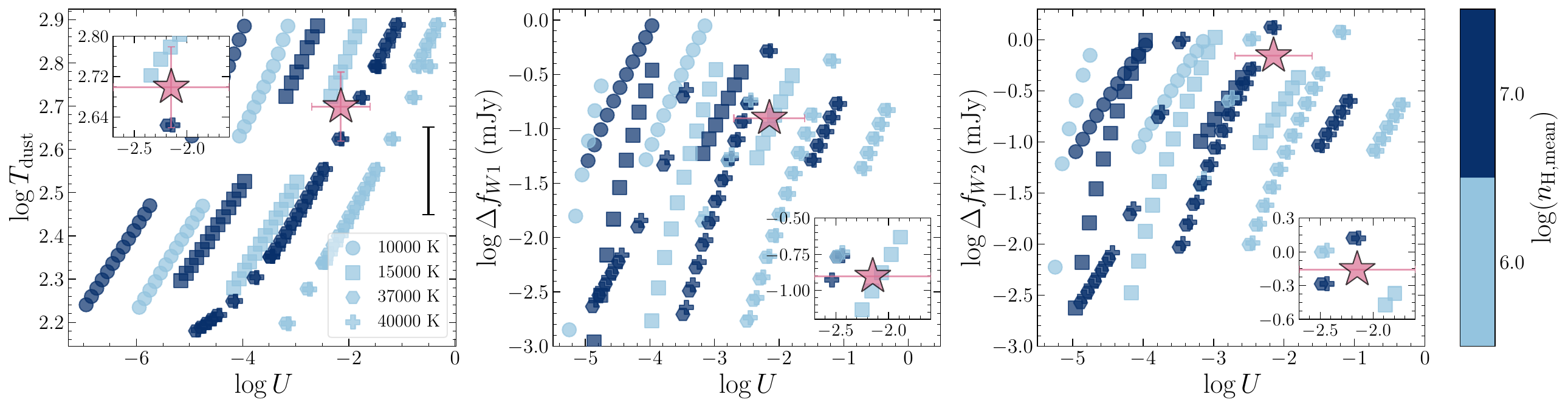}
\caption{\textit{Left:} \texttt{CLOUDY} predictions of the innermost dust temperature, $T_{\rm{dust}}$. 
The typical $1\sigma$ radial scatter in dust temperature across grain species is indicated in the panel.
\textit{Middle:} The predicted near-infrared (NIR) excess in the W1 band ($\Delta f_{\rm{W1}}$).
\textit{Right:} The predicted NIR excess in the W2 band ($\Delta f_{\rm{W2}}$). 
These predictions are compared with the observational constraints on $T_{\rm{dust}}$ from \cite{Hunt_2014} and \citetalias{Hatano_2026}, and the measured $\Delta f_{\rm{W1}}$ and $\Delta f_{\rm{W2}}$ from \citetalias{Hatano_2026}.
The displayed $\log U$ of our observation (red star) represents the ionization-weighted value from the low- to high-ionization zones (Section \ref{sec:nebular_prop}).
} 
\label{fg:j0337_dust_formations}
\end{figure*}

\subsection{A Coherent Giant Eruption Picture}\label{subsec:picture}
% \textbf{
% Figure \ref{fg:lbv_cartoon} summarizes the proposed LBV outburst scenario that can naturally account for the observed signatures, which result from the shock interaction between the LBV outburst ejecta (CNO-cycled material) and the surrounding CSM formed by pre-outburst stellar winds with $v_w \sim 200 \ \rm{km \ s^{-1}}$ and $\tau_e \sim 10$ (Section \ref{subsec:asymm_broad_balmer}). 
% This shock interaction 
During the giant eruption phase \citepalias[starting $\sim 5$ years ago based on the NIR variability;][]{Hatano_2026}, the shock interaction between the giant eruption ejecta (CNO-cycled material) and the surrounding CSM formed by pre-eruption stellar winds results in a typical forward shock/reverse shock structure \citep[see Figure \ref{fg:lbv_cartoon};][]{DH_2012, Smith_2017_typeii, CF_2017}.
At the contact discontinuity, a Rayleigh-Taylor-unstable cold dense shell (CDS) forms; this CDS is optically thick to UV radiation and likely hosts cold dust \citep[$\sim80\,$K;][]{Mingozzi_2025} and H$_2$ emission \citep{Thompson_2009}. 
Most of the outgoing flux from the reverse shock is absorbed by the CDS.
The region downstream of the reverse shock (with $n_e\sim10^{6}\,$cm$^{-3}$ and $T_e\sim7\times10^3\,$K; Section \ref{subsec:ne_Te}), partially photoionized by the primary cool star ($T_{\rm{eff}} \sim 1.5\times10^4 \ \rm{K}$), is the source of the observed [Fe\,{\footnotesize II}] emission (Section \ref{subsec:iron_lines}). 
% The potential mechanism of this rapid warm dust formation might be similar to the picture proposed in $\eta$ Carinae \citep{Smith_2013, Smith_2018}.
% This high-density CSM shock interaction efficiently converts kinetic energy into radiation through a radiative shock, cooling the gas and facilitating rapid dust formation.
This same region efficiently converts kinetic energy into radiation, cools the gas, and drives efficient warm-dust formation ($\sim400-500\,$K), thereby producing the NIR excess emission (Section \ref{subsec:high_den}). 
In contrast, radiative cooling in the forward shock explains both the luminosities of very-high-ionization lines such as [Fe\,{\footnotesize V}] $\lambda4227$ and luminous X-ray point source (Section \ref{subsec:csm_analytical_model}).
% [Fe\,{\footnotesize IV}] emission may primarily arise from the gas photoionized by the secondary hot star ($T_{\rm{eff}} \sim 3.5-4\times10^4 \ \rm{K}$) and exhibits time variability when the secondary star plunges into the primary's dense wind, resulting in a temporary cutoff of most ionizing UV flux (Section \ref{subsec:iron_lines}).
Furthermore, Thomson scattering within the pre-shock CSM causes a red-excess asymmetry in the H$\alpha$ line profile, with blue and red wings extending to velocities of $-5\,000$~km~s$^{-1}$ and $10\,000$~km~s$^{-1}$, respectively (Section \ref{subsec:asymm_broad_balmer}).
The following sections provide a detailed illustration of each observed signature summarized above. 
% }

\begin{figure*}[htb]
\centering
\includegraphics[width=1.0\linewidth]{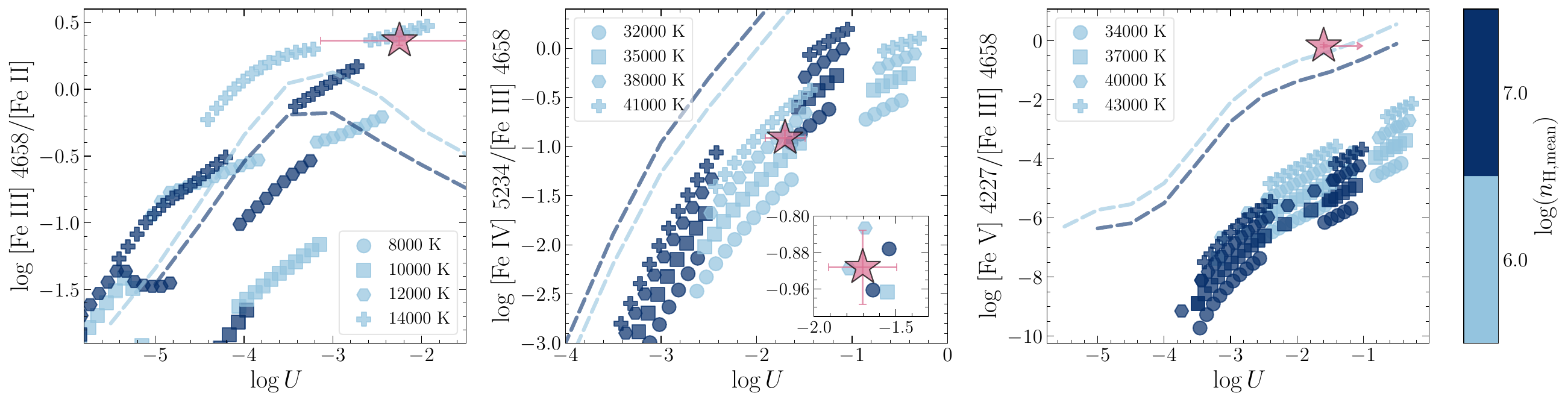}
\caption{Comparison of observed (red star) Fe line ratios--the combined [Fe\,{\footnotesize II}] emission ([Fe\,{\footnotesize II}] $\lambda5262$ plus [Fe\,{\footnotesize II}] $\lambda8617$) (left), [Fe\,{\footnotesize IV}] $\lambda5234$ (middle), and [Fe\,{\footnotesize V}] $\lambda4227$ (right)--relative to [Fe\,{\footnotesize III}] $\lambda4658$. 
Overplotted are \texttt{CLOUDY} model predictions (color-coded by $n_{\rm H}$ and shaped by input blackbody $T_{\rm eff}$) and AGN model curves (dashed lines with the same color coding).
% The coexistence of high-intensity low [Fe\,{\footnotesize II}] and high-ionization ([Fe\,{\footnotesize IV}] and [Fe\,{\footnotesize V}]) lines reveal a binary-system similar to $\eta$ Carinae where the low (dominated by $T_{\rm{eff}} \sim 1.5 \times10^4$ K) and high-ionization (dominated by $T_{\rm{eff}} \gtrsim 3.7 \times10^4$ K) regions are physically distinct.
} 
\label{fg:j0337_fe_ratios}
\end{figure*}

\subsection{High-Density Gas} \label{subsec:high_den}
The detections of the [Fe\,{\footnotesize II}] lines (Figures \ref{fg:j0337_1and2_spectrum} and \ref{fg:j0337_high_den}) reveal the high-density PI region found in SSCs 1\&2. 
Moreover, the detection of Ly$\beta$-pumped O\,{\footnotesize I} $\lambda8446$ also suggests an optically-thick neutral region where Ly$\alpha$ trapping is efficient.
O\,{\footnotesize I} $\lambda8446$ is also observed in the Weigelt knots of $\eta$ Carinae \citep{Johansson_2005}.

This high-density region might arise from the shock interaction between the giant eruption ejecta and the pre-eruption stellar wind material that forms the dense CSM. 
If the observed warm dust \citep[e.g.,][]{Hunt_2014} forms in this same region, we can use the photoionization model \texttt{CLOUDY} \citep[v23.01;][]{CLOUDY_v23} to estimate the stellar and nebular conditions required to reproduce the observed dust temperature and the corresponding NIR excess.
We vary the blackbody (BB) effective temperature, $T_{\rm{eff}}$, from $8 \times 10^3$ to $4.3 \times 10^4$ K, the total luminosity ($L_{\rm tot}$) from $10^{39}$ to $10^{40} \ \rm{erg \ s^{-1}}$ \citep[based on the S Dor instability strip on the HR diagram;][]{Wolf_1989, HD_1994, WB_2020}, the hydrogen density, $n_{\rm{H}}$, from $10^6$ to $10^7 \ \rm{cm^{-3}}$ (constraints derived from the [Fe\,{\footnotesize II}] ratio), and the inner radius, $r_{\rm{in}}$ (from the central ionizing source to the illuminated face of the cloud), at $10^{16}$, $10^{17}$, and $10^{18} \ \rm{cm}$.

Figure \ref{fg:j0337_dust_formations} shows that the \texttt{CLOUDY} model with $T_{\rm{eff}} = 1.5\times10^4$ K, $\log U \sim -2$, $n_{\rm{H}} = 10^6 \ \rm{cm^{-3}}$, and $r_{\rm{in}} = 10^{16} \ \rm{cm}$ best reproduce the warm-dust temperature ($T_{\rm{dust}} \sim 400 - 500$ K) and the observed NIR excess in the W1 and W2 bands.
Here, $T_{\rm{dust}}$ is constrained by fitting a \texttt{DUSTY} model \citep{IE_1997}, which analytically solves the radiative transfer of a spherical dust shell heated by a stellar point source, in \cite{Hunt_2014} and the derived $W1 - W2$ color temperature in \citetalias{Hatano_2026}.
This result holds when adopting a CNO-cycle chemical composition \citep{Mehner_2010}.
% \citep[e.g., $\log \rm{(N/C)} = 2.42$ and $\log \rm{(N/O)} = 2.10$ in][]{Verner_2005}
% Our observations cannot constrain the detailed dust‐grain composition or size distribution, which lie beyond the scope of this Letter.
% $\tau_{\rm{H}\alpha} \sim 3$

The \texttt{CLOUDY} model best matching the observed O\,{\footnotesize I} $\lambda8446$/H$\alpha$ ratio has $\tau_{\rm{Ly}\alpha} \sim 10^8$, supporting the existence of high-density neutral gas within SSCs 1\&2.
This inferred $\tau_{\rm Ly\alpha}$ corresponds to a hydrogen column density of $\log (N_{\rm H} / \rm cm^{-2}) \sim 22$, since $\tau_{\rm Ly\alpha} \sim \sigma_{\rm Ly\alpha} N_{\rm H} \sim (1.5-7.5)\times10^{8}$ for Doppler broadening parameters in the range $b = 10-50\,\rm km\,s^{-1}$.
Notably, this column density is consistent with that inferred from the [Fe\,{\footnotesize II}] diagnostics and best-fit \texttt{CLOUDY} model (i.e., $N_{\rm{H}}\sim n_{\rm H}\,r_{\rm in}\sim (10^{6}\,\rm{cm^{-3}})\,(10^{16}\,{cm})\sim10^{22}\,\rm{cm^{-2}}$).
The agreement between the column densities independently inferred from O\,{\footnotesize I} and [Fe\,{\footnotesize II}] emission suggests that both lines likely originate from the same high-density region (see Figure \ref{fg:lbv_cartoon}).

\subsection{Low and High Ionization Fe Lines} \label{subsec:iron_lines}
% \textbf{
The simultaneous detection of iron in ionization stages from Fe$^{+}$ to Fe$^{+4}$ in SSCs 1\&2 is rarely seen in other low-metallicity star-forming galaxies \citepalias[e.g.,][]{berg_characterizing_2021}. Such detections require not only the presence of hard ionizing photons with energies $\gtrsim55$ eV to produce [Fe\,{\footnotesize V}] emission, but also extremely dense gas capable of self-shielding against these photons to preserve [Fe\,{\footnotesize II}] emission.
% }

% Besides the detection of these low-ionization [Fe\,{\footnotesize II}] lines, we also observe the VH-ionization iron line [Fe\,{\footnotesize V}] $\lambda 4227$ within SSCs 1\&2\footnote{Although \citet{Izotov_2009} report a detection of [Fe\,{\footnotesize VI}] $\lambda5146$ in SSCs 1\&2, we find only a marginal ($\sim1.5\sigma$) detection and therefore exclude this line from our \texttt{CLOUDY} modeling. }.  
% and the best‐fit velocity centroid is redshifted by $\sim30-50\ \mathrm{km\,s^{-1}}$ relative to other emission lines. 
% Therefore, we exclude this line from our analysis.
To gain more insights into the excitation mechanism of each ionization stage of iron, we compare the observed Fe line ratios ([Fe\,{\footnotesize III}] $\lambda 4658$ (left), [Fe\,{\footnotesize IV}] $\lambda 5234$ (middle), and [Fe\,{\footnotesize V}] $\lambda 4227$ (right), relative to [Fe\,{\footnotesize II}] $\lambda 5262$ plus [Fe\,{\footnotesize II}] $\lambda 8617$) with \texttt{CLOUDY} model predictions (Figure \ref{fg:j0337_fe_ratios}). We find that the relatively low-ionization ratio ([Fe\,{\footnotesize III}]/[Fe\,{\footnotesize II}]) is consistent with a photoionization solution for a cool star ($T_{\rm{eff}} = 14\,000$ K and $L_{\rm tot}/L_{\odot} = 10^{5.70 \pm 0.20}$). 
In contrast, a high $T_{\rm{eff}}$ solution ($T_{\rm{eff}} \sim 3.5-4.0\times10^4$ K and $L_{\rm tot}/L_{\odot} = 10^{6.20 \pm 0.40}$) at $n_{\rm{H}} = 10^6 \ \rm{cm^{-3}}$ (the preferred $n_{\rm{H}}$ in Figure \ref{fg:j0337_dust_formations}) can potentially reproduce the [Fe\,{\footnotesize IV}]/[Fe\,{\footnotesize III}] ratio.
The inferred $T_{\rm{eff}}$ for the low and high ionization lines are similar to those of $\eta$ Carinae which exists in a binary system \citep{Mehner_2010, Davidson_2012}.
However, even the hottest models ($T_{\rm{eff}} = 4.3 \times 10^4$ K) underpredict the observed [Fe\,{\footnotesize V}]/[Fe\,{\footnotesize III}] ratio by $3-4$ orders of magnitude.

To explore whether AGN photoionizes the high ionization lines (see details in Appendix \ref{sec:appendix_agn_models}), we run a set of AGN models through \texttt{CLOUDY} based on the nebular properties of SSCs 1\&2, shown as dashed lines in Figure \ref{fg:j0337_fe_ratios}.
These AGN models underestimate the [Fe\,{\footnotesize III}]/[Fe\,{\footnotesize II}] and overestimate the [Fe\,{\footnotesize IV}]/[Fe\,{\footnotesize III}] ratios, but can potentially reproduce the observed [Fe\,{\footnotesize V}]/[Fe\,{\footnotesize III}] ratio.
However, the AGN models that can potentially reproduce the [Fe\,{\footnotesize V}]/[Fe\,{\footnotesize III}] ratio have $n_{\rm H} = 10^{6} \ \rm{cm^{-3}}$, which is comparable to the critical density of [Fe\,{\footnotesize V}] $\lambda 4227$, where collisional de-excitation becomes significant. 
Consequently, the predicted [Fe\,{\footnotesize V}] line luminosity is substantially lower than observed; specifically, the modeled [Fe\,{\footnotesize V}] $\lambda 4227$/H$\beta$ ratio ($\sim 10^{-4}$) is approximately a dex smaller than the observed value ($\sim 10^{-3}$).
% Assuming the VH-ionization [Fe\,{\footnotesize V}] line originates in the BLR, we estimate its luminosity using the BLR radius, $r_{\rm{BLR}}$, from the scaling relation of \citet{Greene_2005}, yielding $r_{\rm{BLR}} \sim 10^{16} \ \rm{cm}$.  
% The predicted [Fe\,{\footnotesize V}] luminosity for this $r_{\rm{BLR}}$ is $5$–$6$ orders of magnitude lower than the observed value ($\sim 3 \times 10^{37}\,\rm{erg\,s^{-1}}$). 
We propose that this VH-ionization line arise from the radiative cooling of the CSM shock interaction (see Section \ref{sec:discussions}). 

\begin{figure}[!htb]
\centering
\includegraphics[width=1\linewidth]{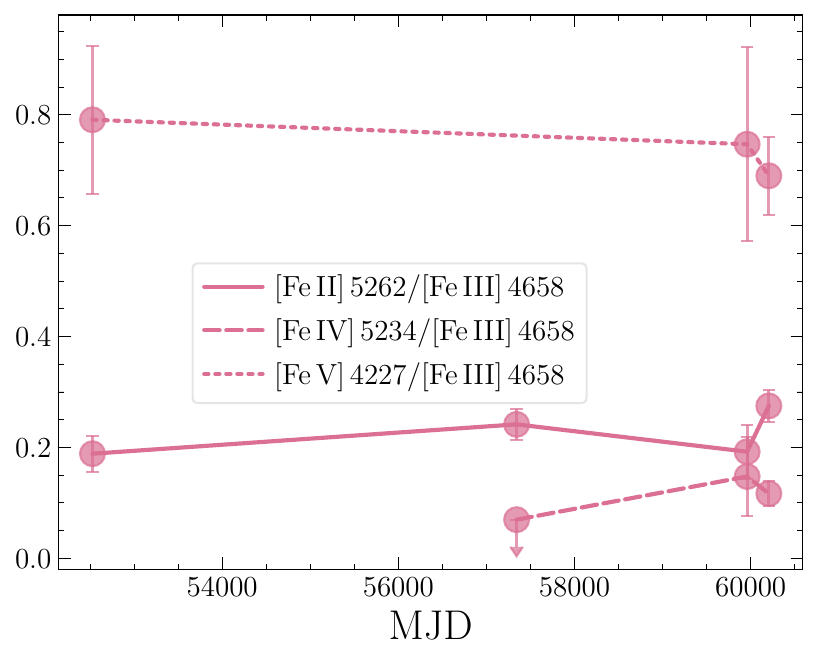}
\caption{
% \textit{Top}: Illustration of the potential geometry of the binary system near periastron and apastron (adapted from the $\eta$ Carinae picture proposed in \citealt{Mehner_2011phd} and \citealt{Gull_2022}). 
% \textit{Bottom}: 
Temporal evolution of forbidden Fe line ratios--[Fe\,{\footnotesize II}]/[Fe\,{\footnotesize III}] (solid), [Fe\,{\footnotesize IV}]/[Fe\,{\footnotesize III}] (dashed), and [Fe\,{\footnotesize V}]/[Fe\,{\footnotesize III}] (dotted)--based on VLT/FORS (MJD 52527; \citealt{Izotov_2009}), VLT/MUSE (MJD 57342; \citealt{Herenz_2023, Herenz_2023_data}), Keck/ESI (MJD 59959; \citealt{Peng_2025}), and our KCWI/KCRM data (MJD 60205; this work). 
% Approximate epochs near periastron (coinciding with the giant eruption, based on NIR variability in the W1 and W2 bands reported by \citetalias{Hatano_2026}) and apastron (assuming a current orbital period of $\sim 5.5$ years, similar to $\eta$ Carinae) are indicated.
}
\label{fg:iron_ratios_vs_time}
\end{figure}

To further explore the excitation mechanisms for these iron lines, we also present the temporal evolution of their line ratios for SSCs 1\&2 over the past $\sim 20$ years in Figure \ref{fg:iron_ratios_vs_time}.
% based on VLT/FORS (MJD 52527; \citealt{Izotov_2009}), VLT/MUSE (MJD 57342; \citealt{Herenz_2023, Herenz_2023_data}), Keck/ESI (MJD 59959; \citealt{Peng_2025}), and our KCWI/KCRM data (MJD 60205; this work) in  
% Figure \ref{fg:iron_ratios_vs_time}. 
The [Fe\,{\footnotesize II}]/[Fe\,{\footnotesize III}] (solid) and [Fe\,{\footnotesize V}]/[Fe\,{\footnotesize III}] (dotted) ratios remain consistent within 1$\sigma$ over $\sim$20 years. 
However, the [Fe\,{\footnotesize IV}] (dashed) detection in the VLT/MUSE  spectrum (MJD 57342; \citealt{Herenz_2023, Herenz_2023_data}) is below 3$\sigma$ (and is not reported for VLT/FORS; \citealt{Herenz_2023, Herenz_2023_data}), so we only derive upper limits. 
% \textbf{
The tentative variability of [Fe\,{\footnotesize IV}] (i.e., increases by a factor of $\gtrsim2$ over the past eight years) can be qualitatively explained if the giant-eruption candidate in SSCs 1\&2 exists in a binary system analogous to $\eta$ Carinae \citep[e.g., references in][]{DH_2012}.
In this scenario, the secondary hot star ($T_{\rm{eff}} \sim 3.5-4.0\times10^4$ K) should provide the ionizing photons responsible for the [Fe\,{\footnotesize IV}] emission, while the primary cool star ($T_{\rm{eff}} \sim 1.5\times10^4$ K), having recently undergone eruptions, sustains the [Fe\,{\footnotesize II}] emission.
During the binary system's periastron (roughly when the proposed giant eruption phase starts), the ionizing UV radiation from the secondary star is temporarily suppressed as it plunges into the primary's dense wind, leading to a significant decline in the [Fe\,{\footnotesize IV}] luminosity.

% \textbf{
Nevertheless, we emphasize that a binary system has not been shown to be strictly necessary for an $\eta$ Carinae-like giant eruption \citep[e.g.,][]{Davidson_2020}.
Hence, higher-cadence observations will be necessary to confirm any periodic variability in [Fe\,{\footnotesize IV}] and to distinguish it from enhanced emission driven by radiative cooling in the proposed CSM interaction during the giant-eruption phase (i.e., non-periodic; see Section \ref{subsubsec:vh_lines_ulx} for details).
% }

\begin{figure*}[htb]
\centering
\includegraphics[width=1.0\linewidth]{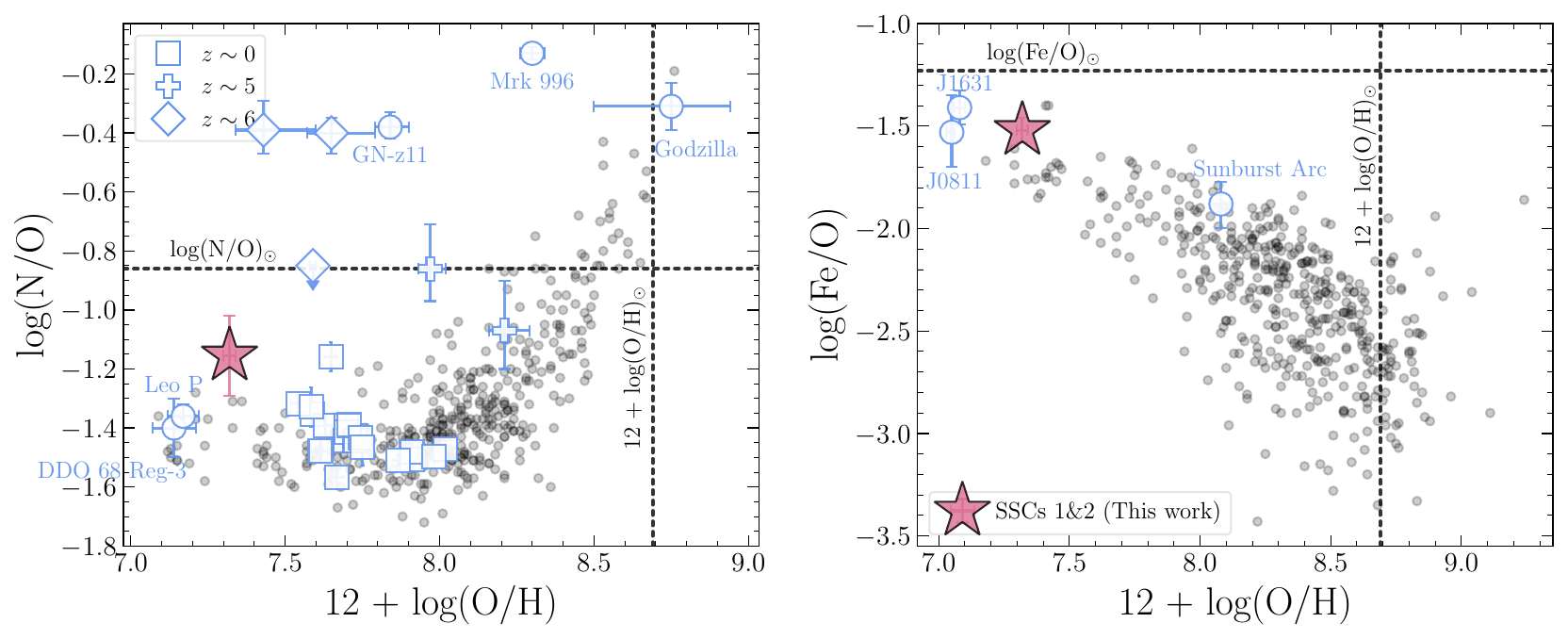}
\caption{Nitrogen (\textit{left}) and iron (\textit{right}) abundances of J0337-0502 SSCs 1\&2 (red star) are illustrated in the N/O-O/H and Fe/O-O/H panels, respectively. 
% In both panels, the measurements for J0337 SSCs 1\&2 (red star) are compared with its observation $\sim 15$ years ago \citep[blue star;][]{Izotov_2009}.
\textit{Left:} We compare our nitrogen measurement to the local H\,{\footnotesize II} regions (gray circles) from \cite{Pilyugin_2012}, local metal-poor dwarf galaxies at $z<0.04$ (blue squares) from \cite{Berg_2019}, high N/O galaxies at $z \sim 5$ \citep[blue crosses;][]{AC_2024} and $z \sim 6$ \citep[blue diamonds;][]{Yanagisawa_2024} observed by JWST, and several well-known high N/O galaxies at both low and high redshifts, including DDO 68 Region-3 \citep{Annibali_2019}, Leo P \citep{Skillman_2013}, GN-z11 \citep{Senchyna_2024}, Mrk 996 \citep{James_2009}, and Godzilla of the Sunburst Arc \citep{PD_2024}.
\textit{Right:} Our iron measurement is compared to the galaxy sample compiled by \cite{MD_2024} (gray circles), including two local extremely low-metallicity galaxies \citep[J0811+4730 and J1631+4426;][]{Izotov_2018, Kojima_2021}
and a known Lyman-continuum emitter in the Sunburst Arc \citep[$z = 2.37$;][]{Welch_2025}.
% and composite spectra of galaxies at $z \sim 2 - 3$ \citep{Steidel_2016, Cullen_2021} as well as at $z \sim 6$ \citep{Harikane_2020}.
}
\label{fg:j0337_NO_FeO_OH_ratios}
\end{figure*}
% (TODO: Needs some stromgren radius measurment here)
\subsection{Nitrogen and Iron Abundances: Comparison with Literature Values}\label{subsec:no_feo_analysis}

\subsubsection{$\rm{N/O}$ Abundance Ratio}\label{subsubsec:no_analysis}
We find that the N/O abundance ratio in SSCs 1\&2, $\log(\rm{N/O})\sim -1.30$ to $-1.0$ (accounting for systematic ICF(N$^+$) uncertainties; Section \ref{subsec:abundance}), is approximately $0.2-0.5$ dex higher than that of typical local H\,{\footnotesize II} regions \citep{Pilyugin_2012} at the extreme low-metallicity end (12 + log(O/H) $\lesssim 7.5$) in the N/O-O/H diagram (left panel of Figure \ref{fg:j0337_NO_FeO_OH_ratios}), where previous studies have reported a ``tight plateau'' ($\rm{log(N/O)_{plat}} \sim -1.6$ to $-1.4$) at this end \citep[e.g.,][]{IT_1999, Nava_2006}.
% over the low-metallicity range of 12 + log(O/H) $\sim 7.0$–7.5 \citep[e.g.,][]{IT_1999}. 

This observed high N/O ratio is more extreme when compared with known local metal-poor dwarf galaxies with mild nitrogen enrichments ($\rm{\log{N/O}} - \rm{log(N/O)_{plat}} \lesssim 0.1$), such as Region-3 (Reg-3) of DDO 68 \citep{Annibali_2019}, Leo P \citep{Skillman_2013}, and targets like J1044+0353 presented in \cite{Berg_2019}. 
The broad-emission component of Mrk 996 ($z = 0.0054$) is the sole exception, where a similar nitrogen enrichment is observed in a relatively metal-rich environment (12 + log(O/H) $\sim 8.3$; \citealt{James_2009}). 

Nitrogen enhancements are more commonly observed in high-redshift galaxies, e.g., Godzilla of the Sunburst Arc \citep[$z \sim 2.4$;][]{PD_2024}, several $z \sim 5$ \citep{AC_2024} and $z \sim 6$ \citep{Yanagisawa_2024} galaxies observed by JWST, and GN-z11 \citep[$z \sim 10.6$;][]{Senchyna_2024}. 
Among the high N/O targets, potential giant eruption and/or LBV candidates have been identified in Reg-3 of DDO 68 \citep{IT_1999} and Godzilla \citep{Choe_2025}. 
\citet{Choe_2025} also qualitatively claim that the giant-eruption candidate Godzilla likely has a hotter companion star, a scenario we propose for SSCs 1\&2 in Section \ref{subsec:iron_lines}.

% In contrast to a primary element, synthesized directly from primordial compositions \citep[e.g.,][]{Tinsley_1980}, like oxygen that mainly ejected by core-collapse supernovae (CCSN) in massive stars, nitrogen has both primary and secondary production (e.g., produced from CNO burning in stars with existing carbon and oxygen) channels and often originates in intermediate-mass stars with delayed release via winds.
% For example, the CNO burning predicts that the secondary-to-primary abundance ratio is proportional to the primary abundance \citep[i.e., $Z_{\rm{sec}}/Z_{\rm{prim}}\propto Z_{\rm{prim}}$;][]{Pagel_2009}, which is supported by the observed positive correlation between N/O and O/H at $\rm{12 + log(O/H)} \gtrsim 7.5$.

The conventional primary and secondary production channels of nitrogen cannot explain the large scatter of N/O values observed at the low-metallicity end \citep[e.g.,][]{Tinsley_1980, Pagel_2009}, which suggests the need for earlier nitrogen enrichment prior to the onset of CCSNe from massive stars (i.e., $\lesssim 3 \ \rm{Myr}$). 
For example, an explanation based on the ejection of nitrogen-rich material during the Wolf-Rayet (WR) phase (typically $\sim 3 - 6 \ \rm{Myr}$) cannot reproduce such high N/O ratios at these early times given the extremely young stellar populations of SSCs 1\&2 ($\sim 3 \ \rm{Myr}$; see Table \ref{table:ssc_properties}); additionally, the characteristic broad helium emission lines and WR bumps associated with WR winds are not observed in SSCs 1\&2.

We argue that the nitrogen excess observed in SSCs 1\&2 results from the ejection of CNO-cycled material during giant eruptions \citep[i.e., nitrogen is naturally enriched relative to oxygen;][]{Davidson_1982, Davidson_1986, Hillier_2001, Przybilla_2010, Maeder_2014}.
Previous studies \citep{Timmes_1995, Maeder_2001} have demonstrated that this process can be important in the low-metallicity environment where the efficiency of rotational mixing in massive stars is significantly enhanced, facilitating the transport of CNO-processed material from the stellar core to the outer envelope well before the end of the main-sequence phase (i.e., $\lesssim 3 \ \rm{Myr}$). 

An immediate question is how many giant eruptions like $\eta$ Carinae ($N_{\rm GE}$) are needed to account for the observed nitrogen enrichment. 
Adopting an initial abundance ratio of $\log(\mathrm{N/O})_i \simeq -1.4$,
corresponding to the low-metallicity ``tight plateau'' without primary nitrogen enrichment,
and a final abundance ratio of $\log(\mathrm{N/O})_f \simeq -1.3$ (the lower limit of our measurements), we estimate that $N_{\rm GE}\sim 6 - 25$ (see Appendix \ref{sec:nitrogen_enrich_lbv} for derivation details) is sufficient to explain the observed nitrogen enrichment.
A similar analysis has been done in NGC 5253 \citep{Kobulnicky_1997}, where they find $\sim$ 12 giant eruptions are needed to reproduce the localized nitrogen enrichment. 
% if they assume $y_{\mathrm{N}} =2 \times 10^{-3}$ and $M_{\rm ej} = 6 \ M_{\odot}$.
% We caution, however, $N_{\rm GE}$ may easily vary by a factor of a few, given that $R_{\mathrm{N/O}}$, $y_{\mathrm{N}}$, and $M_{\rm{ej}}$ are not well constrained by observations or simulations.
% If $\log \left(\rm{N/O} \right)_f \sim -1.45$ \citep{Izotov_2009}, then $R_{\mathrm{N/O}}\sim1.4$ and $N_{\rm{LBV}} \sim 40$.
The inferred $N_{\rm GE}$ should be regarded as an upper limit, as we do not account for other physical mechanisms \citep[e.g., winds from very massive stars with zero-age main-sequence masses of $100-1000\,M_{\odot}$;][]{Vink_2023} that may also contribute to the observed nitrogen enhancement.

\subsubsection{$\rm{Fe/O}$ Abundance Ratio}\label{subsubsec:feo_analysis}
In addition to nitrogen enrichment, SSCs 1\&2 also exhibit a relatively high Fe/O abundance ratio ($\log \rm{Fe/O} \sim -1.5$) compared to typical H\,{\footnotesize II} regions at similar low metallicities \citep[][see right panel of Figure \ref{fg:j0337_NO_FeO_OH_ratios}]{MD_2024}. Nevertheless, SSCs 1\&2 might be relatively iron-poor compared to other SSCs (see Section \ref{subsec:abundance}). 
This potential iron depletion indicates that CNO-cycled material from giant eruptions promotes the formation of abundant metallic dust (dominated by FeSi and Fe) within the CSM \citep[e.g.,][]{Gail_2025}.

The $\log \rm{Fe/O}$ value for SSCs 1\&2, as well as for other SSCs, is comparable to those observed in two local extremely low-metallicity galaxies, J0811+4730 and J1631+4426 \citep{Izotov_2018, Kojima_2021, MD_2024}, where 12 + log(O/H) $\lesssim 7.1$. This may indicate iron enrichment from very massive stars ($M_\ast > 300 \ M_\odot$; \citealt{Kojima_2021}) and/or pair-instability supernovae, as seen in high N/O systems such as GN-z11 \citep{Nakane_2024}, occurring before the onset of CCSNe.
Nevertheless, we caution that such comparisons are subject to the choice of Fe ICF adopted in the literature. For example, the \citet{MD_2024} sample uses the \citet{RR_2005} Fe ICF, which can differ by $\sim 0.2$ dex from direct estimates based on all iron ionization states.

\defcitealias{HC_2018}{HC18}
\begin{figure*}[htb]
\centering
\includegraphics[width=1\linewidth]{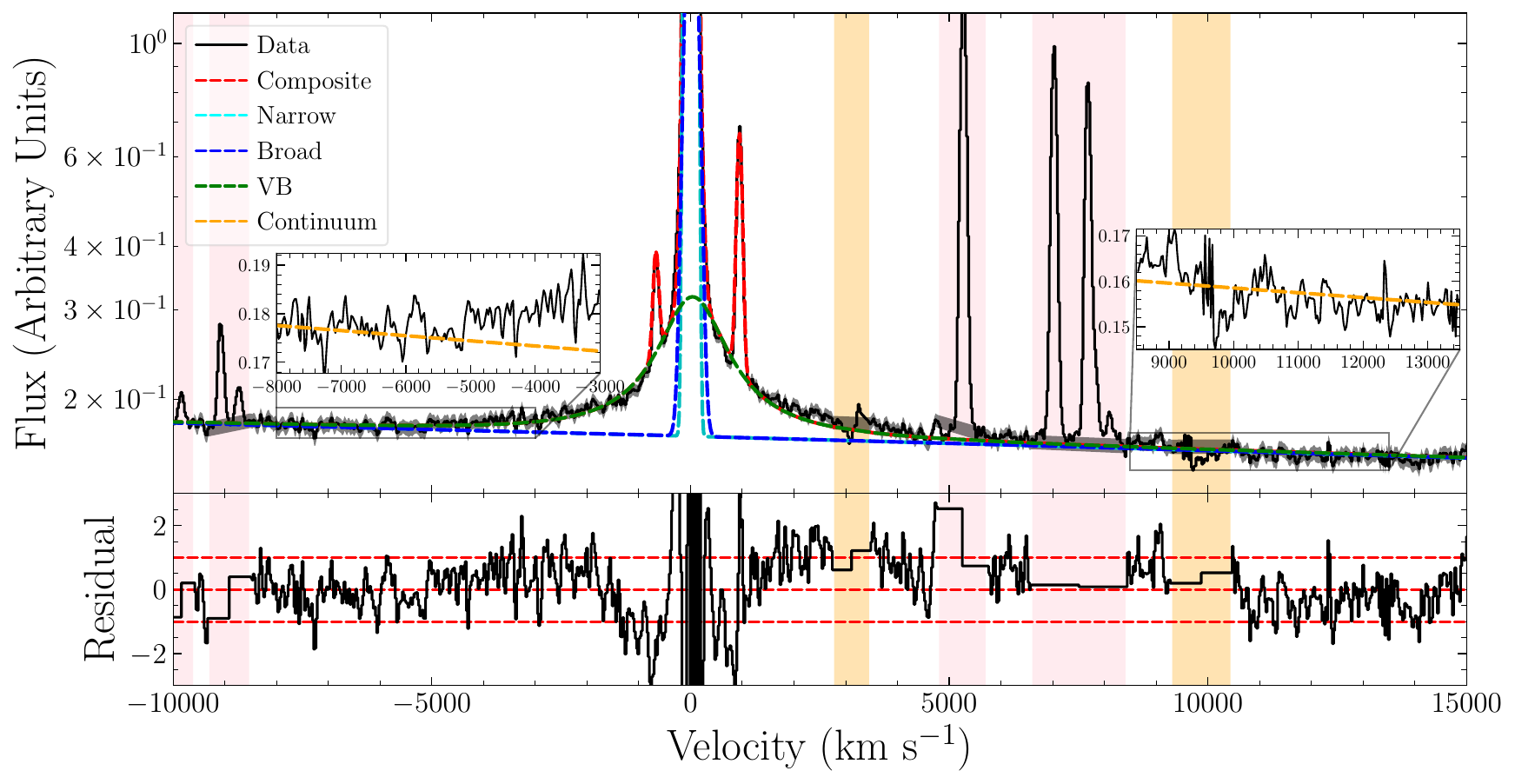}
\includegraphics[width=1\linewidth]{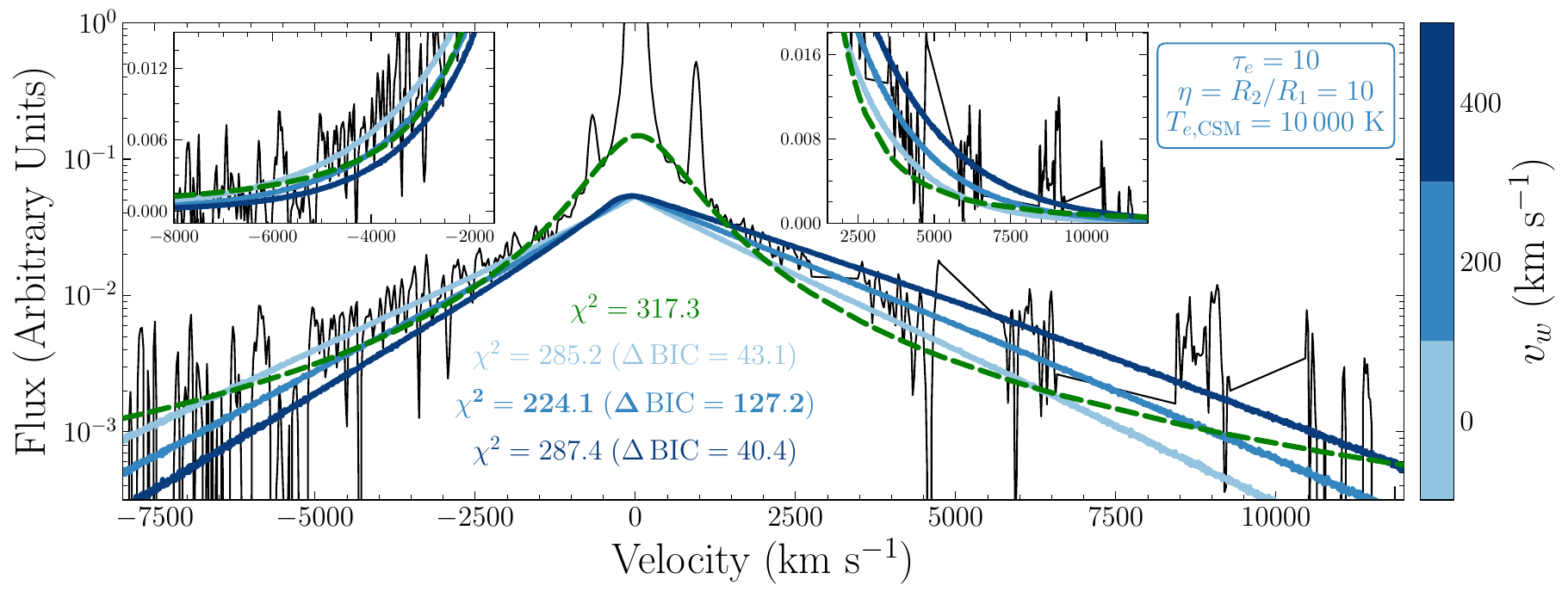}
\caption{Asymmetric H$\alpha$ broad wings with a red excess. \textit{Top}: The best-fitting dGL model (red dashed line) is plotted alongside the observed line profile (black solid line). The narrow, broad, and VB emission velocity components are shown as cyan, blue, and green-dashed lines. Grey shading indicates the 1$\sigma$ uncertainties of the raw data. Strong emission lines (or bad sky-subtraction features) next to the intended lines are masked during line fittings and are shaded in pink (or yellow). 
By visual inspection, the blue and red wings reach the best-fit continuum (orange dashed line) at $\sim -5\,000 \ \rm{km \ s^{-1}}$ and $\sim 10\,000 \ \rm{km \ s^{-1}}$, respectively.
Perturbing the line profile 1\,000 times based on the error spectrum and adopting a 1$\sigma$ detection threshold yields $v_{\rm{max}} = -5\,370_{-70}^{+140} \ \rm{km\,s^{-1}}$ (blue) and $9\,940_{-90}^{+50} \ \rm{km\,s^{-1}}$ (red), consistent with the visual estimate.
The residual plot (data - model / error) is displayed in the second row.
The 1$\sigma$ residuals are indicated by red dashed lines.
\textit{Bottom:} Comparison between the best-fit Lorentzian profile (green dashed) for the VB component and the \citetalias{HC_2018} models color-coded at different $v_w$ (see Section \ref{subsec:asymm_broad_balmer}).
% The $v_w = 200 \ \rm{km \ s^{-1}}$ model is indicated by the solid blue line, while the blue-shaded region represents the range of \citetalias{HC_2018} models spanning $v_w = 0$ to $400 \ \rm{km \ s^{-1}}$. 
The $\chi^2$ values for the Lorentzian profile and the \citetalias{HC_2018} models, along with the corresponding difference in the Bayesian Information Criterion ($\Delta \rm{BIC} = \rm{BIC}_{\rm{Lorentzian}} - \rm{BIC}_{\rm{HC18}}$; \citealp{lmfit}), are displayed in the panel.
% \textbf{HC: Can you apply different line styles for $v_w=0$ and $v_w=400$? The current plot provides an incorrect illusion that the shaded region indicates the 1$\sigma$ uncertainty. It is almost impossible for the reader to realize that the $v_w=0$ comprises the upper boundary of the blue wing and the lower boundary of the red wing.}
% \textit{Bottom Right:} Observed blue‑to‑red flux ratio in the broad wings (red line) compared to the \citetalias{HC_2018} models (blue points).
}
\label{fg:j0337_asymmetric_balmer_line}
\end{figure*}

% * Also plot the Fe/O ratio versus O/H in SSCs1\&2? (dust-depletion due to efficient dust formation) \\ 

\subsection{Asymmetric Broad Wings in Balmer Lines}\label{subsec:asymm_broad_balmer}
The asymmetric broad wings of the Balmer line profiles in J0337-0502 were first discovered in the integrated Keck/ESI spectrum from \cite{Peng_2025}. 
They find that the broad wings of H$\beta$ exhibit a pronounced red excess--i.e., the red wing extends further in velocity space than the blue wing--and cannot be adequately fitted with either a Gaussian or a Lorentzian function. 
Our KCWI/KCRM observations confirm this result. 
Because the red wing of H$\beta$ is blended with the blue wing of [O\,{\footnotesize III}] $\lambda4959$, we focus on the H$\alpha$ broad wings with a higher SNR.
% , whose higher SNR more clearly shows the red excess. 
Figure \ref{fg:j0337_asymmetric_balmer_line} shows that the H$\alpha$ red wing extends to $\sim10\,000\ \mathrm{km\,s^{-1}}$, whereas the blue wing reaches only $\sim-5\,000\ \mathrm{km\,s^{-1}}$.
The observed $v_{\rm{max}}$ far exceeds the typical values ($v_{\rm{max}}\sim1-2\times10^3\ \rm{km\,s^{-1}}$) predicted by both single-phase \citep{Thompson_2016} and multiphase \citep{Fielding_2022} galactic wind models. 
% which also produce symmetric emission profiles (e.g., \citealp{Peng_2025}). Consequently, additional mechanisms are needed to explain the high-velocity, asymmetric broad wings.

% \textbf{
Previous studies of $\eta$ Carinae and of SNe IIn \citep[e.g.,][]{Davidson_1995, Hillier_2001, Mehner_2015, Dessart_2009, Humphreys_2012, Fransson_2014, Davidson_2020} have demonstrated that the high-velocity, asymmetric broad wings arise from Thomson scattering. 
% }
% Photons can acquire large Doppler shifts through multiple scattering events before escaping from the optically thick CSM. 
% The radial expansion of the CSM preferentially shifts photons toward longer wavelengths, producing the observed red excess in the broad wings \citep[see Section 2.3 in][]{HC_2018}.
The observed red excess in the broad wings arises from the spherical divergence of the radially expanding flow \citep[e.g., see Section 2.3 in][]{HC_2018}.
To test this hypothesis, we employ the Monte Carlo Thomson-scattering framework of \cite{HC_2018} (hereafter HC18)\footnote{The publicly available code provided by \citetalias{HC_2018} can be accessed at \url{https://github.com/Huang-CL/elsc/}}. 
\citetalias{HC_2018} models adopt the Thomson scattering differential cross-section for each scattering event that occurs in a fully ionized medium with uniform temperature $T_{e,\rm CSM}$, in the shape of a spherical shell spanning radii $R_1$ to $R_2$. 
The medium has an Thomson-scattering optical depth $\tau_e$ measured in the radial direction. 
We adopt an $r^{-2}$ density profile for the scattering medium, representing a constant pre-eruption mass outflow.  
We apply a constant outflow velocity by ignoring the radiative shock acceleration ($v_{\rm sh}\sim 0$), which is only significant shortly after the shock breakout.
%\citetalias{HC_2018} models adopt the Thomson scattering differential cross‑section for each scattering event spanning radii $R_1$ to $R_2$, and treat the scattering medium as fully ionized hydrogen with uniform temperature, $T_{e,\rm CSM}$ (fixed at 10\,000\,K), and electron‑scattering optical depth, $\tau_e$. 
%We adopt an $r^{-2}$ density profile for the scattering medium and assume that radiative shock acceleration is negligible at the time of observation ($v_{\rm sh}\sim 0$), becoming significant only during shock breakout.

We explore the parameter space in $\tau_e$ and shell-radius ratio $\eta = R_2/R_1$ from 1 to 10, and CSM expansion velocity $v_w$ from 0 to 1\,000 km s$^{-1}$ to identify the combination that best reproduces the observed H$\alpha$ broad-wing profile. 
% Compared to the symmetric Lorentzian fit of the VB component, 
The bottom panel of Figure \ref{fg:j0337_asymmetric_balmer_line} shows the \citetalias{HC_2018} Thomson-scattering models at $v_w$ from 0 to 400 $\rm{km\,s^{-1}}$. 
These models are normalized to the peak flux in the broad-wing region, using the blue-wing velocity interval $\sim -5\,000$ to $\sim -1\,000\,$km\,s$^{-1}$ and the red-wing interval $\sim1\,500$ to $\sim10\,000\,$km\,s$^{-1}$, thereby avoiding contamination by [N\,{\footnotesize II}] $\lambda\lambda6548,6583$.
We find that the \citetalias{HC_2018} model with $v_w=200\,$km s$^{-1}$, $\tau_e=10$, and $\eta=10$ yields the lowest $\chi^2$ within the broad-wing region. 
% The $\chi^2$ calculation uses the blue‐wing velocity range $-5\,000$ to $-1\,500\,$km s$^{-1}$ and the red‐wing range $1\,500$ to $10\,000\,$km s$^{-1}$, avoiding contamination from [N\,{\footnotesize II}] $\lambda\lambda6548,83$.  

\begin{figure}[htb]
\centering
\includegraphics[width=1\linewidth]{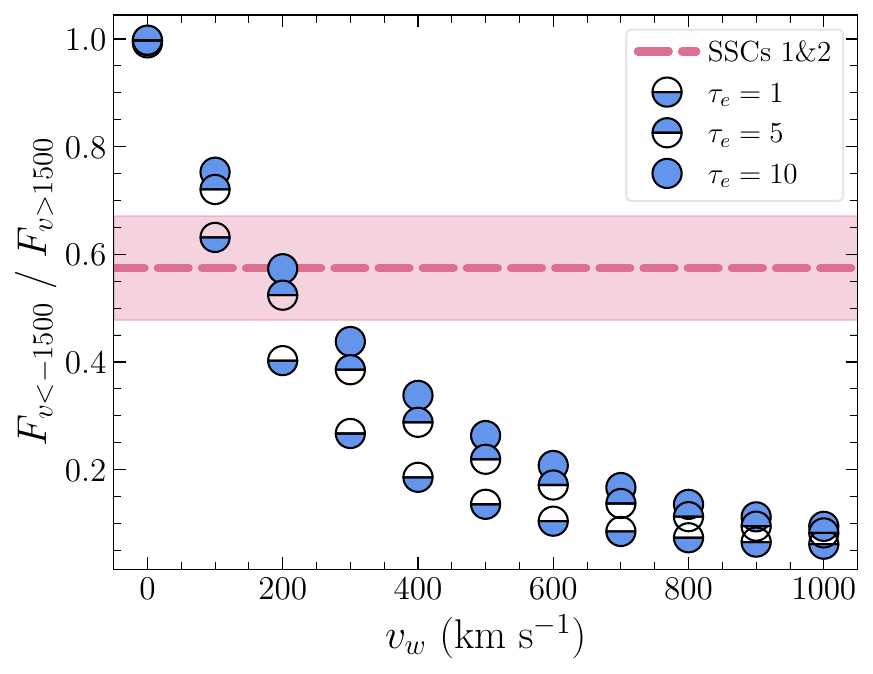}
\caption{Observed blue-to-red flux ratio in the broad wings (red line) compared to those of \citetalias{HC_2018} models (blue points) at $\tau_e$ = 1, 5, and 10 with $\eta = 10$ and $T_{e,\rm{CSM}} = 10\,000$ K.
}
\label{fg:j0337_asymmetric_balmer_line_ratio}
\end{figure}

This conclusion is further supported by the comparison between the observed blue-to-red flux ratio ($F_{v<-1500} / F_{v>1500}$) in the broad wings ($0.57 \pm 0.10$), defined as the integrated flux for $v<-1\,500\ \rm{km\,s^{-1}}$ relative to $v>1\,500\ \rm{km\,s^{-1}}$, and those of \citetalias{HC_2018} models at different $v_w$.
Figure \ref{fg:j0337_asymmetric_balmer_line_ratio} shows that $F_{v<-1500}/F_{v>1500}$ decreases with increasing $v_w$, and that the setup $v_w=200\ \rm{km\,s^{-1}}$ best reproduces the observed ratio, being insensitive to the choice of $\tau_e$ in the case of large $\eta$ considered in this study.
%and that the setup $v_w=200\ \rm{km\,s^{-1}}$ with $\tau_e=10$ best reproduces the observed ratio. 
This result is in agreement with our \(\chi^2\) analysis shown in Figure~\ref{fg:j0337_asymmetric_balmer_line}.

One caveat of the \citetalias{HC_2018} Thomson scattering model is its assumption of spherical symmetry. In contrast, observations of systems such as $\eta$ Carinae indicate the presence of a dense, bipolar-shaped CSM \citep[e.g.,][]{DH_2012, Davidson_2020}. 
However, the physical mechanism responsible for producing the red excess in emission-line profiles does not inherently require spherical symmetry, but rather a radially outward-expanding CSM like $\eta$ Carinae. 
Therefore, our inferred $\tau_e$ from the best-fit \citetalias{HC_2018} model may only reflect the densest regions of the CSM if the CSM shares a similar non-spherically symmetric geometry to $\eta$ Carinae, and $\tau_e$ could be significantly less than 10 in other directions (i.e., $\tau_{e,\,\rm max} \sim 10$).
% A detailed analysis of such asymmetric electron scattering models is beyond the scope of this work.

% if these extremely high-velocity H$\alpha$ broad wings arise from electron scattering within the high-density CSM, we would expect [O\,{\footnotesize III}] $\lambda5007$ to exhibit a similar $v_{\rm max}$. 
% Because [O\,{\footnotesize III}] $\lambda5007$ is saturated (Section \ref{sec:observation_and_data}), we extract the Keck/ESI spectrum \citep{Peng_2025} of SSCs 1\&2 (shown in Appendix \ref{sec:esi_ssc1a2_o3_ha}) to investigate this point.
Furthermore, we find that the [O\,{\footnotesize III}] $\lambda5007$ line extracted from the Keck/ESI observation of SSCs 1\&2 (see Appendix \ref{sec:esi_ssc1a2_o3_ha}; \citealp{Peng_2025}) exhibits a relatively smaller $v_{\rm max}$ ($\sim 5\,000\ \rm km\,s^{-1}$) than H$\alpha$ ($\sim 10\,000\ \rm km\,s^{-1}$). 
This discrepancy likely suggests that the broad wings of [O\,{\footnotesize III}] $\lambda5007$ do not arise from Thomson scattering within the dense CSM, since in that case one would expect a similar $v_{\rm max}$ to H$\alpha$. 
% \textbf{
Moreover, the forbidden [O\,{\footnotesize III}] $\lambda5007$, with a critical density of $n_{\rm crit}\sim 10^6\,\rm{cm^{-3}}$ \citep{Osterbrock_2006}, would become strongly collisionally de-excited and approaches saturation within the dense CSM, whereas permitted lines such as H$\alpha$ (i.e., with a much higher $n_{\rm crit}$ of $\sim 10^{15} \, \rm{cm^{-3}}$) continue to increase in emissivity with density in this regime \citep{Osterbrock_1988}.
% }

% \textbf{
Therefore, the broad wings of [O\,{\footnotesize III}] $\lambda5007$ likely originate from lower-density ejecta 
% ($300 \lesssim n_e \lesssim 2500$ in the high-ionization zone, as constrained by the [S\,{\footnotesize II}] and [Ar\,{\footnotesize IV}] density diagnostics in Section \ref{subsec:ne_Te}) 
located outside the bipolar CSM, analogous to the outer ejecta of $\eta$ Carinae beyond the Homunculus (with an average density $\gtrsim 500\,\rm{cm^{-3}}$), which reach velocities of $\sim 2\,000-6\,000\ \rm km\,s^{-1}$ \citep{Davidson_1984, Smith_2004, Smith_2008}. 
The driving mechanism of this fast outer ejecta remains uncertain. 
For example, \citet{Smith_2008} propose a core-explosion blast wave, possibly triggered by pulsational pair instability, while alternative scenarios suggest that such high velocities could arise from shocks accelerating outward through the steep density gradient in the star's outer layers and wind \citep{Davidson_2012}.
% This difference likely explains the relatively smaller $v_{\rm max}$ observed in [O\,{\footnotesize III}] $\lambda5007$ ($\sim5\,000~\rm{km\,s^{-1}}$) compared to H$\alpha$ ($\sim 10\,000 \ \rm{km\,s^{-1}}$). 
% }
% Given that the CSM density is high enough that the forbidden [O\,{\footnotesize III}] emission starts to become strongly collisionally excited and approaches saturation, we find that the red wing of [O\,{\footnotesize III}] $\lambda5007$ extends only to $\sim5\,000~\rm{km\,s^{-1}}$ (while the blue wing is blended with [O\,{\footnotesize III}] $\lambda4959$).
% The Keck/ESI H$\alpha$ profile exhibits a similar $v_{\rm max}$, from $\sim -5\,000$ to $\sim 10\,000 \ \rm{km\,s^{-1}}$, as that observed in our KCWI/KCRM data.

\section{Discussions} \label{sec:discussions}
Based on the spectroscopically-derived properties (Section \ref{sec:nebular_prop}), we explore the potential giant eruption signatures and propose a coherent picture for the giant eruption in SSCs 1\&2 in Section \ref{sec:lbv_signs}.
To further discuss the proposed giant eruption picture, we present order-of-magnitude estimates of the CSM properties 
% using an analytical model developed for Type IIn supernovae \citep{Moriya_2013} 
in Section \ref{subsec:csm_analytical_model}.
In Section \ref{subsec:agn_possibilities}, we also discuss whether the observed spectroscopic features of SSCs 1\&2 can be reproduced by the existence of an AGN.  
Last but not least, due to the detection of O\,{\footnotesize I} $\lambda8446$, we illustrate the significance of Ly\(\alpha\) radiation pressure in SSCs 1\&2 in Section \ref{subsec:lya_radiation}. 

\subsection{Crude Estimations of CSM Properties}\label{subsec:csm_analytical_model}
% \textbf{
In this section, using an analytical framework for CSM shock interaction \citep{Moriya_2013}, we estimate a shock radius of $R_{\rm{sh}} \sim 10^{16} \ \rm{cm}$ and a bolometric luminosity of the CSM interaction of $L_{\rm{CSM}} \sim 10^{41} \ \rm{erg \ s^{-1}}$ at the observation epoch (Section \ref{subsubsec:csm_radius_lum_mass}).
% and an upper mass limit on the preshock CSM of $M_{\rm CSM, \, pre} \sim 30 \, M_{\odot}$ (Section \ref{subsubsec:csm_radius_lum_mass}). 
We argue that radiative cooling from $L_{\rm{CSM}}$ can naturally explain the luminosities of the observed VH-ionization [Fe\,{\footnotesize V}] $\lambda4227$ line and the \textit{Chandra} observation of the luminous X-ray point source (Section \ref{subsubsec:vh_lines_ulx}). 
We also explore if the high-density CSM can efficiently trap He\,{\footnotesize II} $\lambda304$ photons to enhance He\,{\footnotesize II} $\lambda4686$ emission in Section \ref{subsubsec:heii_ulx_connection}.
% Detailed derivations are presented below.
% }

% \subsubsection{Estimations of $R_{\rm sh}$, $L_{\rm CSM}$, and $M_{\rm CSM}$}\label{subsubsec:csm_radius_lum_mass}
\subsubsection{Estimations of $R_{\rm sh}$ and $L_{\rm CSM}$}\label{subsubsec:csm_radius_lum_mass}

% Previous studies of interacting supernovae \citep[e.g.,][]{Chevalier_1982, CC_2011, Moriya_2013, Fransson_2014} have extensively investigated CSM shock interactions.
% , which we discuss as applicable to LBV outbursts. 
We assume the dense CSM can be described by
a steady wind with a pre-eruption mass-loss rate of $\dot{M}_{w}$ and a constant wind velocity of $v_w$ before the giant eruption. 
This assumption gives a density profile of the CSM as 
\begin{align}
\label{eq:5.0} \rho_{\rm{CSM}} &= \frac{\dot{M}_{w}}{v_w \ 4 \pi r^{2}} = D \ r^{-2} \nonumber \\ 
&\simeq 7.5 \times 10^{16} \ D_{\ast} \ r^{-2} \ \rm{g \ cm^{-3}},
\end{align}
where $D_{\ast} = \dot{M}_{w, 0.3} / v_{w,200}$, $\dot{M}_{w, 0.3} = \dot{M}_{w} / 0.3 \ \rm{M_{\odot} \ yr^{-1}}$, and \(v_{w,200} = v_w / 200\,\mathrm{km\,s^{-1}}\). 
The normalization for \(v_w\) is taken from the best-fit value in our Thomson-scattering model (Section \ref{subsec:asymm_broad_balmer}), and \(\dot{M}_w\) is based on the value of \(\eta\) Carinae \citep{Smith_2013}.
% The resulting $D$ value lies in the ``giant eruption'' regime \textbf{\citep{DH_2012}}.
% To reproduce both the Thomson optical depth required for the asymmetric H$\alpha$ broad wings (Section \ref{subsec:asymm_broad_balmer}) and the luminosity of the very-high-ionization [Fe\,{\footnotesize V}] $\lambda4227$ line ($\sim 3\times10^{37}\,$erg \ s$^{-1}$; see below), we explore $D_\ast$ values from 1 to 10 and found XX. 
% Therefore, we normalize XXX
% $D_{\ast, 2.5} = D_{\ast} / 2.5$ 
% hereafter for our estimations.
% This $D_{\ast}$ value places the LBV candidate in SSCs 1\&2 within the ``giant eruptions'' regime, analogous to \(\eta\) Carinae \citep{Smith_2017}.
The density structure of the giant eruption ejecta can be described as $\rho_{\rm{ej}} \propto r^{-k}$ where we adopt a fiducial value of $k = 7$ that is 
% appropriate for $\eta$ Carinae \citep{Smith_2013} and 
roughly consistent with some SNe IIn candidates \citep[e.g.,][]{Fransson_2014}. 

Following this model setup, 
% we use Equation (8) of \citet{Moriya_2013} to estimate the time required for the outburst ejecta to fully interact with the CSM:
%  \begin{align}
%     \label{eq:5.1} t_i \simeq 2 - 3 \ \frac{M_{\rm{ej,10}}^{3/2}}{E_{\rm{ej,50}}^{1/2}} \frac{v_{w,300}}{\dot{M}_{\rm{w,0.3}}} \ \rm{years}.
% \end{align}
% \begin{align}
%     \label{eq:5.1} t_i \simeq 0.5 - 1 \ \frac{M_{\rm{ej,10}}^{3/2}}{E_{\rm{ej,51}}^{1/2}} D_{\ast}^{-1} \ \rm{years}.
% \end{align}
% \begin{align}
%     \label{eq:5.1} t_i \simeq 0.3 - 0.7 \ \frac{M_{\rm{ej,10}}^{3/2}}{E_{\rm{ej,51}}^{1/2}} D_{\ast}^{-1} \ \rm{years}.
% \end{align}
% \begin{align}
%     \label{eq:5.1} t_i \simeq 0.2 - 0.4 \ \frac{M_{\rm{ej,10}}^{3/2}}{E_{\rm{ej,51}}^{1/2}} D_{\ast,2.5}^{-1} \ \rm{years}.
% \end{align}
% Here we normalize the LBV outburst ejecta mass ($M_{\rm{ej},10} = M_{\rm{ej}} / 10 \ M_{\odot}$) as the values based on $\eta$ Carinae \citep{Smith_2013}.
% and mechanical energy ($E_{\rm{ej}} / 10^{50} \ \rm{erg}$) . 
% The wind velocity is expressed as \(v_{w,300} = v_w / (300\,\mathrm{km\,s^{-1}})\), based on the best‐fit value from our electron‐scattering model (Section \ref{subsec:asymm_broad_balmer}). 
% The ejecta kinetic energy is normalized as \(E_{\rm ej,51} = E_{\rm ej}/10^{51}\,\mathrm{erg}\), chosen to reproduce the observed [Fe\,{\footnotesize V}] \(\lambda4227\) luminosity (see discussion below).
the asymptotic solution of the CSM shell position, $R_{\rm{sh}}$, can be derived as: 
% \citep[Equation (19) in][]{Moriya_2013}:
% \begin{align}
%     \label{eq:5.2} R_{\rm{sh}} &\simeq 3.2\times10^{16} \ \frac{v_{w,300} M_{\rm{ej,10}}}{\dot{M}_{\rm{w,0.3}}} \nonumber \\
%     &\left[-1 + \left(1 + 0.2 \sqrt{\frac{E_{\rm{ej,50}}}{M_{\rm{ej,10}}^{3}}} \frac{\dot{M}_{\rm{w,0.3}}}{v_{w,300}} t_1\right)^{1/2} \right].
% \end{align}
% \begin{align}
%     \label{eq:5.2} R_{\rm{sh}} &\simeq 3.2\times10^{16} \ \frac{M_{\rm{ej,10}}}{D_{\ast}} \nonumber \\
%     &\left[-1 + \left(1 + 0.6 \sqrt{\frac{E_{\rm{ej,51}}}{M_{\rm{ej,10}}^{3}}} D_{\ast} t_1\right)^{1/2} \right].
% \end{align}
\begin{align}
    \label{eq:5.2} R_{\rm{sh}} &\simeq 2\times10^{16} \ \frac{M_{\rm{ej,10}}}{D_{\ast}} \nonumber \\
    &\left[-1 + \left(1 + 0.9 \sqrt{\frac{E_{\rm{ej,51}}}{M_{\rm{ej,10}}^{3}}} D_{\ast} t_1\right)^{1/2} \right] \ \rm{cm}.
\end{align}
% \begin{align}
%     \label{eq:5.2} R_{\rm{sh}} &\simeq 2\times10^{16} \ \frac{M_{\rm{ej,10}}}{D_{\ast}} \nonumber \\
%     &\left[-1 + \left(1 + 0.3 \sqrt{\frac{E_{\rm{ej,50}}}{M_{\rm{ej,10}}^{3}}} D_{\ast} t_1\right)^{1/2} \right] \ \rm{cm}.
% \end{align}
Here, we normalize the giant eruption ejecta mass ($M_{\rm{ej},10} = M_{\rm{ej}} / 10 \ M_{\odot}$), adopting the value based on the giant eruption of $\eta$ Carinae \citep[][]{Davidson_2012}, and the ejecta kinetic energy as $E_{\rm ej,51} = E_{\rm ej}/10^{51}\,\rm{erg}$. 
% \citep[the kinetic energy distribution exhibits double peaks at $\sim10^{50}$ and $\sim10^{51}$ erg for Type-IIn supernovae;][]{Hiramatsu_2024}. 
% \textbf{HC: This sentence is a bit ambiguous. Do you suggest the $10^{50}$ erg is the peak of the kinetic energy distribution for LBV? You may want to clarify it. And if so, why do you normalize the energy with $10^{51}$ erg, rather than $10^{50}$ erg? }
% The ejecta kinetic energy is normalized as \(E_{\rm ej,50} = E_{\rm ej}/10^{50}\,\rm{erg}\), consistent with the kinetic energy range of \(\sim 1 - 2 \times 10^{50} \ \mathrm{erg}\) estimated for $\eta$ Carinae \citep{Smith_2008, Smith_2013}.
% chosen to reproduce the observed [Fe\,{\footnotesize V}] \(\lambda4227\) luminosity (see discussion below).
Assuming the shock interaction has persisted for \(t \sim 5\) years ($t_1 \sim 5$), based on the observed duration of the NIR plateau \citepalias{Hatano_2026}, we estimate a shock radius of $R_{\rm sh} \sim 1-1.5 \times 10^{16}\,\rm{cm}$ for $E_{\rm{ej},51} \sim 0.1-0.2$, comparable to that of $\eta$ Carinae's giant eruption \citep{Davidson_2012}. This shock radius is consistent with the inner radius of our best-fit \texttt{CLOUDY} models (Section \ref{subsec:high_den}). 
% The thickness of the CSM shell, \(\Delta R_{\rm sh}\), can be constrained by the electron‐density estimate from [Fe\,{\footnotesize II}] lines via
% \begin{align}
%     \label{eq:5.3} \overline{n}_{e} &\simeq \int_{R_{\rm{sh}}}^{R_{\rm{sh}} + \Delta R_{\rm{sh}}} \frac{\rho_{\rm{CSM}}}{\overline{m}_H} \ d r / \Delta R_{\rm{sh}}
% \end{align}
% where 
% % $\rho_{\rm{CSM}} = 5\times10^{16} \left(\dot{M}_{\rm{w,0.3}} / {v_{w,300}}\right) \ r^{-2} \ \rm{g \ cm^{-3}}$ and 
% \(\overline m_H \simeq 0.7 m_p\) is the mean particle mass for a partially ionized medium. Adopting \(\overline{n}_e\sim10^{6.5}\ \mathrm{cm^{-3}}\) at $T_e$([Fe\,{\footnotesize II}])$_{\rm{CL}}$ $\sim7\,500$ K (Figure \ref{fg:j0337_high_den}), we find 
% \(\Delta R_{\rm sh}\sim7\times10^{17}\)\,cm. The corresponding Thomson optical depth is then
% \begin{align}
%     \label{eq:5.4} \tau_{e} = \int_{R_{\rm{sh}}}^{R_{\rm{sh}} + \Delta R_{\rm{sh}}} \frac{\rho_{\rm{CSM}}}{\overline{m}_H} \ \sigma_{\rm{Th}} \ dr \simeq 1.5,
% \end{align}
% where $\sigma_{\rm{Th}}$ is the Thomson scattering cross section.
% , which is consistent with our best-fit electron-scattering model (double-chek).

% To quantitatively evaluate this argument, we first need to estimate 
The bolometric luminosity of the CSM interaction at the observation epoch can be estimated as:
% based on Equation (29) in \citet{Moriya_2013}: 
% \begin{align}
%     \label{eq:5.5} L_{\rm{CSM}} &\simeq 1.5 \times 10^{42} \ \epsilon_{0.1} D_{\ast} \left( \frac{E_{\rm{ej,51}}}{M_{\rm{ej,10}}}\right)^{3/2} \nonumber \\ 
%     &\left[ 1 + 0.9 D_{\ast} \left( \frac{E_{\rm{ej,51}}}{M_{\rm{ej,10}}^{3}}\right)^{1/2} t_1 \right]^{-3/2} \ \rm{erg \ s^{-1}},
% \end{align}
\begin{align}
    \label{eq:5.5} L_{\rm{CSM}} &\simeq 7.5 \times 10^{42} \ \epsilon_{0.5} D_{\ast} \left( \frac{E_{\rm{ej,51}}}{M_{\rm{ej,10}}}\right)^{3/2} \nonumber \\ 
    &\left[ 1 + 0.9 D_{\ast} \left( \frac{E_{\rm{ej,51}}}{M_{\rm{ej,10}}^{3}}\right)^{1/2} t_1 \right]^{-3/2} \ \rm{erg \ s^{-1}},
\end{align}
% \begin{align}
%     \label{eq:5.5} L_{\rm{CSM}} &\simeq 1.6 \times 10^{41} \ \epsilon_{0.5} D_{\ast} \left( \frac{E_{\rm{ej,50}}}{M_{\rm{ej,10}}}\right)^{3/2} \nonumber \\ 
%     &\left[ 1 + 0.3 D_{\ast} \left( \frac{E_{\rm{ej,50}}}{M_{\rm{ej,10}}^{3}}\right)^{1/2} t_1 \right]^{-3/2} \ \rm{erg \ s^{-1}},
% \end{align}
where $\epsilon_{0.5} = \epsilon / 0.5$ is the conversion efficiency from kinetic energy to radiation for the CSM shock interaction \citep[appropriate for the derived $v_w$ and the CSM to giant eruption ejecta mass ratio like $\eta$ Carinae;][]{Marle_2010}. 
Therefore, $L_{\rm{CSM}} \sim 0.6 - 1.3\times10^{41} \, \rm{erg \,s^{-1}}$ for $t_1 \sim 5$ and $E_{\rm{ej},51} \sim 0.1-0.2$.

\subsubsection{Estimated Luminosities of [Fe\,{\footnotesize V}] and X-ray Point Source Emission}\label{subsubsec:vh_lines_ulx}
Since both stellar and AGN models fail to reproduce the luminosity of the [Fe\,{\footnotesize V}] $\lambda4227$ line ($\simeq 3\times10^{37}\,$erg \ s$^{-1}$), we propose that this emission may instead arise from the radiative cooling of the CSM interaction. 
By integrating the [Fe\,{\footnotesize V}] $\lambda4227$ emissivity—peaking at $\sim10^5\,$K—over $10^4-10^8\,$K and comparing it to the total radiative cooling rate across the same range, we estimate that $\sim0.02 - 0.03\%$ of the cooling luminosity is emitted in [Fe\,{\footnotesize V}] $\lambda4227$ \citep[][]{PS_2020, chianti_v10}. 
Therefore, we estimate the CSM interaction can produce an [Fe\,{\footnotesize V}] $\lambda4227$ luminosity of  $\sim(1-4)\times10^{37}\ \mathrm{erg\,s^{-1}}$ when $t_1 \sim 5$ and $E_{\rm{ej},51} \sim 0.1-0.2$, in good agreement with the observed line luminosity. 
% Therefore, to reproduce the [Fe\,{\footnotesize V}] $\lambda4227$ line luminosity at $t_1 \sim 5$, the ejecta kinetic energy is estimated as \(E_{\rm{ej},51} \sim 0.2\), which is roughly consistent with the kinetic energy released by the giant eruption of $\eta$ Carinae \citep{Smith_2008}.
% When $t_1 \sim 5$, we have $L_{\rm{CSM}} \sim 6 \times 10^{41} \ \rm{erg \ s^{-1}}$.
% Therefore, we infer that the CSM interaction can radiate $\sim1.2$–$2.5\times10^{37}\ \mathrm{erg\,s^{-1}}$ in [Fe\,{\footnotesize V}] $\lambda4227$, in approximate agreement with the observed line luminosity. 
On the contrary, radiative cooling of this CSM interaction can only contribute to less than 50\% of the observed [Fe\,{\footnotesize IV}] $\lambda 5234$ line luminosity ($\sim 6 \times 10^{36} \ \rm{erg \ s^{-1}}$). 
Consequently, the majority of [Fe\,{\footnotesize IV}] $\lambda 5234$ emission should originate from stellar photoionization (e.g., from the secondary hot star if the giant eruption candidate exists in a binary system), consistent with our \texttt{CLOUDY} modeling in Section \ref{subsec:high_den}.

Similarly, to compare with the \textit{Chandra} observations \citep{Thuan_2004}, we predict the X-ray luminosity in the $0.5-10\,$keV band (corresponding to $T\sim6\times10^6-10^8\,$K), which is dominated by free-free bremsstrahlung cooling \citep{Fransson_2014, PS_2020}, contributes to $\sim3\%$ of the total cooling luminosity. 
% (i.e., $L_{0.5-10\,\rm{keV}}\sim2-4\times10^{39}\,\mathrm{erg\,s^{-1}}$).
This yields $L_{0.5-10\,\rm keV} \sim (2-4)\times10^{39}\,\mathrm{erg\,s^{-1}}$, in reasonable agreement with the \textit{Chandra} measurement of $(2.8-3.5)\times10^{39}\,\mathrm{erg\,s^{-1}}$ \citep{Thuan_2004} for the X-ray point source.
% This estimation is in reasonable agreement with the \textit{Chandra} measurement of $(2.8-3.5)\times10^{39}\,\mathrm{erg\,s^{-1}}$ \citep{Thuan_2004} for the X-ray point source.
This estimation supports the interpretation that the X-ray point source originates from SSC 2 \citep{Thuan_2004, Kehrig_2018}, as the CSM interaction naturally reproduces its observed luminosity.
The differing column densities inferred from the Thomson-scattering wings ($N_{\rm H,max} \sim 10^{25} \, \rm{cm^{-2}}$) and the \textit{Chandra} X-ray data ($N_{\rm H} \sim 10^{22} \, \rm{cm^{-2}}$) could be explained by small-scale porosity in the CSM, which allows X-rays to leak through low-density channels \citep{Fransson_2014}.

Furthermore, given our $\tau_{e, \, \rm{max}}$ constraint from Monte-Carlo simulations of asymmetric H$\alpha$ line profile (Section \ref{subsec:asymm_broad_balmer}), we can estimate an upper limit on the X-ray photon energy. 
Each Compton scattering reduces the photon energy (with an initial energy of $E_0$), and after approximately $N (\sim \tau_{e, \, \rm{max}}^2)$ scatterings in the optically thick CSM, the remaining photon energy is roughly $E \approx m_e c^2/\tau_{e, \, \rm{max}}^2 = 511/\tau_{e, \, \rm{max}}^2 \ \rm{keV}$ \citep[when $E \ll E_0$;][]{CF_2017}.
For $\tau_{e, \, \rm{max}} \sim 10$ (Figure \ref{fg:j0337_asymmetric_balmer_line}), this corresponds to an upper energy limit of $E \approx 511 / 10^2 \ \rm{keV} \approx 5 \ \rm{keV}$. 
% consistent with the observed absence of hard X-ray photons in the \textit{Chandra} spectrum \citep{Thuan_2004}.

% Nevertheless, the X-ray point source is located $\sim0\farcs3 - 0\farcs7$ north of SSCs 1\&2 \citep[see Figure \ref{fg:j0337_hst_plot};][]{Thuan_2004, Prestwich_2013, Kehrig_2018}. 
% Previous studies suggest that this X-ray emission may originate from SSC 2 \citep{Thuan_2004, Kehrig_2018}, as the position offset can be attributed to the $\sim0\farcs42$ rms astrometric uncertainty of its position.
% Our estimates support this interpretation, as the CSM interaction in SSCs 1\&2 naturally reproduces its observed luminosity.
% If this X‐ray emission originates from the CSM interaction in SSCs 1\&2, the position offset may be explained by the $\sim0\farcs42$ rms astrometric uncertainty of the ULX position \citep{Thuan_2004}.
% The differing column densities inferred from the electron-scattering wings ($N_{\rm H,max} \sim 10^{25} \, \rm{cm^{-2}}$) and the \textit{Chandra} X-ray data ($N_{\rm H} \sim 10^{22} \, \rm{cm^{-2}}$) could be explained by small-scale porosity in the CSM, which allows X-rays to leak through low-density channels \citep{Fransson_2014}.

Hard X-ray emission can also arise from a dense ($n_e\gtrsim 10^{10}\,\rm{cm^{-3}}$) colliding-wind region in a massive binary system (located near $\sim 10^{13}\,\rm{cm}$ relative to the primary star), as proposed for $\eta$ Carinae; however, its peak X-ray luminosity is only of order $L_{0.5-10\,\rm{keV}} \sim 10^{35}\,\mathrm{erg\,s^{-1}}$ \citep[e.g.,][]{PC_2002, Martin_2006_HeII, CI_2012}. 
Therefore, although hard X-ray emission may originate from colliding-wind binaries, this mechanism might be far too faint to account for the observed luminosity of the X-ray point source.

\subsubsection{The Mystery of the Extreme He\,{\footnotesize II} Emission}\label{subsubsec:heii_ulx_connection}
% \textbf{
\citet{Kehrig_2018} find that the extreme He\,{\footnotesize II} $\lambda4686$ emission of J0337-0502 can only marginally be explained by 
% either single, rotating metal-free stars or 
a \texttt{BPASS} (Binary Population and Spectral Synthesis) \texttt{v2.1} model \citep{BPASS_2017} with a $0.05\%\,Z_{\odot}$ binary population ($\sim1\%$ of the derived gas-phase $Z$ of SSCs 1\&2) and a top-heavy initial mass function (IMF).
% This conclusion does not qualitatively change when using more recent \texttt{BPASS} models that include the contribution of X-ray binaries (XRBs), because the XRB contribution to the He,{\footnotesize II}-ionizing photon budget at $\sim3,\rm{Myr}$ and $Z = 5\%,Z_{\odot}$ is $\lesssim 0.1\%$ compared to that of the stellar population \citep[][]{XBPASS}.
Moreover, \citet{Kehrig_2018} argue that the X-ray point source cannot provide a sufficient He\,{\footnotesize II}-ionizing photon rate, $Q(\rm{He}^{+})$, to account for the observed He\,{\footnotesize II} $\lambda4686$ emission in SSCs 1\&2.
% }

% \textbf{
This conclusion should be reassessed in light of the high-density CSM found in this work. 
In such an extremely dense CSM, the He\,{\footnotesize II} $\lambda4686$ emission can be significantly enhanced by trapped He\,{\footnotesize II} $\lambda304$ ($\rm{He^{+}}\,1s-2p$) photons in the $\rm He^{++}$ zone photoionized by soft X-ray photons from the CSM interaction (i.e., dominated by free-free bremsstrahlung).
This trapping of He\,{\footnotesize II} $\lambda304$ photons substantially increase the population of $\rm He^{+}$ ions in the $n=2$ level \citep[i.e., similar to the effect of trapped Ly$\alpha$ in pumping the $n=2$ level in neutral hydrogen;][]{Martin_2006_HeII}.
We adapt the \citet{Martin_2006_HeII} model (originally for the colliding-wind region of $\eta$ Carinae) to our CSM parameters (see Appendix \ref{sec:martin_2006_model} for details).
% we can test if photon trapping at He\,{\footnotesize II}\,$\lambda304$ can reproduce the extreme $L_{\lambda4686}$.
The He\,{\footnotesize II}\,$\lambda4686$ enhancement factor is quantified using the ratio of the line flux to the soft X-ray flux in the 54--550 eV band ($F(\rm{He\,{\footnotesize II}}\,\lambda4686)\,/\,F(54\!-\!550\,\rm{eV})$), which increases with $n_e$ (see their Figure 9). 
The intrinsic $L_{\rm 54-550\,eV}$ can be extrapolated from the fixed-$N_{\rm H}$ ($\sim10^{22}\,\rm cm^{-2}$) 
% power-law ($L_{\rm 54-550\,eV} \sim 4.5 \times 10^{39}\,\rm erg\,s^{-1}$) or 
optically thin plasma model ($L_{\rm 54-550\,eV} \sim 6 \times 10^{38}\,\rm erg\,s^{-1}$) for the hard X-ray point source in \citet{Thuan_2004}. 
% (the limited photon counts cannot distinguish which model provides a better fit).
% $\sim0.07$ ($\sim0.5$)
To reproduce the observed He\,{\footnotesize II} $\lambda4686$ luminosity within SSCs 1\&2 ($L_{4686}\simeq 3\times 10^{38}\ \mathrm{erg\ s^{-1}}$, comparable to the value derived for the similar region ``Knot B'' in \citealp{Kehrig_2018}), $F(\rm He\,{\footnotesize II}\,\lambda4686)/F(54-550\,\rm eV)$ must be $\sim0.5$. 
% $\sim3$ ($\sim10$)
This value exceeds the model prediction by factors of $\sim10$ given the estimated parameter values illustrated in Appendix \ref{sec:martin_2006_model}.
% assuming $\eta \sim 10^{-10}\,\rm s^{-1}$ and $A_{2c} \sim 10$, as illustrated above.
Moreover, since the predicted $F(\rm{He\,{\footnotesize II}}\,\lambda4686)\,/\,F(54\!-\!550\,\rm{eV})$ ratios should be regarded as upper limits \citep{Martin_2006_HeII}, the extreme He\,{\footnotesize II} emission remains unexplained. Nevertheless, since many parameters in Appendix \ref{sec:martin_2006_model} are poorly constrained, future radiation-hydrodynamic simulations could self-consistently determine the He\,{\footnotesize II}\,$\lambda4686$ enhancement factor in the high-density CSM.

\subsection{AGN Possibilities}\label{subsec:agn_possibilities}

% \textbf{
We assess whether the emission-line properties of SSCs 1\&2 can be explained by an AGN by examining three commonly invoked BLR/NLR mechanisms: self-Balmer absorption in dense clouds, collisional excitation affecting neutral helium lines, and optical forbidden iron emission from an AGN narrow-line region (NLR). 
% For each mechanism, we test whether the required physical conditions are consistent with the observed line profiles, line ratios, and continuum properties. 
As shown below, none of these scenarios simultaneously reproduces all observational constraints, and an AGN origin for the emission-line properties in SSCs 1\&2 is therefore disfavored. \\
% } 

\noindent
\underline{\textit{Self-Balmer Absorptions}}: Asymmetric Balmer-line profiles are typically observed in AGN. Various models are proposed to explain these asymmetric profiles, including deviation from a Keplerian motion, a binary super-massive BH, and an asymmetric distribution of the BLR clouds. In addition, \cite{Gaskell2018} pointed out that partial obscuration of the BLR by compact dusty clumps can also produce these asymmetries. Similarly, for a dust-free cloud where a sufficiently high $n_{\rm H}$ can populate the $n = 2$ level of neutral hydrogen through collisional excitation,
%H\,{\footnotesize I} column density populates the $n = 2$ level of neutral hydrogen,
leading to significant self-Balmer absorption \citep[e.g.,][]{Netzer_1976, Ferland_1979, Ferland_1979_b, Lawrence_1982, Hall_2007, Juodzbalis_2024, IM_2025}.
The intensity of the Ly$\beta$-fluorescence line O\,{\footnotesize I} $\lambda8446$ strongly depends on the population of the $n = 2$ level in H\,{\footnotesize I} and is commonly observed in the BLR \citep[][]{Netzer_1976}. Nevertheless, this AGN interpretation implies a non-stellar Balmer break caused by significant absorption at wavelengths just blueward of the Balmer limit \citep[rest wavelength at $3646 \ \rm{\AA}$;][]{IM_2025}, which is not observed in our spectrum (Figure \ref{fg:j0337_1and2_spectrum}). \\

\noindent
\underline{\textit{Collisional Excitations in He\,{\footnotesize I} lines}}: In addition, if we attribute the asymmetric Balmer-line profiles to the collisional excitation that populates the H\,{\footnotesize I} $n=2$ state, we would also see enhancement in He\,{\footnotesize I} $\lambda5876$ and He\,{\footnotesize I} $\lambda7065$ due to collisional excitation from the $2\,^3\!S$ state of neutral helium. 
This picture is commonly proposed to explain the observed large He\,{\footnotesize I}/H$\beta$ intensity ratio in the BLR \citep{Almog&Netzer1989}. 
However, we do not observe any broad component in our spectrum for He\,{\footnotesize I}$\,\lambda7065$, and He\,{\footnotesize I}\,$\lambda5876$ lies on the gap in the spectrograph.

To what extent will the collisional excitation affect the narrow components of neutral helium emission if they originate from the NLR? 
For a fiducial ionization parameter of $\log U=-2$ (see Appendix \ref{sec:appendix_agn_models} for the AGN models run by \texttt{CLOUDY}) and $\log \rm{(He/H)}=0.83\log \rm{(He/H)}_{\odot}$ (i.e., $Y\simeq0.25$), 
% assuming negligible He enhancement from primordial helium abundance
the \texttt{CLOUDY} model yields He\,{\footnotesize I} $\lambda5876$/H$\beta = 0.13$ for density of $n_{\rm H}=10^5 \rm~cm^{-3}$ and increased slightly to 0.15 for $n_{\rm H}=10^7 \rm~cm^{-3}$. 
Similarly, for the same density range, the He\,{\footnotesize I} $\lambda7065$/H$\beta$ ratio increased from 0.10 to 0.12, consistent with that computed in \cite{DelZanna2022}. 
% Our model result is consistent with that computed in \cite{DelZanna2022}, where the fractional enhancement in emissivity is $\sim 20\%$ as the density $n_{\rm H}$ increases by two dex for a fixed $T_e=10^4~\rm{K}$. 
% This gives a lower limit on the He\,{\footnotesize I}/H$\beta$ intensity ratio in a sense where this ratio increases as He enhancement occurs due to pollution from stars. 
As a reference, we measure He\,{\footnotesize I} $\lambda7065$/H$\beta = 0.05$ \citep[0.04 for][]{Izotov_2009} and \cite{Izotov_2009} measure He\,{\footnotesize I} $\lambda5876$/H$\beta = 0.10$. 
This implies that collisional excitation does not have a major effect on these neutral helium emission lines in SSCs 1\&2, assuming a typical NLR density ($10^5 < n_{\rm H} < 10^7~\rm cm^{-3}$). \\
% Since the emitting clouds in the BLR are optically thick to the Balmer lines, the observed red excess in this AGN interpretation may be attributed to most Balmer line photons originating in outflowing clouds on the far side of the ionizing source (i.e., most near-side photons are absorbed) \citep[see, e.g.,][]{Lawrence_1982}. 
% Here we propose that the asymmetric Balmer-line profiles are due to a P-Cygni absorption component, as also detected in $\eta$ Carinae \citep{Smith_2018}.

\noindent
\underline{\textit{Optical Forbidden Fe Lines}}: Optical [Fe\,{\footnotesize II}] emissions and high-ionization iron forbidden lines are commonly observed in Type I quasars \citep{Osterbrock1977}, and are occasionally observed in Type II (obscured) quasars \citep{VillarMartin2015}. 
\cite{RodriguezArdila} also reports the coronal lines that trace the gas outflow from a very highly ionized nuclear region of Seyfert galaxies. 
However, in Section \ref{subsec:iron_lines}, we demonstrate that our AGN models fail to reproduce the observed iron line ratios (e.g., [Fe\,{\footnotesize III}]/[Fe\,{\footnotesize II}] and [Fe\,{\footnotesize IV}]/[Fe\,{\footnotesize III}]) and significantly underestimate the luminosities of VH-ionization iron lines (e.g., [Fe\,{\footnotesize V}]).

In addition to the density and the ionization parameter (Figure \ref{fg:j0337_fe_ratios}), are these iron line ratios sensitive to the Fe abundance (if our iron abundance measurement suffers from significant systematic uncertainties)? 
To assess its impact, we explore a wide range of Fe abundance from $12 + \log (\rm{Fe/H}) \simeq 5.0$ to 7.0.
% [Fe/H]=-2.4 to -0.4.
We find that our estimates for the iron line ratios are fairly robust regardless of the iron abundance we choose for the NLR cloud.
Consequently, the inability of AGN models to reproduce the observed iron line ratios cannot be attributed to Fe abundance uncertainties.
% , and thus the LBV picture proposed here is strongly preferred.  
% Additionally, we note for our highest density model ($n_{\rm{H}} = 10^7~\rm{cm^{-3}}$), collisional de-excitation weakens the intensity of [Fe\,{\footnotesize V}] $\lambda 4227$, and the resulting [Fe\,{\footnotesize V}] $\lambda 4227/\rm H\beta$ intensity ratio is too small ($5\times10^{-5}$). 
% This is inconsistent with the observed ratio of 0.002. 
% Figure \ref{fgapp:ironfraction} shows the ionic fraction of Fe as a function of depth in the NLR cloud. 
% As the depth reaches $2\times10^{16}~\rm cm$, the ionization transitions from higher ionization stages to predominantly singly ionized Fe. 
% This transition happens at
% [Fe\,{\footnotesize II}] $\lambda 8617$

\subsection{Ly$\alpha$ Radiation Pressure}\label{subsec:lya_radiation}
We argue that Ly$\alpha$ radiation pressure may be crucial for SSCs 1\&2, where Ly$\alpha$ trapping is significant, as evidenced by the detection of the O\,{\footnotesize I} $\lambda8446$ line and the inferred $\tau_{\rm{Ly}\alpha} \sim 10^8$ from our \texttt{CLOUDY} model discussed in Section \ref{subsec:high_den}.
In this scenario, the momentum injection rate from Ly$\alpha$ would dominate over the momentum budgets from Lyman Continuum (LyC) and UV photons prior to the onset of CCSNe and be approximately one dex higher than that from CCSNe in a low-metallicity ($Z \lesssim 0.05 Z_{{\odot}}$) environment during the first 10 Myr \citep{Kimm_2018}.
 % with high neutral-hydrogen column densities (i.e., $\log N_{\rm{HI}} \ (\rm{cm^{-2}}) \gtrsim 21$)

We provide a crude comparison of momentum injection rates from Ly\(\alpha\) radiation pressure and supernova‐driven hot winds for a $\sim 3$ Myr stellar population like SSCs 1\&2.
Following \citet{CC85}, which describes a steady-state, adiabatic, spherically symmetric hot wind, we parameterize the total mass and energy injection rates as $\dot{M}_{\rm{{hot}}} = \eta_{\rm{M}} \ \rm{SFR}$ and $\dot{E}_{\rm{{hot}}} \simeq 10^{40} \ \rm{erg \ s^{-1}} \ \eta_{\rm{E}} \ \rm{SFR}/(M_{\odot} \ \rm{yr^{-1}})$, respectively. 
$\eta_{\rm{M}}$ and $\eta_{\rm{E}}$ are the mass loading factor and the thermalization efficiency factor. 
The normalization of \(\dot{E}_{\rm hot}\) assumes a simple stellar population (SSP) with a total stellar mass of $\sim 10^6 \ M_{\odot}$ and each CCSN deposits \(E_{\rm SN}=10^{51}\) erg \citep[see Appendix B of][]{Peng_2025}.
The momentum injection of hot wind, $\dot{p}_{\rm{hot}}$, can then be derived as 
\begin{align}
    \label{eq:5.6} \dot{p}_{\rm{hot}} = \left( 2 \dot{E}_{\rm{hot}} \dot{M}_{\rm{hot}} \right)^{1/2} \simeq 0.5 \ (\eta_{\rm{E}}\eta_{\rm{M}})^{1/2} \ \frac{L}{c},
\end{align}
where $L \simeq 2000 L_{\odot} M_{\ast}/M_{\odot}$ is the bolometric luminosity for a $\lesssim 4 \ \rm{Myr}$ SSP assuming a \cite{Kroupa_2001} IMF with a 300\,$M_{\odot}$ high-mass cutoff \citep{TH_2024}.
% where $L \simeq 10^{10} L_{\odot} \rm{SFR}/(M_{\odot}\rm{yr^{-1}})$ is the bolometric luminosity from continuous star formation \citep{TH_2024}. 
If we assume $(\eta_{\rm{E}} \eta_{\rm{M}})^{1/2} \sim 0.5 - 1$, similar to M82, $\dot{p}_{\rm{hot}} / (L / c) \sim 0.25 - 0.5$.

Within a dustless and static medium, the Ly$\alpha$ momentum injection rate, $\tau_{\rm{Ly}\alpha}$, can be estimated based on Equation (34) in \citet{TH_2024}:
\begin{align}
\label{eq:5.7} \dot p_{\rm{Ly}\alpha} &\simeq \tau_{\rm{eff}} \,\frac{L_{\rm{Ly}\alpha}}{c} \simeq \tau_{\rm{eff}}\, \frac{0.68\, L_{\rm{ion}}\, f_{\rm{abs}}}{c} \nonumber \\ 
&\simeq 4 \,\left(\frac{L}{c}\right)\,\frac{f_{\mathrm{abs}}\, \xi_{0.1}\, N_{\mathrm{H},21}^{1/3}\,}{T_{4}^{1/3}\,},
\end{align}
where $\tau_{\rm{eff}} \simeq 100 \, N_{\rm{H},21}^{1/3} \, T_{4}^{-1/3}$ is the Ly$\alpha$ effective optical depth \citep[equivalent to the Ly$\alpha$ force multiplication factor $M_F$;][]{Dijkstra_2008, Kimm_2018, TH_2024, Nebrin_2025} and is valid for $\tau_{\rm{Ly\alpha}} \gtrsim 10^6$. 
Here $f_{\rm{abs}} = 1 - f_{\rm{esc}}$ denotes the Ly$\alpha$ absorption fraction, and $\xi_{0.1} = L_{\rm{ion}}/0.1L \simeq 3 - 5$ for the $\sim 3\, \rm{Myr}$ stellar population, where $ L_{\rm{ion}}$ is the ionizing photon luminosity. 
Therefore, for $f_{\rm{abs}} \simeq 1$, $\log(N_{\rm H}/{\rm cm^{-2}})\simeq22$, and $T_4 \simeq 1$, $\dot p_{\rm{Ly}\alpha} / (L/c) \simeq 20 - 40$, which is a factor of $\sim 50 - 150$ larger than $\dot p_{\rm{hot}}$.

Based on our analysis in Section \ref{sec:lbv_signs}, we know that SSCs 1\&2 have efficient dust formations and the dust-formation site (CSM) might radially expand at $\sim 200 \ \rm{km \ s^{-1}}$, suggesting the assumption of a dustless and static medium might break down.
For a static, dust-filled medium where Ly$\alpha$ can be absorbed by
dust grains, Equation (35) in \citet{TH_2024} shows that $\dot p_{\rm{Ly}\alpha}$ would decrease to
\begin{align}
\label{eq:5.8} \dot p_{\rm{Ly}\alpha} \simeq 1 \,\left(\frac{L}{c}\right)\,\frac{f_{\mathrm{abs}}\, \xi_{0.1}}{f_{\rm{dg,MW}}^{1/4} \, T_{4}^{1/3}\,},
\end{align}
where $f_{\rm{dg,MW}} = f_{\rm{dg}} / 10^{-2}$ is the dust-to-gas mass ratio normalized to the Milky-Way value \citep[$f_{\rm{dg,MW}} \simeq 0.1$ for J0337-0502;][]{Hunt_2014}.
For the fiducial parameter values adopted above, $\dot p_{\rm{Ly}\alpha}$ would decrease by a factor of four to $\dot p_{\rm{Ly}\alpha} / (L/c) \simeq 5 - 10$.
Moreover, for a dust‐free, radially expanding medium with $\log(N_{\rm H}/{\rm cm^{-2}})\simeq22$, extrapolating the relation among $\tau_{\rm eff}$, $v_w$, and $N_{\rm H}$ yields $\tau_{\rm eff}\sim10$ \citep{Dijkstra_2008}, or equivalently $\dot p_{\rm{Ly}\alpha} / (L/c) \simeq 1 - 2$. Considering both effects, $\dot p_{\rm{Ly}\alpha}$ would only be a factor of a few or even comparable to $\dot p_{\rm{hot}}$ in a radially expanding, dust-filled medium. 

Leveraging the cutting-edge analytical Ly$\alpha$ radiative transfer solution by \cite{Nebrin_2025}, which incorporates major Ly$\alpha$ suppression mechanisms (e.g., continuum absorption, gas velocity gradients), we find that $M_F$ ranges from $\sim 1$ (uniform source) to $\sim 60$ (point source) given the observed nebular properties of SSCs 1\&2 (e.g., $N_{\rm H}$ and $f_{\rm{dg}}$). This estimate is roughy consistent with the range of $\tau_{\rm{eff}}$ derived from the back-of-envelope calculations above.

In general, dust grains can be destroyed by intense UV radiation \citep{Madden_2006} and shocks \citep{McKee_1989} after the onset of CCSNe. Conversely, grains can survive--or even grow--when they are embedded in $T\lesssim10^4\,$K ionized clouds entrained in the hot outflow \citep{Chen_2024}. 
To fully assess the role of Ly$\alpha$ radiation pressure in a dusty, expanding medium, high-resolution multiphase simulations that self-consistently solve the dust dynamics are required \citep[e.g.,][]{Hu_2019}.

The significant role of Ly$\alpha$ radiation pressure in SSCs 1\&2 may offer valuable insights into the ``luminosity-deficit'' problem for the VB velocity component (FWHM $\sim 600 - 2500 \ \rm{km \ s^{-1}}$), which originates from fast-outflowing galactic winds \citep{Peng_2025}.
\cite{Peng_2025} demonstrate that while supernova-driven galactic wind models \citep{Thompson_2016, Fielding_2022} can explain the $v_{\rm{max}}$ of the VB component observed in strong lines like H$\alpha$ and [O\,{\footnotesize III}] $\lambda5007$, these wind models cannot account for the observed VB luminosities. 
Since we quantitatively show that Ly$\alpha$ resonant scattering can impart substantial momentum to the gas--potentially surpassing other feedback processes (e.g., CCSNe), particularly in extremely young, metal-poor systems--it is essential to incorporate Ly$\alpha$ radiation pressure when attempting to explain the VB luminosities in systems like SSCs 1\&2.

% They argue that the additional energy could originate from stellar photoionization. 

% for SSCs 1\&2 due to the observed significant $N_{\rm{H\textsc{I}}} \sim $ $\tau_{\rm{Ly}\alpha}$ 

\section{Summary and Conclusion} \label{sec:conclusion}
Recent studies \citepalias[e.g.,][]{Hatano_2026} argue for the presence of an active massive BH in the blue compact dwarf galaxy SBS 0335-052 E based primarily on its observed NIR variability and broad H$\alpha$ wings. 
Leveraging our recent KCWI/KCRM observation, we challenge this view by proposing that SSCs 1\&2 of this galaxy show $\eta$ Carinae-like giant eruption signatures instead. 
% In this scenario, the rapid and efficient dust formation resulting from the shock interaction between the fast outflowing LBV-outburst ejecta and the pre-outburst stellar wind material that forms the CSM can account for the observed NIR variability. 
The main signatures of SSCs 1\&2 can be summarized as follows:

\begin{enumerate}[leftmargin=*] 
    \item The electron density constraint ($n_e \sim 10^6 \ \rm{cm^{-3}}$) derived from the [Fe\,{\footnotesize II}] emission-line ratio reveal a dense paritally ionized region in SSCs 1\&2.
    % The resulting shock interaction between the LBV ejecta and this CSM facilitates efficient warm dust formation, accounting for the observed NIR excess.
    This dense gas resides in a CSM formed by pre-eruption stellar winds, which contains substantial warm dust ($\sim400-500$ K) and accounts for the observed NIR excess.
    \item 
    % The detection of low-ionization [Fe\,{\footnotesize II}] and high-ionization [Fe\,{\footnotesize IV}] lines suggests that 
    The giant eruption candidate may reside in a binary system similar to $\eta$ Carinae, where a secondary hot star ($T_{\rm{eff}} \sim 3.5-4.0\times10^4$ K) might provide the ionizing photons responsible for the [Fe\,{\footnotesize IV}] emission, while the primary, cool star ($T_{\rm{eff}} \sim 1.5\times10^4$ K) sustains the [Fe\,{\footnotesize II}] emission. 
    The [Fe\,{\footnotesize IV}] emission also exhibits time variability.
    Higher-cadence observations will be necessary to confirm any periodic variability in [Fe\,{\footnotesize IV}] associated with the binary system and to distinguish it from enhanced emission driven by radiative cooling in the proposed CSM interaction during the giant-eruption phase (non-periodic).
    % as the majority of UV ionizing flux from the secondary is temporarily suppressed when the binary system moves from apastron to periastron, during which the secondary plunges deeply into the primary's dense wind.
    \item Compared to other SSCs, the $\sim 0.1 - 0.2$ dex increase in $\log \rm{N/O}$ for SSCs 1\&2 suggests enrichment by CNO-cycled material ejected during multiple giant eruptions.
    The potential iron depletion in SSCs 1\&2 ($\sim -0.3$ to $-0.1$ dex in $\log \rm{Fe/O}$ relative to other SSCs) is subject to systematic uncertainties in the adopted iron ICF; if confirmed, this depletion may indicate that CNO-cycled material facilitates the formation of abundant iron-rich dust.
    % and $\sim -0.3 - -0.1$ dex decrease in $\log \rm{Fe/O}$ 
    % and facilitates the formation of abundant ironic dust.
    \item We propose that the observed red-excess broad wings of H$\alpha$, extending from $\sim -5\,000\ \mathrm{km\,s^{-1}}$ (blue) to $\sim 10\,000\ \mathrm{km\,s^{-1}}$ (red) with a blue-to-red wing flux ratio of $0.57\pm0.10$, can be explained by Thomson scattering in the radially expanding ($v_w \sim 200\ \rm{km\,s^{-1}}$), optically thick ($\tau_e \sim 10$) CSM of size $\sim10^{17}$\,cm.
    \item We find that radiative cooling from the proposed CSM shock interaction can naturally explain the luminosities of the observed VH-ionization [Fe\,{\footnotesize V}] $\lambda4227$ emission (and potentially contribute to the extreme He\,{\footnotesize II} $\lambda4686$ emission), as well as the \textit{Chandra} observation of the luminous X-ray point source. 
    % \item The detection of O\,{\footnotesize I} $\lambda8446$ and the inferred $\tau_{\rm{Ly}\alpha} \sim 10^8$ from our best-fit \texttt{CLOUDY} model suggest that the momentum injection rate of Ly$\alpha$ radiation can dominate over other momentum budgets (e.g., LyC, UV, and CCSNe), especially in a low-metallicity ($Z \lesssim 0.05 Z_{{\odot}}$) environment during the first 10 Myr.
    \item The detection of O\,{\footnotesize I} $\lambda8446$ and the inferred $\tau_{\rm{Ly}\alpha} \sim 10^8$ from our best-fit \texttt{CLOUDY} model suggest that the momentum injection rate of Ly$\alpha$ radiation
    can dominate over other momentum budgets (e.g., LyC, UV, and CCSNe), especially in a low-metallicity ($Z \lesssim 0.05 Z_{{\odot}}$) environment during the first 10 Myr.
\end{enumerate}

% Using an analytical framework for CSM interaction \citep{Moriya_2013}, we estimate the shock radius $R_{\rm{sh}} \sim 10^{16} \ \rm{cm}$ (consistent with our \texttt{CLOUDY} modelings) and a bolometric luminosity of the shock interaction $L_{\rm{CSM}} \sim 10^{41} \ \rm{erg \ s^{-1}}$ at the observation epoch. 
% We find that the radiative cooling luminosity from $L_{\rm{CSM}}$ can naturally explain the luminosities of the observed VH-ionization [Fe\,{\footnotesize V}] $\lambda4227$ line and the \textit{Chandra} X-ray observation of the ULX. 

Besides SSCs 1\&2, the [Fe\,{\footnotesize II}] line ratio of SSC 3 also reveals a high-density PI region ($n_e\sim10^6\rm\ cm^{-3}$), although these lines are detected at $<3\sigma$. 
SSC 3 further exhibits a Wolf-Rayet blue bump \citep{Papaderos_2006, Kehrig_2018} and hosts an older stellar population ($\sim7\,$Myr), suggesting it appears to represent an analogous system to SSCs 1\&2 but is observed at a more advanced evolutionary stage. 
In this scenario, the original donor star in the SSC 3 binary may have lost its hydrogen envelope through the binary interaction, becoming a WR star at the observation epoch. At the same time, the mass gainer evolves into an LBV and could ultimately explode as a Type IIn supernova \citep{KG_1985, Smith_2017, WB_2020}.

The recently approved JWST Cycle-4 proposal CLASSY+IR (CLASSYIR; PI: Danielle Berg) will obtain MIRI spectra covering 5 to 28~$\mu$m for 31 local star-forming galaxies.  
% By combining these data with archival FUV and optical spectra of the CLASSY galaxies \citep{Berg_2022}, future studies can perform full-spectrum fitting on aperture-matched UV+optical+MIR spectra to constrain the temperature of potential warm dust and to detect IR time variability through comparison with archival IR photometry, thereby identifying systems analogous to SSCs 1\&2 in J0337-0502.
Future studies can investigate IR time variability from warm dust formation by comparing with archival IR photometry and identify high-density CSM via the [Fe\,{\footnotesize II}] emission-line ratio in extremely young ($\lesssim 3 \ \rm{Myr}$) SSCs of the CLASSY galaxies \citep{Berg_2022}, thereby revealing systems analogous to SSCs 1\&2 in J0337-0502.

% \clearpage % Forces a new page
\begin{acknowledgments}
Z.P. appreciates the insightful discussions with Adam Carnall, Zirui Chen, Jared Goldberg, Daichi Hiramatsu, Kantapon Jensangjun, John Martin, and the ENIGMA group at UC Santa Barbara and Leiden University. 
Z.P. sincerely acknowledges support for this work from  NASA FINESST (Future Investigators in NASA Earth and Space Science and Technology) grant 80NSSC23K1450.
Z.Z. acknowledges financial support from NASA FINESST grant 80NSSC22K1755.
T.W. acknowledges and thanks support by the National Science Foundation through grant PHY-2309135, and by the Gordon and Betty Moore Foundation through grant GBMF5076.
% \textbf{
The authors are grateful to the anonymous referee for providing constructive suggestions.
% }
The data presented herein were obtained at the W. M. Keck Observatory, which is operated as a scientific partnership among the California Institute of Technology, the University of California, and the National Aeronautics and Space Administration. 
The Observatory was made possible by the generous financial support of the W. M. Keck Foundation. 
The authors wish to recognize and acknowledge the very significant cultural role and reverence that the summit of Maunakea has always had within the indigenous Hawaiian community. 
We are most fortunate to have the opportunity to conduct observations from this mountain.
\end{acknowledgments}

\vspace{0.2mm}
\begin{contribution}
%%This section gives authors the space to recognize author contributions. The text inside this environment is NOT counted towards the total word quanta. At a minimum, manuscripts are expected to include this text:
Z.P. and C.M. proposed the observing run.
Z.P. reduced the data, formulated the initial research idea, performed the formal analysis and validation, and wrote the manuscript. 
C.M. helped significantly improve the writing and structure of the manuscript.
J.H. and J.F.H. contributed to discussions regarding the possibility of an AGN in this target. 
N.P. developed the \texttt{KCWIKIT} package and Z.Z. developed the \texttt{KSkyWizard} package, both of which are essential for reducing the KCWI/KCRM data.
C.H. helped fine-tune the Thomson-scattering model to simulate the asymmetric broad H$\alpha$ wings.
Y.L. drizzled the archival \textit{HST} images and created the color-composite image.
% J.Y. helped fine-tune the \texttt{BAGPIPES} model.
T.W. provided input on the validity of the stellar physics involved in this work. 
All authors discussed the results and provided comments on the manuscript.

% All authors contributed equally to the Terra Mater collaboration.

%% But authors are expected to provide more specific details, e.g. 
%%
%%SC was responsible for writing and submitting the manuscript.
%%WWM came up with the initial research concept and edited the manuscript.
%%OTS obtained the funding and edited the manuscript.
%%EBF provided the formal analysis and validation. He also edited the manuscript.
%%GEH Supervised the undergraduates, wrote the software and administers the project github and Zenodo repositories.
%%
%% Authors can use the Contributor Role Taxonomy (CRediT) at
%% for ideas on how write a good statement tailored to their needs.

\end{contribution}

\facilities{Keck:II (KCWI)}
% \facilities{Keck:II (ESI)}

%%%%% software usage
\software{{\tt\string Astropy} \citep{astropy:2022},
% {\tt\string BEAGLE} \citep{chevallard_modelling_2016},
% {\tt\string BAGPIPES} \citep{Carnall_2018,Carnall_2019},
% {\tt\string BPASS} \citep{BPASS_2018},
{\tt\string CHIANTI} \citep{chianti_v10},
{\tt\string LMFIT} \citep{lmfit},
{\tt\string Matplotlib} \citep{Matplotlib:2007},
{\tt\string NumPy} \citep{numpy},
{\tt\string pandas} \citep{reback2020pandas},
% {\tt\string PARSEC} \citep{Costa_2025},
% {\tt\string Photutils} \citep{larry_bradley_2022_6825092},
{\tt\string pyCloudy} \citep[][]{pyCloudy}, 
{\tt\string PyNeb} \citep[][]{pyneb,pyneb_2015}, 
{\tt\string PypeIt} \citep{pypeit:zenodo},
{\tt\string SciPy} \citep{2020SciPy-NMeth},
% {\tt\string statmorph} \citep{statmorph}, 
{\tt\string VerEmisFitting} \citep{veremisfit}}


\appendix 

\section{Nebular Properties of SSCs}\label{sec:appendix_ssc_neb_prop}
\begin{figure*}
\includegraphics[width=0.49\linewidth]{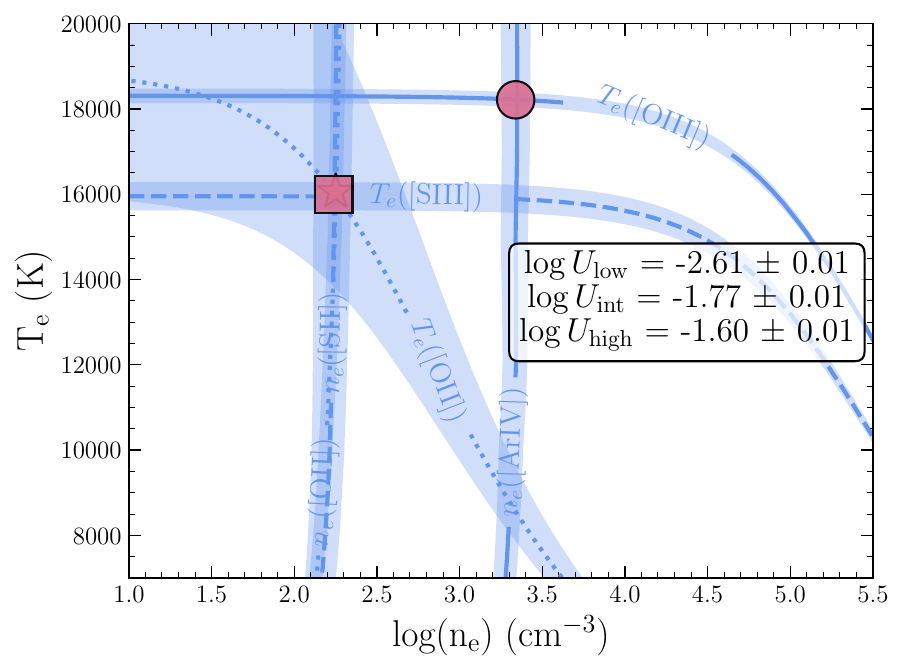}
\includegraphics[width=0.49\linewidth]{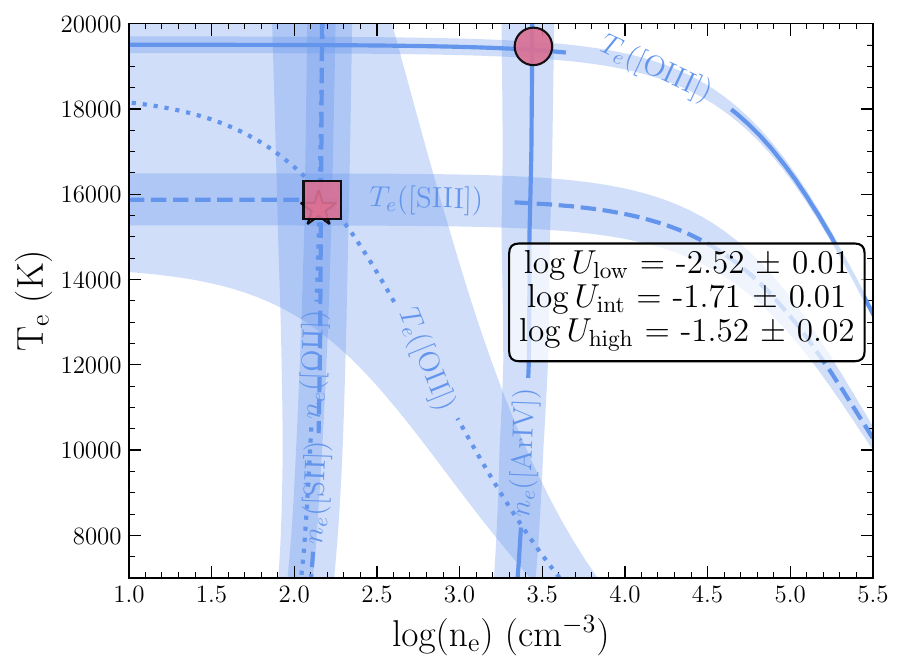}
\includegraphics[width=0.49\linewidth]{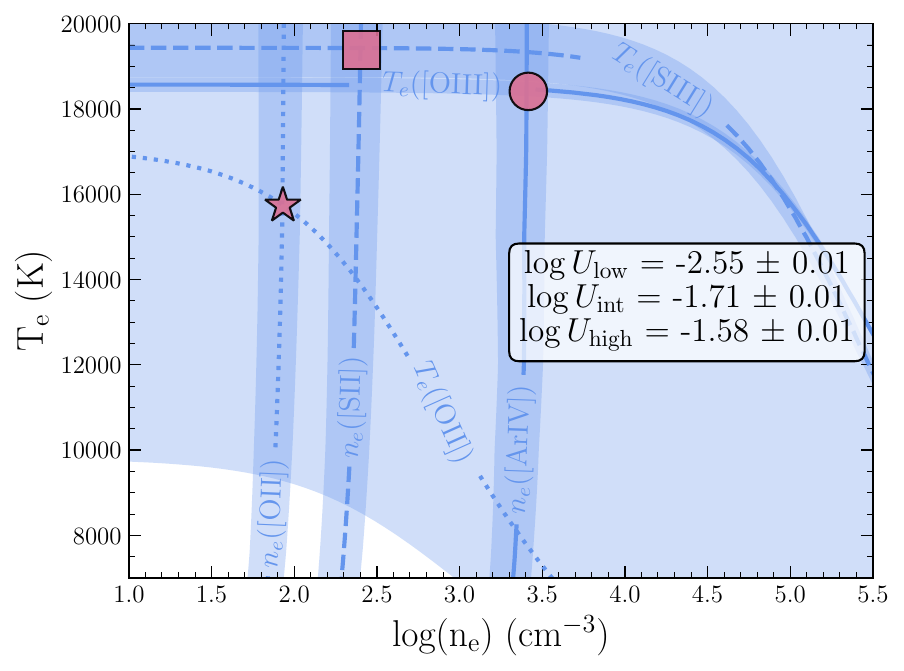}
\includegraphics[width=0.49\linewidth]{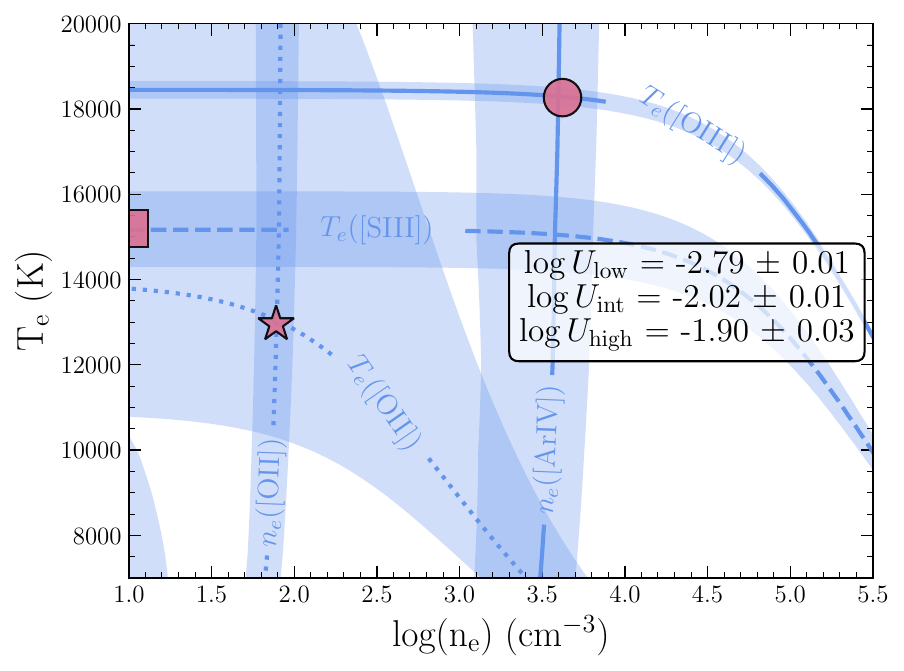}
\caption{$T_e$ and $n_e$ solutions for SSC 3 (\textit{top left}), SSCs 4\&5\&7 (\textit{top right}),
SSC 6 (\textit{bottom left}), and SSC 8 (\textit{bottom right}). 
The plotting style is identical to that of Figure \ref{figure:ne_Te_j0337_1and2}.
For SSC 8, the $n_e$ in the intermediate-ionization zone, traced by the [S\,{\footnotesize II}] $\lambda\lambda$6716, 6731 doublet, approaches its low-density limit ($10 \ \rm{cm^{-3}}$).
} 
\label{figure:ne_Te_j0337}
\end{figure*}

Table \ref{table:nebular_properties_ssc} shows SSCs' nebular properties derived from our extracted KCWI/KCRM spectra using the adopted methodologies described below.
The best-fit $n_e$, $T_e$, and $\log U$ values of low, intermediate, and high ionization zones for SSCs 3 to 8 are summarized in Figure \ref{figure:ne_Te_j0337} (those of SSCs 1\&2 are summarized in Figure \ref{figure:ne_Te_j0337_1and2}).

\subsection{Extinction}\label{subsec:extinction}
To correct for Galactic foreground extinction, we obtain $E(B-V)_{\rm{MW}} = 0.0402 \pm 0.002$ from the dust maps of \cite{schlafly_measuring_2011} and apply the correction using the Galactic extinction curve interpolated from Table 3 of \cite{Fitzpatrick_1999}. 
We then model the intrinsic dust attenuation using the SMC extinction law \citep{Gordon_2003}.
The SNR-weighted intrinsic color excess, $E(B-V)_{\rm int}$, is derived by comparing the observed recombination line ratios—specifically, $\rm H\delta/\rm H\gamma$, $\rm H9/\rm H\gamma$, $\rm H10/\rm H\gamma$, and Paschen lines Pa\,9 through Pa\,18 relative to H$\gamma$ to the theoretical Case B ratios from \citet{storey_recombination_1995}.
These chosen relatively high-order Balmer and Paschen lines are free from contamination by nearby strong lines or obvious Thomson-scattering broad wings that are seen in H$\alpha$ and H$\beta$ (Section \ref{subsec:asymm_broad_balmer}).
The measured $E(B-V)_{\rm int}$ value for SSCs 1\&2 is $0.06 \pm 0.01$, whereas the remaining SSCs are consistent with zero internal reddening (i.e., $E(B-V)_{\rm int} = 0$). 
% The zero $E(B-V)_{\rm int}$ value for other SSCs is confirmed by our full-spectrum fittings using \texttt{BAGPIPES} (see Appendix \ref{appendix:full_spec_fit} for details).

\subsection{Ionization-Zone Definitions and Diagnostics}\label{subsec:ionization_zone}
We use the \textsc{getCrossTemDen} method in \texttt{PyNeb} \citep{pyneb,pyneb_2015} to derive $n_e$ and $T_e$ for each ionization zone defined by \citetalias{berg_characterizing_2021}. 
The ions used to constrain $n_e$ and $T_e$ in each ionization zone are summarized below:  
% \underline{\textbf{Partially-ionized zones}}: 
% \renewcommand{\labelitemi}{}
% \begin{itemize}[leftmargin=*] 
% \setlength{\itemsep}{0pt}
%     \item $\sim 8 - 15 \ \rm{eV}$: $n_e$ ([Fe\,{\footnotesize II}]); $T_e$ ([Fe\,{\footnotesize II}])$_{\rm{CL}}$ 
% \end{itemize}
% \underline{\textbf{Ionization zones}}:
\renewcommand{\labelitemi}{\textbullet}
\begin{itemize}[leftmargin=*] 
\setlength{\itemsep}{0pt}
    \item Low ($14.5 - 20.6 \ \rm{eV}$): $n_e$ ([O\,{\footnotesize II}]); $T_e$ ([O\,{\footnotesize II}])
    \item Intermediate ($23.3 - 34.8 \ \rm{eV}$): $n_e$ ([S\,{\footnotesize II}]); $T_e$ ([S\,{\footnotesize III}])
    \item High ($35.1 - 54.9 \ \rm{eV}$): $n_e$ ([Ar\,{\footnotesize IV}]); $T_e$ ([O\,{\footnotesize III}])
    \item Very-High (VH; $ > 54.4 \ \rm{eV}$): $n_e$ ([Ar\,{\footnotesize IV}]); $T_e$ ([Fe\,{\footnotesize V}])$_{\rm{CL}}$
\end{itemize}
Notably, we use the ratio of the [S\,{\footnotesize II}] $\lambda\lambda$6716, 6731 doublet to estimate the electron density of the intermediate ionization zone, as the [Cl\,{\footnotesize III}] $\lambda\lambda$5517, 5537 doublet \citepalias{berg_characterizing_2021} falls outside the wavelength coverage of our KCWI/KCRM spectra (i.e., it lies too close to the dichroic).

Moreover, since the temperature-sensitive [Ne\,{\footnotesize III}] $\lambda3342$ line of the VH ionization zone lies outside our wavelength coverage, we adopt the ion-fraction-weighted temperature of [Fe\,{\footnotesize V}] from the best-fitting \texttt{CLOUDY} models \citep[][Section \ref{sec:lbv_signs}]{CLOUDY_v23} to determine $T_e$ ([Fe\,{\footnotesize V}])$_{\rm{CL}}$ (e.g., $\sim$ 23\,000 for SSCs 1\&2).

Besides these zones with ionization potential energies $\gtrsim 15$ eV, we include an additional partially ionized (PI) zone—traced by [Fe\,{\footnotesize II}] emission ($\sim 8 - 15 \ \rm{eV}$; see Figure \ref{fg:j0337_1and2_spectrum})—which originates in the densest, most shielded neutral regions near ionization and shock fronts \citep{BP_1994}.

\subsection{Line-Ratio Constraints on $\log U$}\label{subsec:logU_constraint}
We estimate the ionization parameter, $\log U$, of low-to-intermediate ($\log U_{\rm{low}}$), intermediate-to-high ($\log U_{\rm{int}}$), and high-to-very-high ionization zones ($\log U_{\rm{high}}$) using the [S\,{\footnotesize III}]/[S\,{\footnotesize II}], [O\,{\footnotesize III}]/[O\,{\footnotesize II}], and [Ar\,{\footnotesize IV}]/[Ar\,{\footnotesize III}] ratios, respectively.
These estimates are based on the polynomial relations between these line ratios and theoretical predictions of $\log U$ from \texttt{CLOUDY} models at different gas-phase metallicities, as presented in \cite{Berg_2019} and \citetalias{berg_characterizing_2021}.

The ionization-averaged $\log U$, $\log U_{\rm{ave}}$, can characterize the degree of ionization in the nebulae within each SSC. 
$\log U_{\rm{ave}}$ is obtained by weighting $\log U_{\rm{low}}$ and $\log U_{\rm{high}}$ by their respective oxygen ionization fractions measured in Appendix \ref{subsec:abundance_details}.
%  = -1.69 \pm 0.06
We derive $\log U_{\rm{ave}}$ value ranges from $\sim -1.60$ (SSCs 4\&5\&7) to $\sim -2.10$ (SSC 8).

\subsection{Ionic Abundances from Multiple Ionization Stages}\label{subsec:abundance_details}
We apply the \textsc{getIonAbundance} method in \texttt{PyNeb} to derive ionic abundances from the detected emission lines and apply ICFs to account for ionization stages that are not directly observed.

For oxygen, we assume that $\rm O/H = \left(O^{+} + O^{+2}\right) / H^{+}$ and derive O$^+$ and O$^{+2}$ ionic abundances using the [O\,{\footnotesize II}] $\lambda\lambda3726,29$ doublet and [O\,{\footnotesize III}] $\lambda4959$ emission line, because contributions from O$^0$ and O$^{+3}$ are negligible in galaxies with physical properties (e.g., $\log U_{\rm{ave}}$ and gas-phase metallicity) similar to J0337-0502 \citepalias[e.g., J1044+0353;][]{berg_characterizing_2021}.
% \begin{align}
%     \label{eq:3.1}
%     \rm{\frac{O}{H} = \frac{O^{+} + O^{+2}}{H^{+}}} 
% \end{align}

For nitrogen, we 
derive the N$^+$ ionic abundance from [N\,{\footnotesize II}]~$\lambda6583$ and 
apply 
% the N$^{+}$ ionization correction factors 
$\rm{ICF(N^{+})}$ to account for higher ionization stages as the high-ionization UV N\,{\footnotesize III}] and N\,{\footnotesize IV}] lines at $\sim1485\,$\AA\ and $\sim1750\,$\AA, respectively, lie outside the wavelength coverage of our KCWI spectrum.
% \begin{align}
%     \label{eq:3.2}
%     &\rm{\frac{N}{H}} = \rm{\frac{N^{+}}{H^{+}}} \times \rm{ICF(N^{+})}
% \end{align}
% \begin{align}
%     \label{eq:3.2}
%     &\rm{\frac{N}{H}} = \rm{\frac{N^{+}}{H^{+}}} \times \rm{ICF(N^{+})}, \nonumber \\  
%     &\rm{ICF(N^{+})} = \left\{\begin{array}{ll}
%     f(\log U, Z) \\ 
%     \rm{O}^{+} / \left(\rm{O}^{+} + \rm{O}^{+2}\right) = \rm{O}^{+} / \rm{O} \\ 
%    0.39 + 1.19 \times \left(\rm{O} / \rm{O}^{+}\right)
% \end{array} \right.
% \end{align}
We use the following three approaches to estimate $\rm{ICF(N^{+})}$: 
(1) the relation between ICF(N$^+$), $\log U$, and $Z$ derived in \citetalias{berg_characterizing_2021} (ICF(N$^+$)\,$\simeq$\,30 for $\log U_{\rm ave}=-1.67$ and $Z\simeq0.05\,Z_\odot$),
(2) the simple assumption that N$^+$ and O$^+$ share the same ionic fraction \citep[ICF(N$^+$)\, = $\rm{O}/\rm{O}^{+}$;][]{PC_1969},
and (3) the empirical fit based on Milky Way H\,{\footnotesize II} regions \citep[$\rm{ICF(N^{+})} = 0.39 + 1.19 \times \left(\rm{O} / \rm{O}^{+}\right)$;][]{Esteban_2020}.

% The derived nitrogen abundance expressed as 12 + log(N/H) for the three approaches are $6.37 \pm 0.09$, $5.99 \pm 0.12$, and $6.08 \pm 0.12$, respectively.
% The derived \(\log(\mathrm{N/O})\) values, obtained via three different ICF prescriptions, can vary by up to \(\sim0.4\) dex. 

For iron,
we determine the Fe$^+$ ionic abundance using both [Fe\,{\footnotesize II}] $\lambda5262$ and [Fe\,{\footnotesize II}] $\lambda8617$, for which Ly$\alpha$ or UV-continuum pumping effects are negligible.
Systematic uncertainties arising from the choice of [Fe\,{\footnotesize II}] lines are included in the errors listed in Table \ref{table:nebular_properties_ssc}.
We then use [Fe\,{\footnotesize III}] $\lambda4658$ and [Fe\,{\footnotesize IV}] $\lambda5234$, which are de-blended from other strong emission lines such as H$\beta$ and the [O\,{\footnotesize III}] doublet, to derive the ionic abundances of Fe$^{+2}$ and Fe$^{+3}$, respectively. 
Finally, to constrain the Fe$^{+4}$ ionic abundance, we use [Fe\,{\footnotesize V}] $\lambda4227$, which has been detected in other extreme emission-line galaxies like J1044+0353 and J1418+2102 \citepalias{berg_characterizing_2021}.
For the other SSCs without detections for all ionization stages from Fe$^{+}$ to Fe$^{+4}$, we use ICF(Fe$^{+2}$+Fe$^{+4}$) for SSCs 3 to 7 and ICF(Fe$^{+2}$) for SSC 8, which are both functions of metallicity and $\log U$ \citepalias[Figure 7 in][]{berg_characterizing_2021}.

\section{AGN Models}\label{sec:appendix_agn_models}
We run a set of \texttt{CLOUDY} models with the ionizing photons follow the SED with the ``strong UV bump'' in \cite{Nagao2006}. We adopt a wide range for the gas density, $n_{\rm{H}} = 10^5 - 10^7~\rm{cm^{-3}}$, assuming that the narrow (FWHM $\lesssim 100~\rm km~s^{-1}$) emission lines originate from the Narrow-line region (NLR) of an AGN. For the NLR cloud, we use a plane-parallel model with constant density with column density of $N_{\rm H}=10^{23}~\rm cm^{-2}$ \citepalias{Hatano_2026}. We vary the ionization parameter from $\log U=-6.0$ to $-0.5$. For consistency with previous analysis, we assume $Z=0.04 \ Z_{\odot}$ (Section \ref{subsec:abundance}) and the $\alpha$-abundance metallicity scales with the solar abundance.
% and scales as the oxygen abundance $12+\log\rm(O/H)=7.35$ derived from Section \ref{subsec:abundance}. 
% This gives $Z=0.04Z_{\odot}$ for our \texttt{CLOUDY} models.
% In addition, we use [He/H]=-0.08 and [Fe/H]=-1.75.
We also set the heavy element mass ratio and micro-turbulence velocity to zero for simplicity.

% To assess its impact, we manually change the iron abundance from [Fe/H]=-2.4 to -0.4. To first order, as we keep density and ionization parameter fixed, the iron lines [Fe\,{\footnotesize III}] $\lambda 4658$, [Fe\,{\footnotesize IV}] $\lambda 5234$, and [Fe\,{\footnotesize V}] $\lambda 4227$ increases linearly with iron abundance. However, the intensity of [Fe\,{\footnotesize II}] $\lambda 8617$ is less sensitive to the Fe abundance, and the line ratio presented in
% This discrepancy is also seen in the [O\,{\footnotesize III}] $\lambda 4959$ line, where our highest density model gives $\lambda 4959/\rm H\beta = 0.06$, largely inconsistent with the observed ratio of 1.06. This implies that these forbidden lines originate from a lower density region, so that the line intensities cannot be explained from our single zone model with high density. However, our goal is to explore the possibility of having the observed high-ionization lines produced by AGN photoionization, instead of fitting the line intensity ratios with a more sophisticated model. In reality, the emission from the NLR is the sum of emissions from many clouds with a distribution of densities and ionization parameters. We leave advanced NLR modeling for future work.

\section{Crude Estimations of $N_{\rm GE}$}\label{sec:nitrogen_enrich_lbv}

We estimate $N_{\rm GE}$ required to produce the observed nitrogen enrichment in SSCs 1\&2 using a simple order-of-magnitude argument, explicitly stating the assumptions below.
\begin{enumerate}[leftmargin=*] 
    \item Oxygen abundances remain relatively constant before the onset of CCSNe, as the newly-formed oxygen produced by CNO-cycled material from giant eruptions is negligible compared to the initial gas-phase oxygen mass (e.g., the nitrogen-to-oxygen stellar yield ratio $y_{\rm{N}} / y_{\rm{O}} \sim 50$ for a $60 \ \rm{M_{\odot}}$ star in CNO-cycle equilibrium; \citealp{Lamers_2001}).
    \item The CNO-cycled ejecta mix uniformly with the surrounding ISM within $\sim3$ Myr, which is justified by the characteristic giant eruption ejecta temperatures of $10\,000$ K and rapid mixing \citep{Kobulnicky_1997}. 
\end{enumerate}
Under these assumptions, the change in nitrogen-to-oxygen number density is approximately equal to the change in nitrogen mass density relative to hydrogen ($Z_{\rm{g, N}}$):
\begin{align}
\label{eq:4.0} 
&\Delta \log \left(\rm{N/O}\right) = \log \left(\rm{N/O} \right)_f - \log \left(\rm{N/O} \right)_i = \log R_{\rm{N/O}} \sim \nonumber \\
& \Delta \log Z_{\rm{g, N}} = \log \left(1 + \frac{y_{\rm{N}} \  M_{\rm{ej,tot}}}{M_{\rm{g,N}}}\right) = \log \left(1 + \frac{y_{\rm{N}} \ N_{\rm{GE}} \ M_{\rm{ej}}}{Z_{g, \rm{N}} \ M_{g}}\right).
\end{align}
% Here, $Z_{\rm{g, N}} = 14 \left(\rm{O/H} \right) \left(\rm{N/O} \right) \sim 7\times10^{-6}$
% Here, $Z_{\rm{g, N}}$ is the gas‐phase nitrogen mass fraction relative to hydrogen,
Here, $M_{g}$ is the total gas mass, $y_{\mathrm{N}}\sim 10^{-3}$ from the CNO cycle \citep{Johnson_1992,MM_2002, Prantzos_2018}, 
and $M_{\mathrm{ej,tot}} = N_{\rm GE}\,M_{\rm ej}$ is the cumulative ejecta mass from multiple giant eruptions. 

Assuming $\log(\mathrm{N/O})_i \simeq -1.4$ and $\log(\mathrm{N/O})_f \simeq -1.3$ (Section \ref{subsec:no_feo_analysis}),
we obtain $R_{\rm N/O} \simeq 1.25$.
Solving Equation \ref{eq:4.0} then yields
\begin{align}
\label{eq:4.1} 
N_{\rm GE} \sim \frac{(R_{\rm N/O}-1)\, Z_{\rm g,N}\, M_g}
{y_{\rm N}\, M_{\rm ej}} \sim 6 - 25,
\end{align}
where we adopt $M_{\rm ej} \sim 10-40\,M_{\odot}$ \citep{Davidson_2012} and
$Z_{\rm g,N} M_g \sim 1$ for a gas fraction $\mu \simeq 0.95$ \citep{Thompson_2009}
and $M_{\ast} \sim 10^6\,M_{\odot}$ (Table \ref{table:ssc_properties}).
We caution, however, $N_{\rm GE}$ may easily vary by a factor of a few, given that $R_{\mathrm{N/O}}$, $y_{\mathrm{N}}$, and $M_{\rm{ej}}$ are not well constrained by observations or simulations.

\section{\texorpdfstring{Broad Wings in H$\alpha$ and [O\,{\footnotesize III}] $\lambda5007$ from the ESI Spectrum of SSCs 1\&2}{Broad Wings in H-alpha and [O III] 5007 from the ESI Spectrum of SSCs 1\&2}}
\label{sec:esi_ssc1a2_o3_ha}
\begin{figure*}[!htb]
\centering
\includegraphics[width=1\linewidth]{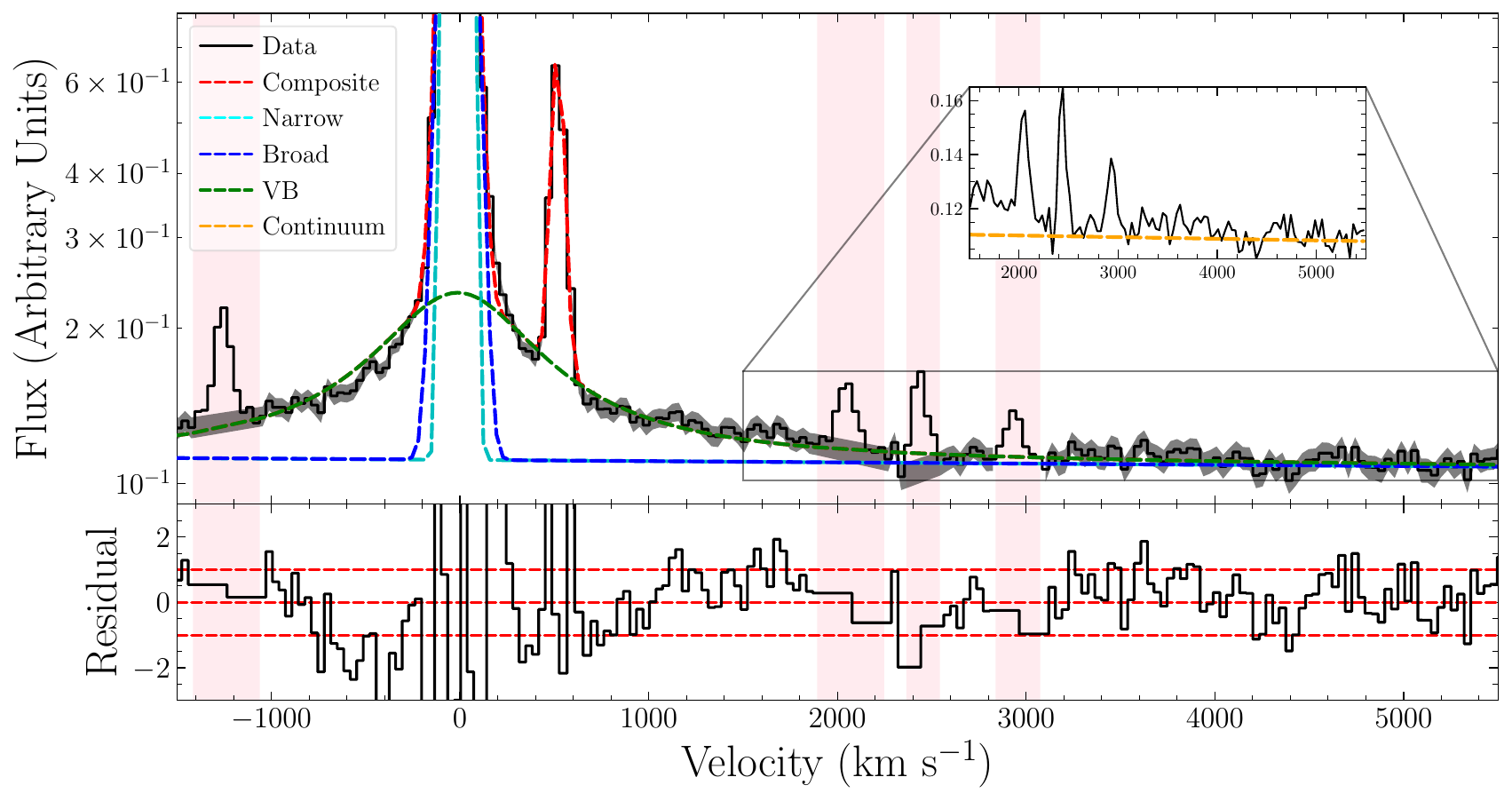}
\includegraphics[width=1\linewidth]{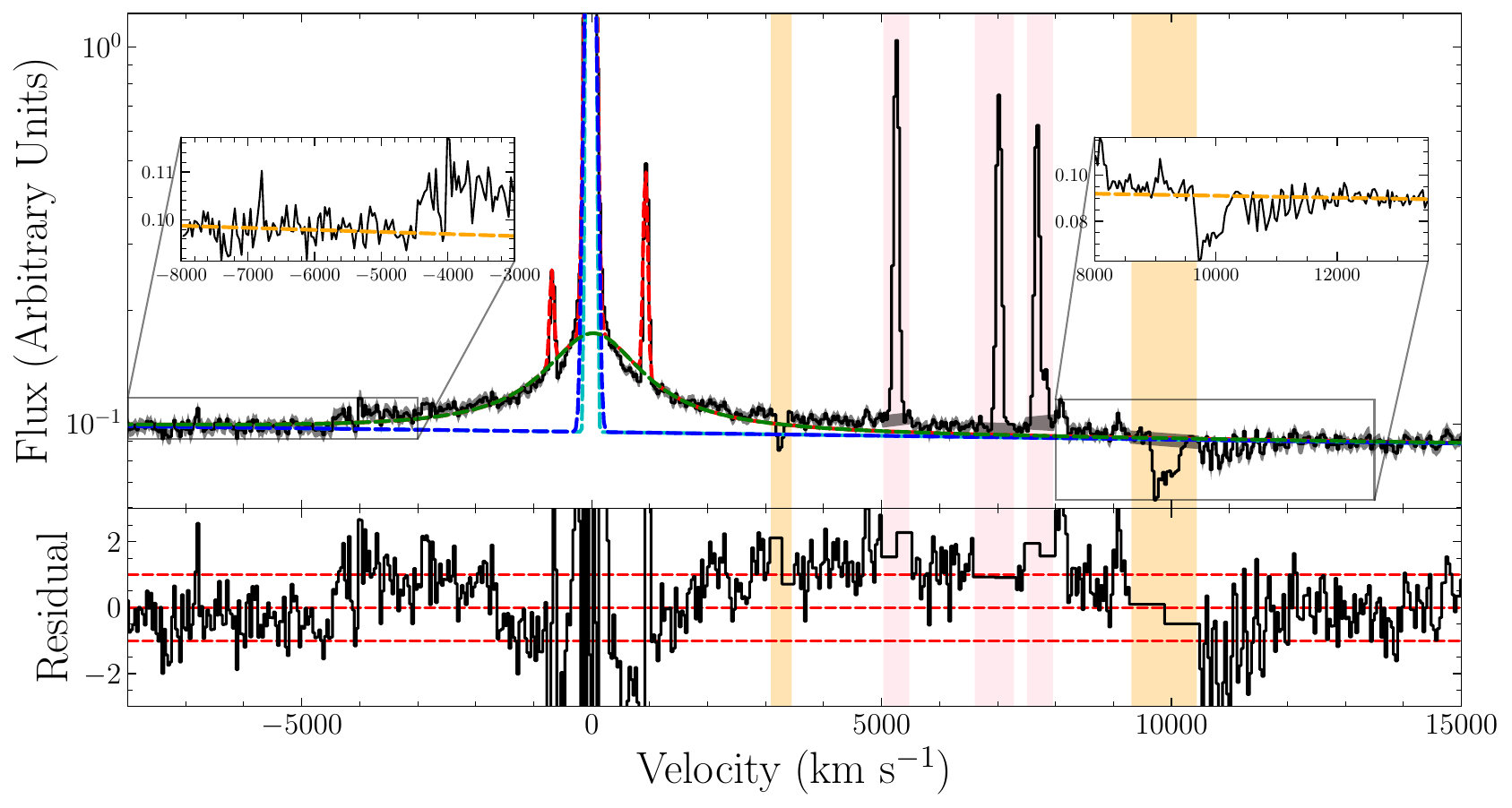}
\caption{Broad Wings in [O\,{\footnotesize III}] $\lambda5007$ (\textit{top}) and H$\alpha$ (\textit{bottom}) from the ESI Spectrum of SSCs 1\&2.
The plotting styles are consistent with those in Figure \ref{fg:j0337_asymmetric_balmer_line}.
If we perturb the [O\,{\footnotesize III}] $\lambda5007$ line profile 1\,000 times based on the error spectrum and adopting a 1$\sigma$ detection threshold yields $4\,850_{-180}^{+140} \ \rm{km\,s^{-1}}$ for the red wing (the blue wing is blended with [O\,{\footnotesize III}] $\lambda4959$).
The H$\alpha$ profile exhibits a similar $v_{\rm max}$ as that observed in our KCWI/KCRM data (see Figure \ref{fg:j0337_asymmetric_balmer_line}).
}
\label{fg:j0337_broad_wings_esi}
\end{figure*}

Figure \ref{fg:j0337_broad_wings_esi} presents the best-fit dGL models for the [O\,{\footnotesize III}] $\lambda5007$ (top) and H$\alpha$ (bottom) line profiles from the Keck/ESI spectrum \citep{Peng_2025} of SSCs 1\&2. 
% The ESI H$\alpha$ profile exhibits a similar $v_{\rm max}$ to that observed in our KCWI/KCRM data (Figure \ref{fg:j0337_asymmetric_balmer_line}), with the blue (red) wing extending to $\sim10\,000$ ($5\,000$) $\rm{km\,s^{-1}}$. 

\section{Overview of the Enhanced He\,{\footnotesize II} $\lambda4686$ Model}\label{sec:martin_2006_model}
% \textbf{
In the framework of \citet{Martin_2006_HeII}, the He\,{\footnotesize II} $\lambda4686$ emission in the He$^{++}$ zone can be strongly amplified by three coupled radiative pumping mechanisms once the He\,{\footnotesize II} $\lambda304$ transition becomes optically thick. 
First, resonant trapping of $\lambda304$ photons overpopulates the He$^{+}$ $n=2$ level, enabling efficient photoionization from $n=2$ by UV photons. 
Second, at large optical depth in the He\,{\footnotesize II} $\lambda 1215$ emission, almost every decay from $n=4$ goes through $n=3$, increasing He\,{\footnotesize II} $\lambda4686$ emission. 
Third, trapped Ly$\alpha$ photons can excite He$^{+}$ from $n=2$ to $n=4$ at $v_{\rm He\,{\footnotesize II}}\approx-120\,\rm{km\,s^{-1}}$ (this contribution is negligible at $n_e \lesssim 10^{9.5}\,\rm{cm^{-3}}$). 
% Resonant trapping of $\lambda304$ photons overpopulates the He$^{+}$ $n=2$ level, enabling efficient photoionization from $n=2$ by UV photons and thereby enhancing the effective He$^{++}$ ionization rate. 
% At large optical depth in the $2\rightarrow4$ transition, recombination cascades are forced through $n=4\rightarrow3$, increasing the fraction of recombinations that produce He\,{\footnotesize II} $\lambda4686$. 
% In addition, trapped H\,{\footnotesize I} Ly$\alpha$ photons can fluorescently excite He$^{+}$ from $n=2$ to $n=4$ at $v_{\rm He\,{\footnotesize II}}\approx-120 \, \rm{km\,s^{-1}}$, but this effect is negligible at $n_e \lesssim 10^{9.5} \, \rm{cm^{-3}}$.
We refer readers to Section 8 and Appendix B in \cite{Martin_2006_HeII} for details.
% }

% \begin{figure*}[!htb]
% \centering
% \includegraphics[width=1\linewidth]{figures/appendix/j0337_heii4686_softXray_ratio_ff_pow_Fint=1e21_v3.pdf}
% \caption{He\,{\footnotesize II} $\lambda4686$ to soft X-ray (54--550 eV) flux ratio based on the \citet{Martin_2006_HeII} models for different $A_{2c}$ (\textit{left}, with fixed $\log_{10}\eta\,(\rm s^{-1}) = -10$) and $\eta$ (\textit{right}, with fixed $\log_{10} A_{2c}\,(\rm s^{-1}) = 4$) values, assuming $F_{54-550\,\rm eV} = 10^{21}\,\rm eV\,cm^{-2}\,s^{-1}$ with a power-law (solid) or bremsstrahlung (dashed) spectrum shape (see Appendix \ref{sec:martin_2006_model} for details).
% }
% \label{fg:M06_heii_soft_xray}
% \end{figure*}

% \textbf{
The key parameters that influence the enhancement factor (i.e., $F(\rm{He\,{\footnotesize II}}\,\lambda4686)\,/\,F(54\!-\!550\,\rm{eV})$) include:
\begin{enumerate}[leftmargin=*]
    \item Assuming homologous expansion (i.e., $v\propto r$), the expansion parameter $\eta \sim v_w / R_2 \sim 10^{-10}\,\rm s^{-1}$ determines the optical depth of He\,{\footnotesize II} $\lambda304$ emission under the Sobolev approximation (i.e., $\tau_{\lambda304} \propto 1/\eta$).
    \item The corresponding incident energy flux of photons between 54 and 550 eV,
    $F_{\rm 54-550\,eV} \sim L_{\rm 54-550\,eV} / (4\pi f_{\Omega} r^2) \sim 10^{20} \,\rm eV\,cm^{-2}\,s^{-1}$, for the optically thin plasma model. Here we assume a fiducial $r$ of $0.1\,R_{\rm sh} \sim 10^{15}\,\rm cm$, corresponding to the approximate width between the CDS and the forward shock \citep{Chevalier_1994}. We define $f_\Omega \equiv \Omega / 4\pi$ as the solid-angle covering fraction of the densest component of the CSM \citep[$f_\Omega \sim 0.2$ for $\eta$ Carinae;][]{Davidson_2001, Smith_2006}.
    \item The UV photoionization rate from the $n=2$ level of He$^{+}$ ions, $A_{2c}$, which is positively correlated with the enhancement factor when $n_e \lesssim 10^{10}\,\rm cm^{-3}$. In \citet{Martin_2006_HeII}, $A_{2c}\approx10^4 \, \rm{s^{-1}}$ is determined solely by stellar UV photons from the binary system; however, in the CSM scenario considered here, the stellar contribution to $A_{2c}$ is reduced by a factor of $\sim10^{5}$ as the characteristic radius increases from $\sim2.5$ AU (for the colliding-wind case) to $R_{\rm sh}\sim10^{16}$ cm. 
    We estimate $A_{2c}$ arising from X-ray photoionization associated with the CSM interaction by running \texttt{CLOUDY} models with a bremsstrahlung (optically thin plasma) incident spectrum, based on the fixed-$N_{\rm H}$ model of \citet{Thuan_2004}. We assume $r_{\rm in} = 10^{15}\,\rm cm$, a $r^{-2}$ density profile (Equation \ref{eq:5.0}) with an initial density of $10^{9}\,\rm cm^{-3}$ (corresponding to $\tau_{e,\,\rm max} \sim 10$), a total luminosity of $10^{41}\,\rm erg\,s^{-1}$ (Section \ref{subsubsec:csm_radius_lum_mass}), and a CNO-cycled abundance pattern. 
    We find that $A_{2c} \sim 10$ within the He$^{++}$ zone.
\end{enumerate}
% }

% \textbf{
% Figure \ref{fg:M06_heii_soft_xray} shows a representative set of \citet{Martin_2006_HeII} models for different values of $A_{2c}$ and $\eta$. 
% To reproduce the observed He\,{\footnotesize II} $\lambda4686$ luminosity, $F(\rm He\,{\footnotesize II}\,\lambda4686)/F(54-550\,\rm eV)$ must be $\sim0.07$ ($\sim0.5$) for the power-law (optically thin plasma) model. 
% These values exceed the model predictions by factors of $\sim3$ ($\sim10$) given the estimated parameter values illustrated in Appendix \ref{sec:martin_2006_model}.
% assuming $\eta \sim 10^{-10}\,\rm s^{-1}$ and $A_{2c} \sim 10$, as illustrated above.
% }

% \textbf{
% Figure \ref{fg:M06_heii_soft_xray} shows a representative set of \citet{Martin_2006_HeII} models for different values of $A_{2c}$ and $\eta$. To reproduce the observed He\,{\footnotesize II} $\lambda4686$ luminosity, $F(\rm He\,{\footnotesize II}\,\lambda4686)/F(54$–$550\,\rm eV)$ must be $\sim0.07$ ($\sim0.5$) for the power-law (optically thin plasma) model. 
% These values exceed the model predictions by factors of $\sim3$ ($\sim10$), assuming $\eta \sim 10^{-10}\,\rm s^{-1}$ and $A_{2c} \sim 10$, as illustrated above.
% }

\bibliography{sample7}{}
\bibliographystyle{aasjournal}


\end{CJK*}
\end{document}